\documentclass[notitlepage, onecolumn,
superscriptaddress,
 amsmath,amssymb,
floatfix,
]{revtex4-2}

\usepackage{graphicx, overpic}
\usepackage{dcolumn}
\usepackage{bm, color,colortbl,xcolor,cancel}
\usepackage[normalem]{ulem}
\usepackage{amsmath}
\usepackage{amsthm}
\usepackage{nameref}
\usepackage[hidelinks]{hyperref}
\usepackage{setspace}
\newcommand{\ket}[1]{\vert#1\rangle}
\newcommand{\bra}[1]{\langle #1\vert}
\newcommand{\norm}[1]{\|{#1}\|}

\newcommand{\Eq}[1]{Eq.~(\ref{#1})}

\newcommand{\Fig}[1]{Figure~\ref{#1}}

\newcommand{\ddt}[1]{\ensuremath{\frac{\partial #1}{\partial t}}}
\newcommand{\trans}{\text{T}}
\newcommand{\IdQ}{\ensuremath{\mathbb{I}_{Q\times Q}}}
\newcommand{\IdnQ}{\ensuremath{\mathbb{I}_{nQ\times nQ}}}

\newcommand{\feq}{\ensuremath{f^\text{eq}}}
\newcommand{\fnm}[1]{#1}
\newcommand{\sur}[1]{#1}
\renewcommand{\vec}[1]{\ensuremath{\mathbf{#1}}}
\renewcommand{\bm}[1]{\vec{#1}}
\renewcommand{\O}{\ensuremath{{O}}}
\newcommand{\poly}{\ensuremath{\operatorname{poly}}}
\newcommand{\kn}{\ensuremath{\text{Kn}}}
\newcommand{\ma}{\ensuremath{\text{Ma}}}
\newcommand{\re}{\ensuremath{\text{Re}}}

\newcommand{\jccomment}[1]{} 

\usepackage{soul}

\newif\ifarXiv
\arXivtrue

\newif\ifKrovi
\Krovitrue

\graphicspath{{./fig/}{./QLBM_heatDiffusion/fig/}}

\begin{document}

\title{Potential quantum advantage for simulation of fluid dynamics}

\author{\fnm{Xiangyu} \sur{Li}}\email{xiangyu.li@pnnl.gov}

\affiliation{{Pacific Northwest National Laboratory}, {Richland}, {WA}, {USA}}

\author{\fnm{Xiaolong} \sur{Yin}}
\affiliation{{Petroleum Engineering}, {Colorado School of Mines},
{Golden}, {CO}, {USA}}

\author{\fnm{Nathan} \sur{Wiebe}}
\affiliation{{Department of Computer Science}, {University of Toronto}, {{Toronto}, {ON}, {Canada}}}
\affiliation{{Pacific Northwest National Laboratory}, {Richland}, {WA}, {USA}}
\affiliation{{Canadian Institute for Advanced Research}, {Toronto}, {ON}, {Canada}}

\author{\fnm{Jaehun} \sur{Chun}}
\affiliation{{Pacific Northwest National Laboratory}, {Richland}, {WA}, {USA}}
\affiliation{{Levich Institute and Department of Chemical Engineering, CUNY City College of New York}, 
  {New York}, {NY}, {USA}}

\author{\fnm{Gregory
  K.} \sur{Schenter}}
\affiliation{{Pacific Northwest National Laboratory}, {Richland}, {WA}, {USA}}
  
\author{\fnm{Margaret S.} \sur{Cheung}}
\affiliation{{Pacific Northwest National Laboratory}, {Richland}, {WA}, {USA}}
\affiliation{{Department of Physics}, {University of Washington}, {{Seattle}, {WA}, {USA}}}

\author{\fnm{Johannes} \sur{M\"ulmenst\"adt}}\email{johannes.muelmenstaedt@pnnl.gov}
\affiliation{{Pacific Northwest National Laboratory}, {Richland}, {WA}, {USA}}

\begin{abstract}
  \ifarXiv
Numerical
simulation of turbulent fluid dynamics
\fi
needs to either parameterize turbulence---which
introduces large uncertainties---or explicitly resolve the smallest
scales---which is prohibitively expensive.
Here we provide evidence through analytic bounds and numerical studies that a
potential quantum exponential speedup can be achieved to
simulate the Navier--Stokes equations governing turbulence using quantum computing. Specifically, we provide a
formulation of the lattice Boltzmann equation for which we give evidence that
low-order Carleman linearization is much more accurate than previously believed
for these systems and that for computationally interesting examples.  This is
achieved via a combination of reformulating the nonlinearity and accurately
linearizing the dynamical equations, effectively trading nonlinearity for
additional degrees of freedom that add negligible expense in the quantum solver.
Based on this we apply a quantum algorithm for simulating the Carleman-linearized
lattice Boltzmann equation and provide evidence that its cost scales
logarithmically with system size, compared to polynomial scaling
in the best known classical algorithms.  This work suggests that an exponential
quantum advantage may exist for simulating fluid dynamics,
paving the way for simulating nonlinear multiscale transport phenomena in a wide range of disciplines using quantum computing.  

\end{abstract}

\maketitle

\section{Introduction}

\ifarXiv
Complex systems exhibiting multiscale dynamics are ubiquitous in
nature. Developing the ability to understand, predict, or modify these systems
ranks among the ``grand challenges'' of science and engineering
\citep{Omenn2006,WCRP,NASEM2016}. 
Areas of application for this knowledge
include the Earth's ocean and atmosphere
\cite{Hasselmann1976,Majda2003,Bony2015,Ghil2020,Gupta2022}, 
biological
systems \cite{Bernaschi2019}, and synthesis of advanced materials
\cite{Boles2016,Salzmann2021,DeYoreo2015,DeYoreo2022,Russel}.  
The dynamics of these systems admits no analytic solution, but numerical
simulation needs to either parameterize the small-scale processes---which
introduces large uncertainties---or explicitly resolve the smallest
scales---which is prohibitively expensive. 
\fi

Turbulence as an example of nonlinear fluid dynamics embodies a defining feature of complex systems: the
lack of scale separation \cite{pope2000turbulent, Bodenschatz2010}. As all
scales interact with each other, they cannot be investigated individually.
Nonlinear fluid dynamics contribute to the multiscale behavior of many of the complex systems
mentioned above, including geophysical  \cite{Sherwood2014,Bretherton2015,Donner2016}, astrophysical, and many engineering
fluid dynamics applications.  Because the length and time
scales of the constituent processes can be a minute fraction of the integral
scale of the system (e.g., the turbulent length scale range is $\sim 10^{7}$ for
atmospheric flows), outright scale range-resolving simulations are simply not
feasible on classical computers \cite{Exascale2017,Schneider2017ClimateClouds}.  

Attempts to understand nonlinear fluid dynamics have centered on the Navier--Stokes equations
(NSE).  These equations describe the macroscopic momentum and density evolution
of fluids; multiscale interactions arise from the nonlinear advection term
$\vec{u}\cdot\nabla\vec{u}$ for the fluid velocity field $\vec{u(\vec{x},t)}$
in space $\vec{x}$ and time $t$ (see \Eq{turb} in Methods).  Approaches to solving the NSE
numerically include using ``reduced'' models that are based on
a proper orthogonal decomposition with key representative modes to mimic the
dynamics of the full spectrum \citep{berkooz1993proper,podvin2017few} and averaging over those scales not explicitly
resolved (e.g., with Reynolds averaging) and using a constitutive relation (e.g., eddy
viscosity) to close the Reynolds-averaged NSE
\citep{tennekes1972first, Durbin18}. Such
approaches parameterize the small-scale behavior based on the resolved-scale
state of the model, which produces tractable simulations at the expense of an
artificial scale separation.  This unavoidable model defect results in
persistent uncertainty in projections
of complex systems' emergent behaviors
\cite{Artime2022,Hasselmann1976,Feingold2016}, e.g., the sensitivity of the
climate to anthropogenic perturbations \cite{Wheatcraft1991,Carslaw2018,Muelmenstaedt2018,Ge2019,Sherwood2020,Bellouin2020,Moum2021}. 

A structurally different formulation of fluid dynamics exists in the form of the
Boltzmann equation.  This formulation describes the evolution of particle
distribution functions $f(\vec{x},\vec{v},t)$ in space and particle velocity
$\vec{v}$ (see \Eq{eq:LBMs} and \Eq{eq:feq} in Methods).  These distribution functions are mesoscopic objects that comprise sufficiently large
numbers of molecules for a statistical-mechanics treatment but that are smaller
than the macroscopic fluid elements.  The Boltzmann equation is still nonlinear (see
below), but in an important difference from the NSE, its
nonlinearity does not reside in the advection term \cite{Chen1998LatticeFlows}.
In classical time-marching spectral or spatial discretization algorithms, the
Boltzmann equation is even more prohibitively expensive than the NSE
\cite{Orszag1986}; in quantum computing algorithms, as we will show, the
trade-off goes the other way, largely due to the linear
Boltzmann advection term.

\leavevmode
\begin{figure}[t!]\begin{center}
\begin{overpic}[scale=0.17]{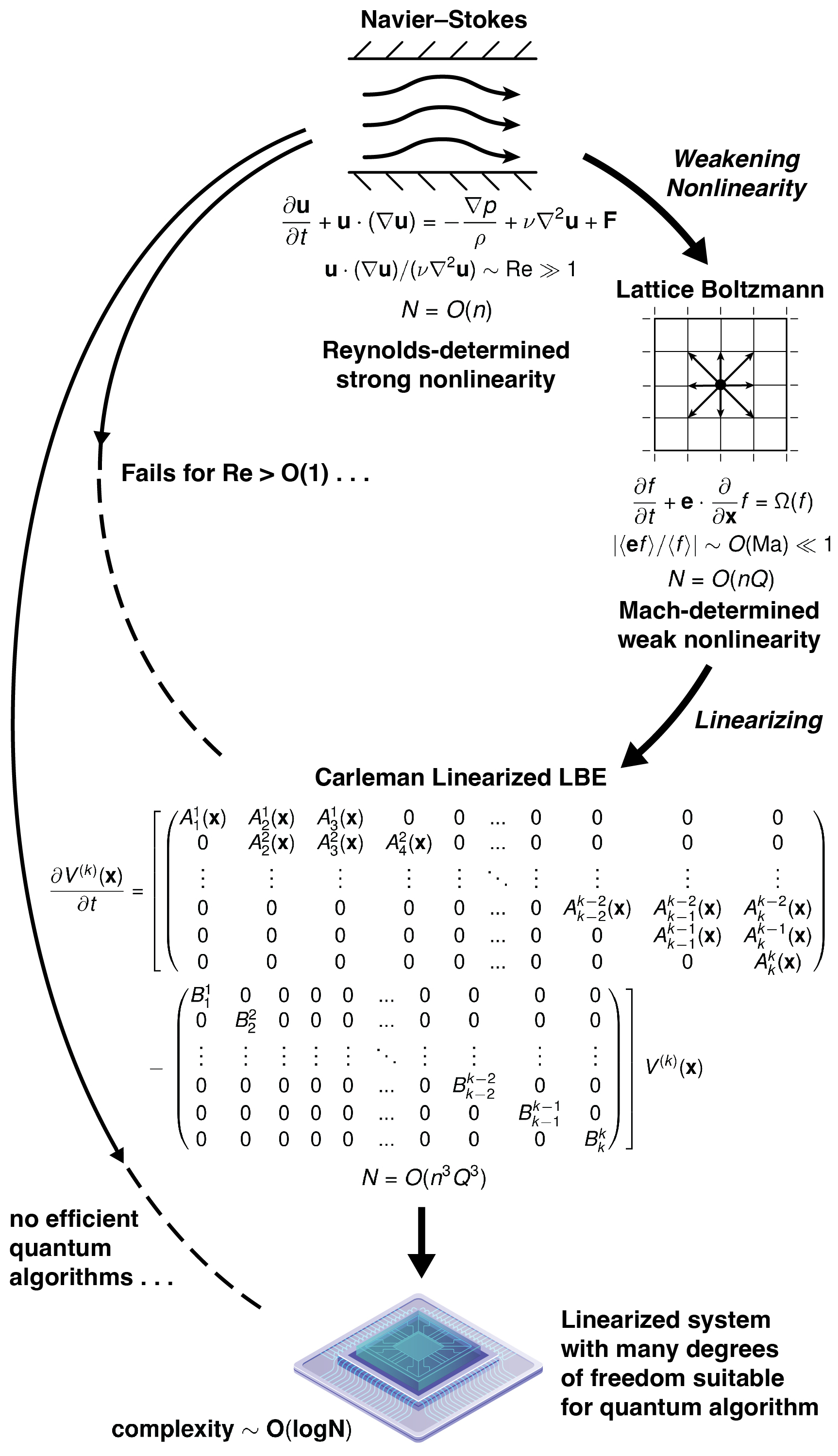}\put(20,97){(a)}\put(49,91){(b)}\put(30,49){(c)}\put(20,15){(d)}\end{overpic}\end{center}
\caption{Illustration of the complexity of solving Navier--Stokes equations
using quantum algorithms: (a) Reynolds number-determined strongly nonlinear
Navier--Stokes equations; (b) Mach number-determined weak nonlinear
lattice Boltzmann form of Navier--Stokes equations; (c) Carleman-linearized lattice Boltzmann equation; (d) exponential quantum advantage in solving the Carleman-linearized lattice Boltzmann equation.
}
\label{PCSD_0590_ILLUS.3}
\end{figure}
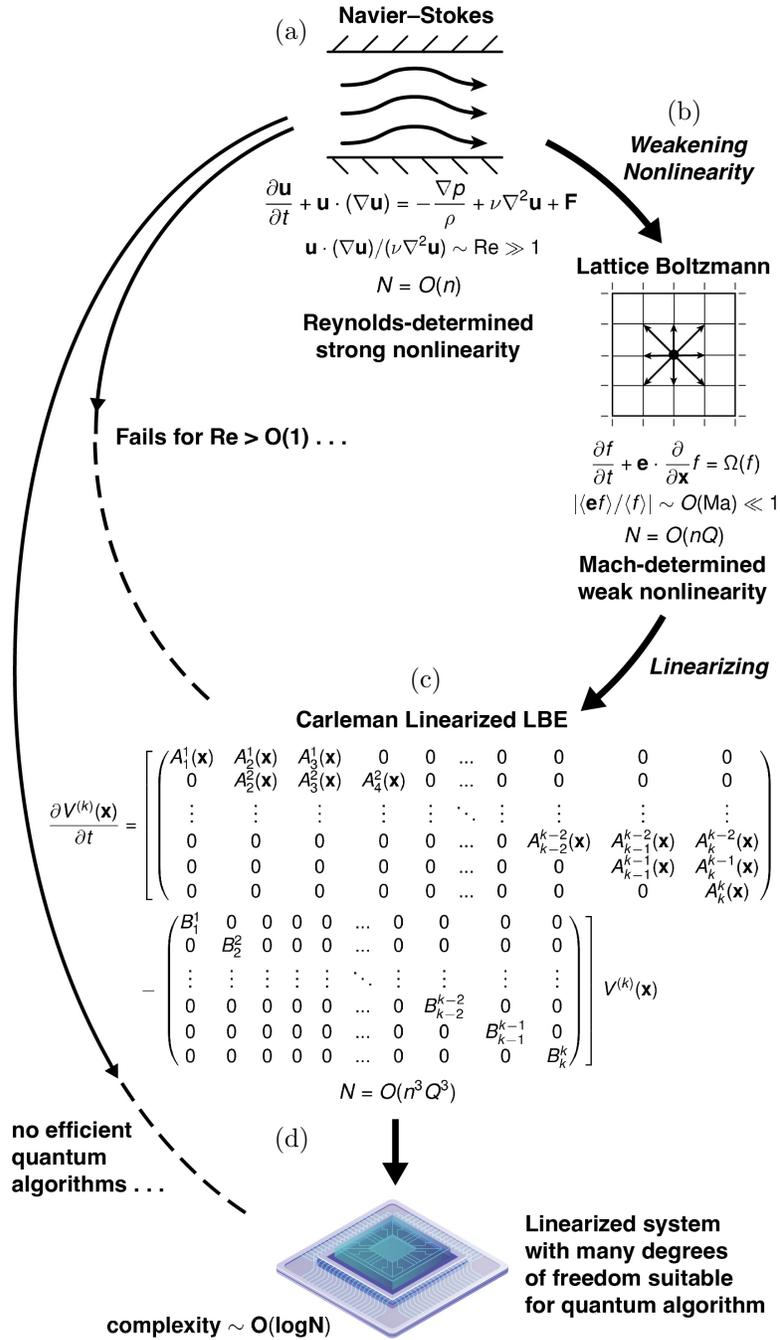

\subsection{Requirements for Quantum Algorithm}
Whether the NSE or Boltzmann equation is chosen, explicit simulation of
fluid dynamics requires large numbers of degrees of freedom $N$.  In both spatial
and spectral discretization, $N$ reflects the ratio between the largest
and smallest scales, i.e., the integral system scale and the Kolmogorov scale.  
As quantum computers inherently grant the ability to manipulate vectors in
exponentially large vector spaces in polynomial time, one may ask whether such
methods could be used to solve fluid dynamics.

Harrow, Hassidim, and Lloyd~\cite{Harrow09} answer this question in
the affirmative for generic linear systems.  This vector manipulation
approach naturally leads to a strategy for achieving a potential
exponential speedup when solving systems of coupled ordinary differential equations~\cite{berry2014high, Berry2017QuantumPrecision}
\ifarXiv
.
Specifically, these quantum linear system  algorithms can
yield a state $\ket{x}$ for any $\ket{b}$ solved for a sparse
Hermitian matrix $A$ such that the equation\begin{equation}
A\ket{x} = \ket{b} \end{equation}
is satisfied up to a constant of proportionality.  The method works by constructing a unitary matrix that encodes $A$ within a block of the original matrix, specifically, $(\bra{0}\otimes I) U_{A^{-1}} (\ket{0} \otimes I) \ket{0}\ket{b} \propto A^{-1} \ket{b}$, with $I$ being the identity matrix. 
 Such a representation is known as a block encoding.  The performance cost scales as $O(d\kappa\log(1/\epsilon))$ accesses to the matrix elements of the $d$-sparse matrix $A$, which we further assume is invertible and has condition number $\kappa$ and desired solution error $\epsilon$~\cite{childs2017quantum}.

The simplest strategy that can be employed for solving systems of differential equations involves using a forward-Euler approach to discretize a differential equation of the form $\partial_t x(t) = A x(t) + b(t)$ and then solve the resulting equation using the quantum linear systems algorithm.  The resulting difference equation takes the form $x(t_{i+1}) = x(t_i) + h Ax(t_i) +hb(t)$ for stepsize $h$.  The central idea behind this approach is to construct a quantum state over both the solution space and the time that the system is evaluated at.  Specifically, if we let our solution be $x(t)$, then we encode our solution as $\sum_i c_i \ket{t_i} \ket{\psi_i}$, where $\ket{\psi_i}$ is the solution at time $t_i$, and $c_i$ is arbitrary constants.  Given a state of this form, we can find the solution state by measuring the $t_i$ register (or using amplitude amplification) to achieve a value that is $t_f$, which is the final time desired for the algorithm.  In practice, the probability of measuring this result is low, so the standard approach to this problem is to extend (typically double) the simulation time but turn off the differential equation to ensure that the solutions are merely copied on all subsequent times.  Since the solution is the same at all such times, this serves to raise the probability of success to a constant without requiring substantial computational overhead.
Specifically,  the solution for a two-time-step result with two further time steps used for padding then reads~\cite{berry2014high}
\begin{equation}
\begin{bmatrix} I &0 &0 &0 &0\\ 
-(I + hA) & I &0&0&0\\
0 &-(I + hA) & I &0&0\\
0&0&-I&I&0\\
0&0&0&-I&I\end{bmatrix} \begin{bmatrix} x(0) \\ x(h) \\x(2h) \\ x(3h) \\ x(4h) \end{bmatrix} = \begin{bmatrix} x_{\text{init}} \\ bh \\ bh \\ 0 \\0 \end{bmatrix}
\end{equation}
In this form, the solution vector over all times can be found by inverting the above matrix.  In practice, this approach is not favorable in classical computing, as it requires a substantial overhead due to the dimension of the space. But, as the cost of the quantum linear systems algorithms does not directly depend on the dimension, this approach can be surprisingly effective in quantum settings.

\fi
Taking the algorithm of \citet{Berry2017QuantumPrecision} as an example, the
overall algorithm complexity, as measured by the number of two-qubit quantum gates and oracle queries, depends on $\log N$, but also on numerous other
properties of the system of equations being solved (expressed through the
$N\times N$ coefficient matrix $\mathcal C$):
\begin{equation}
  \label{eq:ca}
  \text{gate complexity}=O(\Vert \mathcal C\Vert \kappa_J g T s\cdot \text{poly}(\log{(\kappa_J s g \beta T \Vert \mathcal C \Vert N/\epsilon)})).
\end{equation}
These properties include the spectral norm of the coefficient matrix $\|\mathcal C\|$, the condition
number $\kappa_J$ of the eigenvectors of $\mathcal C$, the dissipation parameter $g$, the
evolution time $T$, the sparsity $s$, the norm of the initial state $\beta$, and
the desired solution error $\epsilon$.  If any of $\|\mathcal C\|$, $\kappa_J$, $g$, $T$,
or $s$ has implicit
polynomial (or worse) dependence on $N$, or if
$\beta$ or $\epsilon$ has exponential (or worse) dependence on $N$, the
headline exponential advantage is negated.  

When the underlying system of equations is
nonlinear, the requirements for efficient
solution become more stringent still.  
The complexity of the best known quantum algorithm  that can be directly applicable to nonlinear differential
equations
is exponential in the evolution time $T$ \citep{leyton2008quantum}.
Alternatively, the nonlinear system
can first be approximated by a (larger)
system of linear equations
\cite{Lloyd2020QuantumEquations,Liu21}.
Whether the quantum algorithm is still
efficient after this approximation depends on
the degree of nonlinearity \cite{Liu21}.  The
nonlinearity of NSE is characterized by the
Reynolds number (see Methods).  Quantum
algorithms are unable to simulate NSE
\citep{Liu21} with $R \ge 1$\footnote{$R$ is a parameter characterizing the ratio of the nonlinearity and forcing to the linear dissipation and is analogous to the Reynolds number}, which
is far from the values in many relevant cases
(e.g., atmospheric turbulence
\citep{Siebert06,
Grabowski2013GrowthEnvironment} with $\re
\approx 10^7$). 

We will now analyze whether the Boltzmann form of fluid dynamics, after
linearization, can take advantage of quantum algorithms' efficient handling of large $N$
without violating these algorithms' strict requirements.
Our aim in this manuscript is not quantum algorithm development but, rather, 
showing that nonlinear fluid dynamics is tractable with existing quantum algorithms,
provided the Boltzmann formulation, rather than the NSE formulation, is used.

\section{Methods}

\subsection{Navier--Stokes equations and their lattice Boltzmann form}

The Navier--Stokes equations (NSE) governing fluid dynamics read
\begin{equation}
  \frac{\partial \bm{u}}{\partial t} + \bm{u}\cdot(\bm{\nabla u})  =
  -\frac{{\bm{\nabla}} p}{\rho}
  + \nu \nabla^2 \bm{u},
\label{turb}
\end{equation}
\begin{equation}
  \label{eq:continuity}
  \frac{\partial\rho}{\partial t} + {\bm{\nabla}}\cdot(\rho\bm{u}) = 0 ,
\end{equation}
where $\bm{u}$ is the flow velocity,  
$p$ is the pressure, $\rho$ is the fluid
density, and $\nu$ is the kinematic viscosity. 
The nonlinearity of \Eq{turb} is exhibited in  $\bm{u}\cdot(\bm{\nabla u})$.
This term and the viscosity term $\nu \nabla^2\bm{ u}$ together determine the multiscale property of nonlinear fluid dynamics characterized by the Reynolds number $\text{Re} =
\bm{u}\cdot(\bm{\nabla u})/(\nu \nabla^2\bm{u})$.
Turbulent flow systems in nature are often characterized by high
$\text{Re}$ (e.g., $\text{Re}\approx 10^7$ for atmospheric flows and
$\text{Re}\approx 10^{21}$ for astrophysical flows), which are intractable
to simulate using classical computers, as the best-case computational cost
scales as $\text{Re}^3$ in space and time according to the Kolmogorov
theory\citep{pope2000turbulent}. The compressibility of the flow is
characterized by the Mach number $\text{Ma}=\vert \bm{u} \vert/c_s$ with $c_s$
being the speed of sound.

\ifarXiv
We
\fi
solve the lattice Boltzmann
  form of the NSE to reduce the nonlinearity from
  Re-determined $\bm{u}\cdot(\bm{\nabla u})$ to Ma-determined
  $u^2=\vec{u}\cdot\vec{u}$. The discrete-velocity Boltzmann form of \Eq{turb} with Bhatnagar--Gross--Krook (BGK)
collision \cite{Bhatnagar1954ASystems} leads to:
\begin{equation}
\label{eq:lbm_origin}
\frac{\partial\tilde{f}_m}{\partial\tilde t} +\tilde{\vec{e}}_m \cdot
\tilde{\bm{\nabla}} \tilde{f}_m = -\frac{1}{\tilde\tau}
\left(\tilde{f}_m-\tilde{f}_m^{\rm eq}\right),\quad m \in 1, \dots, Q,
\end{equation}
where $\tilde{f}_m(\tilde{\bm{x}}, \tilde{t})$ is the particle velocity distribution function
describing the probability density of finding a particle around position
$\tilde{\bm{x}}$ at time $\tilde{t}$ with unit discrete velocity
$\tilde{\bm{e}}_m$, and $Q$ is the number of discrete velocities
\citep{Chen1998LatticeFlows}.  If a spatial discretization uses $n$
grid points, the dimensionality of $f$ is $n\times Q$; as we will
discuss below, $Q$ is a fixed, small number ($Q\leq 27$ in widespread
practice), whereas simulations with large domains or fine resolutions
are often desirable, resulting in $n\gg 1$. In our complexity analysis, we
therefore treat $Q$ as a constant and $n$ as a parameter of the
problem whose effects in the large-$n$ limit are of particular
practical interest.
In one, two, and three dimensions, the square or cubic lattices D1Q3,
D2Q9, and D3Q27 are in widespread use \cite{Chen1998LatticeFlows}.
$\tilde{f}^{\rm{eq}}_{m}$ is the local Maxwell equilibrium
that $\tilde{f}_m(\bm{x}, t)$ relaxes to at a single relaxation time of $\tilde{\tau}$. 
$\tilde{\bm{\nabla}}$ represents the spatial gradient
in the $\tilde{x}$, $\tilde{y}$, and $\tilde{z}$ directions. By using a reference
density $\rho_r$ and speed $e_r$, characteristic macroscopic length scale $L$,
and particle collision $t_c$, one obtains the nondimensional form of
\Eq{eq:lbm_origin}, termed the lattice Boltzmann equation (LBE):
\begin{equation}
\partial_t f_m+\bm{e}_m \cdot \bm{\nabla} f_m = -\frac{1}{\tau\, \kn} (f_m-f_m^{\rm eq}), \quad m = 1, \dots, Q,
\label{eq:LBMs}
\end{equation}
where
\begin{align}
  f_m &= \tilde{f}_m/\rho_r, \\
  \feq_m &= \tilde{f}^{\rm{eq}}_m/\rho_r,  \\
  \label{eq:scale-t}
  t &= \tilde{t}/(L/e_r), \\
  \vec{e} &= \tilde{\vec{e}}/e_r, \\
  \label{eq:scale-x}
  \nabla &= L\tilde\nabla, \text{ and} \\
  \tau &= \tilde\tau/t_c. 
\end{align}
(To minimize clutter in the following, variables without tilde are
nondimensional, and variables with tilde are their dimensional physical counterparts.)
The Knudsen number describing the ratio of collision time to flow time is defined as 
\begin{equation}
  \label{eq:epsilon}
  \kn=e_r \frac{t_c}L 
\end{equation}
 and forms the basis for the Chapman--Enskog expansion
\cite{Chen1998LatticeFlows} of the LBE, which recovers the
NSE in the limit $\kn\ll 1$.
The macroscopic fluid density $\rho(\bm{x}, t)$ is retrieved as
\begin{equation}
    \rho(\bm{x}, t)=\sum_{m=1}^Q f_m(\bm{x},t)
\label{eq:rho}
\end{equation}
and the macroscopic fluid velocity as 
\begin{equation}
    \bm{u}(\bm{x}, t)=\frac 1 \rho \sum_{m=1}^Q f_m(\bm{x},t)\bm{e}_m.
\label{eq:u}
\end{equation}
To recover the NSE from \Eq{eq:LBMs}, the equilibrium
distribution function $f_m(\bm{x}, t)$ is chosen as the Taylor
expansion of the Maxwell distribution function:
\begin{equation}
\feq_m = \rho w_m[a+b{\bm e}_m\cdot {\bm u}+c({\bm e_m}\cdot{\bm u})^2+d\vert{\bm u}\vert^2];
\label{eq:feq}
\end{equation}
quadratic degree enables the LBE to reproduce the quadratic NSE to $O(\ma^2)$, and the
constants $a=1$, $b=3$, $c=9/2$, and $d=-3/2$ are the Taylor coefficients \cite{Chen1998LatticeFlows}.

Constants $a$ through $d$ are all $O(1)$ quantities as, in the lattice Boltzmann method (LBM), they are
proportional to the powers of the speed of sound, which is $O(1)$ due to
the velocity rescaling. The lattice constants $w_1,\dots, w_Q$ provide
the Maxwell weighting of the different discrete velocities $\vec
e_1,\dots,\vec e_Q$ (which are not necessarily unit vectors) and satisfy
\begin{equation}
  \label{eq:wm}
\sum_m w_m = 1.
\end{equation}
The lattice constants depend on the dimension $D
\in \{1,2,3\}$ of the simulation and the number $Q$ of discrete velocities used. 
As the speed of sound is $O(1)$, the macroscopic
velocity $\vert \vec u \vert = O(\text{Ma})$, i.e., $\vert \vec u \vert \ll 1$ in weakly
compressible applications. 
Compared to the incompressible NSE ($\text{Ma}=0$), the Chapman-Enskog expansion of LBE leads to an additional ``compressibility error'' of $O(\ma^2)$ as proved in \citet{Sterling95}.
We remark that the LBM is suitable for high Reynolds number flow since the condition of $\text{Kn}=O\text{(Ma/Re)}$ can be well met \citep{Toschi05} with $\text{Re}\gg 1$ (e.g., $\text{Kn} = O(10^{-9})$ considering $\ma \approx 10^{-1}$ and $\re \approx 10^8$ in the Earth atmosphere). This feature of LBM is crucial for the complexity analysis discussed below. 

\subsection{Carleman linearization of the lattice Boltzmann equation}

The LBE (\Eq{eq:LBMs}) contains terms that are nonlinear in $f$
due to the presence of the $\rho u^2$ terms in $\feq_m$ (i.e., \Eq{eq:feq}).  These nonlinear terms
involve the ratio of quadratic combinations of $f_m$ in the numerator and linear
combinations of $f_m$ in the denominator.  Because the deviation of the
denominator (macroscopic density $\rho$) from 1 is small ($\rho \approx 1$) in weakly
compressible flow, we can approximate $1/\rho$ to first-order Taylor expansion as
\begin{equation}
  \label{eq:rhotaylor}
  \frac 1 \rho \approx 2 - \rho, \quad \lvert 1 - \rho \rvert \ll 1
\end{equation}
with $O(\ma^4)$ error; this is acceptable, since the LBE recovers NSE
to $O(\ma^2)$.  Substituting the expansion of $1/\rho$ yields
polynomial combinations of $f_m$ up to cubic degree in 
$\feq_m$.  Cubic degree is therefore the minimum usable Carleman
degree that can recover NSE when $\text{Ma}\ll1$.

We collect the terms of different polynomial degree in \Eq{eq:LBMs} after the expansion of  \Eq{eq:feq}, with linear terms on the first, quadratic on
the second, and cubic terms on the third lines:
\begin{align}
  \ddt{f_m} = 
  &-\vec{e}_m\cdot\nabla f_m - \frac{1}{\kn\tau}f_m +
  \frac 1{\kn\tau} w_m \left[a\sum_{n=1}^Q f_n + b \vec{e}_m \cdot \sum_{n=1}^Q \vec{e}_nf_n
  \right] \nonumber \\
  & + \frac 2 {\kn\tau} w_m \left[c\left(\vec{e}_m\cdot\sum_{n=1}^Q \vec{e}_n f_n\right)^2 +
    d\left(\sum_{n=1}^Q \vec{e}_n f_n\right)^2\right] \nonumber \\
  \label{eq:lbm-grouped}
  & - \frac 1 {\kn\tau} w_m \left[\sum_{n=1}^Qf_n\right]\left[c\left(\vec{e}_m\cdot\sum_{n=1}^Q \vec{e}_n f_n\right)^2 +
    d\left(\sum_{n=1}^Q \vec{e}_n f_n\right)^2\right]
\end{align}
In the following, we adopt the \citet{Forets2017} notation and separate the
linear, quadratic, and cubic terms into products of constant coefficient matrices
$F_1$, $F_2$, and $F_3$ and 1-, 2-, and 3-forms. Note that we will treat the streaming
operator
\begin{equation}
  Sf=\vec{e}_m\cdot\nabla f_m
\end{equation}
separately for reasons that will become apparent in the discussion of $n$-point
grids.  We write
\Eq{eq:lbm-grouped} as 
\begin{equation}
  \label{eq:lbm-carleman}
  \ddt{f} = -Sf + F^{(1)}f + F^{(2)} f^{[2]} + F^{(3)} f^{[3]},
\end{equation}
where
\begin{equation}
f = (f_1, \dots, f_Q)^\trans;\label{eq:def-f}
\end{equation}
the superscript $^{[k]}$ denotes Kronecker
exponentiation:
\begin{equation}
  f^{[k]} = \underset{k \text{ times}}{\underbrace{f\otimes \cdots \otimes f}}
\end{equation}
$f^{[k]}$ is the $Q^k$-length vector of polynomial permutations
of $f_m$ of degree $k$, including terms that are degenerate because of
commutativity under multiplication. The Kronecker product operator $\otimes$
acts on vectors
\begin{equation}
  x\otimes y = (x_1 y_1, x_1 y_2,\ldots, x_1 y_m, x_2 y_1,\ldots,x_2 y_m, \ldots, x_n y_1,\ldots, x_n y_m)^\trans
\end{equation}
and matrices
\begin{equation}
  A \otimes B = \begin{pmatrix}
    a_{11}B &\ldots&a_{1n}B \\
    \vdots & & \vdots \\
    a_{m1}B & \ldots &a_{mn}B 
  \end{pmatrix}
\end{equation}
by forming all product permutations of their elements.  $F^{(k)}$ is a
$Q\times Q^k$ matrix of constant coefficients of the polynomial combinations of
$f_m$.  The procedure of Carleman linearization
\citep{Carleman1932} recasts this system of $Q$ polynomial differential
equations as an infinite system of coupled linear differential equations by
treating $f^{[k]}$ as new degrees of freedom whose time derivatives are
determined from \Eq{eq:lbm-carleman} by the Leibniz product rule.  Denoting
\begin{equation}
  V^{(k)} = (f, f^{[2]}, \dots, f^{[k]})^\trans,
\end{equation}
the infinite-dimensional Carleman form of \Eq{eq:lbm-carleman} reads
\begin{equation}
  \ddt{V^{(\infty)}} = C^{(\infty)} V^{(\infty)},
  \label{eq:cv}
\end{equation}
with 
\begin{equation}
C^{(\infty)} = \begin{pmatrix}
A_1^1 & A_2^1 & A_3^1 & 0 & 0 & 0 & \ldots \\ 
0 & A_2^2 & A_3^2 & A_4^2 & 0 & 0 & \ldots \\
0 & 0 & A_3^3 & A_4^3 & A_5^3 & 0 & \ldots \\   
\vdots &  \vdots &  \vdots &  \vdots  & \vdots &\vdots & \ddots &\\
\end{pmatrix} \label{eq:Cinfdef}
\end{equation}
and 
\begin{align}
A^i_{i+j-1} = \sum_{r=1}^i \overset{i \text{ factors}}{\overbrace{\IdQ \otimes \cdots \otimes \underset{\underset{r\text{th position}}{\uparrow}}{F^{(j)}} \otimes  \cdots  \otimes \IdQ }}. \label{eq:defiBij}
\end{align}
$\IdQ$ is the identity matrix; 
a proof is given in \citet{Forets2017}.  The truncated Carleman
  linearization simply terminates the series at degree $k$, neglecting all
terms of degree $k+1$ and higher:
\begin{equation}
C^{(k)} = \begin{pmatrix}
A_1^1 & A_2^1 & A_3^1 & 0 & 0 & \ldots & 0 & 0 & 0 & 0 \\ 
0 & A_2^2 & A_3^2 & A_4^2 & 0 & \ldots & 0 & 0 & 0 & 0 \\
\vdots &  \vdots &  \vdots &  \vdots  & \vdots & \ddots & \vdots &  \vdots &  \vdots &  \vdots \\
0 & 0 & 0 & 0 & 0 & \ldots & 0 & A_{k-2}^{k-2} & A_{k-1}^{k-2} & A_{k}^{k-2} \\
0 & 0 & 0 & 0 & 0 & \ldots & 0 & 0 & A_{k-1}^{k-1} & A_k^{k-1} \\
0 & 0 & 0 & 0 & 0 & \ldots & 0 & 0 & 0 & A_k^k  \\
\end{pmatrix} \label{eq:Cdef}.
\end{equation}

\subsubsection{Carleman-linearized LBE on spatial discretization grids}

To describe a spatially discretized system with $n$ grid points, we
introduce the following notation.  Greek letters are used to index grid points
(e.g., $\vec{x}_\alpha$, $\alpha\in 1,\dots,n$), to prevent confusion with the
(italic) indices of discrete velocities.  In place of the single-point objects
$f$, $V$, $A$, $F$, and $C$, we use $\phi$, $\mathcal{V}$, $\mathcal{A}$,
$\mathcal{F}$, and $\mathcal{C}$ to denote their $n$-point counterparts.

The distribution functions $f_m(\vec{x}_\alpha)$ at each grid point
are arranged into an $nQ$-length vector
\begin{align}
  \phi(\vec{x}) &= (f(\vec{x_1}), \dots, f(\vec{x_n}))^\trans \nonumber \\
  &= (f_1(\vec{x}_1), \dots, f_Q(\vec{x}_1), f_1(\vec{x}_2), \dots,
  f_Q(\vec{x}_2), \dots, f_1(\vec{x}_n), \dots, f_Q(\vec{x}_n))^\trans.
\end{align}
The $k$-forms $\phi^{[k]}(\vec x)$, then, include both local and nonlocal terms, i.e., products of
distribution functions at different positions.  While the streaming operator $S$ in
\Eq{eq:lbm-carleman}, implemented as finite differences, connects
distribution functions at different locations,  the collision operator is purely
local.  Thus, the $n$-point versions of
$F^{(1)}$, $F^{(2)}$, and $F^{(3)}$ are $n$-fold block repetitions of the single-point
versions; define
\begin{equation}
  \label{eq:F_i_alpha}
  \mathcal F^{(i)}(\vec{x}_\alpha) =  \delta_\alpha^{[i]} \otimes F^{(i)},
\end{equation}
where the local nature of the collision operator is expressed by the vector 
\begin{equation}
  \label{eq:delta-alpha}
  \delta_\alpha = (\underset{n \text{ elements}}{\underbrace{0, \dots,
      \overset{\alpha\text{th position}}{\overbrace 1}, \dots, 0}}).
\end{equation}
Further define
\begin{equation}
  \label{eq:F_i_n}
  \mathcal F^{(i)}(\vec{x}) = (\mathcal F^{(i)}(\vec{x}_1), \dots, \mathcal F^{(i)}(\vec{x}_n)).
\end{equation}
It follows from \Eq{eq:F_i_alpha} that
\begin{equation}
  \label{eq:diagonality-F}
  \mathcal F^{(i)}(\vec x_\alpha) [\underset{i \text{ factors}}{\underbrace{f(\vec x_\beta)
    \otimes \dots \otimes f(\vec x_\gamma)}}] = F^{(i)} f^{[i]}(\vec x_\alpha) \delta_{\alpha\beta}\dots\delta_{\alpha\gamma},
\end{equation}
where $\delta_{\alpha\beta}$ is the Kronecker delta function.
In other words, products of the form
$\mathcal F^{(i)}(\vec{x})\phi^{[i]}(\vec{x})$ only involve operations on
individual grid points even though the $\mathcal F^{(i)}(\vec x)$ matrices and
$\phi(\vec x)$ vector span all $n$ grid points.  We will refer to this property as
diagonality in $\vec x$.

Analogous to \Eq{eq:lbm-carleman}, the $n$-point LBE is
\begin{equation}
  \label{eq:eq:lbm-carleman-n}
  \ddt{\phi(\vec x)} = -S\phi(\vec x) + \mathcal F^{(1)}(\vec x)\phi(\vec x) +
  \mathcal F^{(2)}(\vec x)\phi^{[2]}(\vec x) + \mathcal F^{(3)}(\vec x)\phi^{[3]}(\vec x).
\end{equation}
The $n$-point Carleman matrix involves $n$-point versions of the transfer
matrices $A^i_j$; these are obtained by inserting
\Eq{eq:F_i_n} into the $n$-point equivalent of \Eq{eq:defiBij}:
\begin{equation}
  \label{eq:nptAij}
  \mathcal A^i_{i+j-1}(\vec x) =
  \sum_{r=1}^i \overset{i \text{ factors}}{\overbrace{
      \IdnQ \otimes \cdots \otimes
      \underset{\underset{r\text{th position}}{\uparrow}}{\mathcal F^{(j)}(\vec x)}
      \otimes  \cdots  \otimes \IdnQ }}. 
\end{equation}
It follows from \Eq{eq:diagonality-F} that
\begin{equation}
  \label{eq:diagonality-A}
  \mathcal A^i_{i+j-1}(\vec x_\alpha) [\underset{i+j-1 \text{ factors}}
  {\underbrace{f(\vec x_\beta) \otimes \dots \otimes f(\vec x_\gamma)}}] =
  A^i_{i+j-1} f^{[i]}(\vec x_\alpha)\delta_{\alpha\beta}\dots\delta_{\alpha\gamma}.
\end{equation}
Thus, $\mathcal A^i_{i+j-1}(\vec x_\alpha)$ is also diagonal in $\vec x$.

The streaming operator can be implemented as finite differences between
neighboring points (not necessarily restricted to nearest neighbors if better
than second-order accuracy is desired, but $O(Q)$ for any reasonable finite
difference implementation).  This results in a matrix form for $S$ with
$O(Q)$ off-diagonal bands, and transfer matrices that are obtained from
\Eq{eq:nptAij} by inserting $S$ instead of $\mathcal F^{(1)}$.  We denote
these transfer matrices of $S$ as $\mathcal B^i_i$; since
$S\phi(\vec{x})$ is linear in $\phi(\vec{x})$, $\mathcal B$ only enters on the block diagonal
of $\mathcal C(\vec{x})$.  (Note that, while $\mathcal B$ is diagonal in
Carleman degree, it is not diagonal in $\vec x$.)

The sum of the streaming Carleman matrix

\begin{equation}
  \label{eq:npt-stream}
  \mathcal C_s^{(k)}(\vec x) = - \begin{pmatrix}
    \mathcal B_1^1 & 0 & 0 & 0 & 0 & \ldots & 0 & 0 & 0 & 0 \\ 
    0 & \mathcal B_2^2 & 0 & 0 & 0 & \ldots & 0 & 0 & 0 & 0 \\
    \vdots &  \vdots &  \vdots &  \vdots  & \vdots & \ddots & \vdots &  \vdots &  \vdots &  \vdots \\
    0 & 0 & 0 & 0 & 0 & \ldots & 0 & \mathcal B_{k-2}^{k-2} & 0 & 0 \\
    0 & 0 & 0 & 0 & 0 & \ldots & 0 & 0 & \mathcal B_{k-1}^{k-1} & 0 \\
    0 & 0 & 0 & 0 & 0 & \ldots & 0 & 0 & 0 & \mathcal B_k^k  \\
  \end{pmatrix} 
\end{equation}
and the collision-term $n$-point Carleman matrix
\begin{equation}
  \label{eq:npt-coll}
  \mathcal C_c^{(k)}(\vec x) = \begin{pmatrix}
    \mathcal A_1^1(\vec x) & \mathcal A_2^1(\vec x) & \mathcal A_3^1(\vec x) & 0 & 0 & \ldots & 0 & 0 & 0 & 0 \\ 
    0 & \mathcal A_2^2(\vec x) & \mathcal A_3^2(\vec x) & \mathcal A_4^2(\vec x) & 0 & \ldots & 0 & 0 & 0 & 0 \\
    \vdots &  \vdots &  \vdots &  \vdots  & \vdots & \ddots & \vdots &  \vdots &  \vdots &  \vdots \\
    0 & 0 & 0 & 0 & 0 & \ldots & 0 & \mathcal A_{k-2}^{k-2}(\vec x) & \mathcal A_{k-1}^{k-2}(\vec x) & \mathcal A_{k}^{k-2}(\vec x) \\
    0 & 0 & 0 & 0 & 0 & \ldots & 0 & 0 & \mathcal A_{k-1}^{k-1}(\vec x) & \mathcal A_k^{k-1}(\vec x) \\
    0 & 0 & 0 & 0 & 0 & \ldots & 0 & 0 & 0 & \mathcal A_k^k(\vec x)  
  \end{pmatrix}
\end{equation}
then yields the $n$-point truncated Carleman matrix at degree $k$:
\begin{equation}
  \label{eq:n_point_Carleman}
  \mathcal C^{(k)}(\vec x) = \mathcal C_s^{(k)}(\vec x) + \mathcal C_c^{(k)}(\vec x).
\end{equation}
Similarly, the $n$-point Carleman vector becomes
\begin{equation}
  \mathcal V^{(k)}(\vec x) = (\phi(\vec x), \phi^{[2]}(\vec x), \dots,
  \phi^{[k]}(\vec x))^\trans,
\end{equation}
leading to the Carleman-linearized truncated LBE over $n$ grid points:
\begin{equation}
  \frac{\partial \mathcal V^{(k)}(\vec x)}{\partial t} = \mathcal C^{(k)}(\vec x)
  \mathcal V^{(k)}(\vec x),
  \label{eq:clbm}
\end{equation}
which, like the single-point Carleman LBE, is a system of linear differential
equations in matrix form.  In contrast to the single-point Carleman vector
$V^{(k)}$, which has dimension $\sum_{i=1}^k Q^i$, the dimension of the $n$-point Carleman vector
$\mathcal V^{(k)}(\vec x)$ is $\sum_{i=1}^k n^iQ^i$.  Note that $\mathcal
C_c^{(k)}(\vec x)$, all its transfer-matrix blocks being diagonal in $\vec x$ by
\Eq{eq:diagonality-A}, is also diagonal in $\vec x$, a property we will use
extensively in the quantum-algorithm complexity analysis.

The first problem that we would like to solve involves assuming that we are given an initial distribution $\phi(0)$ that is efficiently computable over our lattice, an evolution time $t$ and an error tolerance $\epsilon_{\rm dyn}$ to generate a quantum state $\ket{\mathcal{V}^{(k)}(x,t)}$ such that $\| \ket{\mathcal{V}^{(k)}(x,t)} - {\mathcal{V}^{(k)}(x,t)}/\|{\mathcal{V}^{(k)}(x,t)}\| \le \epsilon_{\rm dyn}$.  

The problem is the natural problem to first look at, however, it ignores the fact that the observables we are interested in do not act on the extended Hilbert space invoked through the use of Carleman linearization.  The overall problem that we are interested in is then if we define the projector onto the first Carleman block to be $\ket{\phi_0}\!\bra{\phi_0}\otimes I$ then we aim to minimize
\begin{equation}
\|(\bra{\phi_0}\otimes I)(\ket{\mathcal{V}^{(k)}(x,t)} / \|(\bra{\phi_0}\otimes I)(\ket{\mathcal{V}^{(k)}(x,t)}\| - \phi(x)/\|\phi(x)\| \|\le \epsilon_{\rm total}.
\end{equation}
Our aim is then to separate the two sources of error in the problem and use to observation that if $\epsilon_{\rm total} \le \epsilon_{\rm dyn} + \epsilon_{\rm carl}$ then it suffices to make both of these error sources small.  Through existing arguments, we can construct a quantum algorithm that makes the former small.  We will make arguments here based on perturbation theory backed by numerics that $\epsilon_{\rm carl}$ can be made small in regimes of practical interest using small values of $k$.

\ifarXiv
\section{Results}
\label{sec:results}
\fi

\subsection{Carleman truncation error}

We now analyze the effect of truncating the infinite-dimensional Carleman system
of equations at a finite degree.  We show that the zeroth and first moments of the 
lattice Boltzmann distribution functions ($\rho$ and $\rho \vec u$) at $k$th degree
receive corrections from the higher-order Carleman variables at degrees $k+1$ and
$k+2$ and that these corrections are suppressed by $O(\text{Ma}^2)$.

The macroscopic properties of the LBE
fluid are given by moments of the lattice Boltzmann distribution functions, as
can be seen from \Eq{eq:rho}--(\ref{eq:u}): the
zeroth moment gives $\rho$, and the first moment gives $\rho \vec u$.  We introduce the
coefficient vectors
\begin{equation}
  \Phi_\rho = (1, \dots, 1)
\end{equation}
and
\begin{equation}
  \Phi_{\rho u} = (\vec{e}_1, \dots, \vec{e}_Q),
\end{equation}
which recover the macroscopic fluid variables by matrix-multiplying the vector of distribution functions such that
\begin{equation}
\rho = \sum_m (\Phi_\rho)_m f_m = \Phi_\rho f
\end{equation}
and
\begin{equation}
\rho \vec u = \sum_m (\Phi_{\rho u})_m f_m = \Phi_{\rho u} f.
\end{equation}
From the mixed-product property of the
Kronecker product,
\begin{equation}
  \label{eq:kromipro}
  (A \otimes B) (C \otimes D) = (AC) \otimes (BD),
\end{equation}
it follows that 
\begin{align}
  (\rho)^p (\rho \vec u)^q &= (\Phi_\rho f)^{[p]} (\Phi_{\rho u} f)^{[q]}\nonumber \\
                           &= \left(\Phi_\rho^{[p]}\otimes\Phi_{\rho u}^{[q]}\right) f^{[p+q]}
\end{align}
for arbitrary integers $p$ and $q$ (where products of spatial vectors imply the dot product).

Furthermore, inserting the quadratic and cubic terms in
\Eq{eq:lbm-grouped}, we obtain
(introducing the shorthand notation $\zeta_2$ and $\zeta_3$): 
\begin{align}
  \Phi_\rho     F^{(2)}f^{[2]} &= \frac 2 {\kn\tau} \sum_m w_m\left[c (\vec{e}_m\cdot
                                 \rho\vec{u})^2 + d\rho^2 u^2 \right] =
                                 \zeta_2 \rho^2 \nonumber \\
  \label{eq:rhoF2f2}
  &= O((\kn\tau)^{-1}\rho^{2}u^2) \\
  \Phi_\rho     F^{(3)}f^{[3]} &= -\frac 1 {\kn\tau} \sum_m w_m \left[c\rho^3(\vec{e}_m\cdot
                                 \vec{u})^2 + d\rho^3u^2 \right] =
                                 \zeta_3\rho^3 \nonumber \\
  \label{eq:rhoF3f3}
  &= O((\kn\tau)^{-1}\rho^{3}u^2) \\
  \\
  \label{eq:rhouF2f2}
  \Phi_{\rho u} F^{(2)}f^{[2]} &= \frac 2 {\kn\tau} \sum_m w_m\vec{e}_m\left[c (\vec{e}_m\cdot
                                 \rho\vec{u})^2 + d\rho^2 u^2 \right] = 0\\
  \label{eq:rhouF3f3}
  \Phi_{\rho u} F^{(3)}f^{[3]} &= -\frac 1 {\kn\tau} \sum_m w_m \vec{e}_m \left[c\rho^3(\vec{e}_m\cdot
                                 \vec{u})^2 + d\rho^3u^2 \right] = 0,
\end{align}
where the result $\Phi_{\rho u} F^{(2)}f^{[2]} = \Phi_{\rho u} F^{(3)}f^{[3]} =
0$ is a consequence of lattice symmetry dictating that the nonresting particles
occur in pairs of equal weight $w_m$ but opposite direction
$\vec{e}_m$, and the bounds in \Eq{eq:rhoF2f2} and
\Eq{eq:rhoF3f3} stem from $w_m=O(1)$ and $\sum_m w_m=1$.

The lowest-order correction to $\partial V^{(k)}/\partial t$ is given by the
elements of $C^{(k+1)}$ that are not present in $C^{(k)}$.  At this point, for
concreteness, we consider $k=3$ as an example:
\begin{align}
  \label{eq:f2c4}
  \left[\ddt{f^{[2]}}\right]'_{(4)} &= A^2_4 f^{[4]} \\
  \label{eq:f3c4}
  \left[\ddt{f^{[3]}}\right]'_{(4)} &= A^3_4 f^{[4]} ,
\end{align}
where $[\partial/\partial t]'_{(k+1)}$ denotes said correction
.  By the definition
of $A^i_{i + j - 1}$ (\Eq{eq:defiBij}),
\begin{align}
  A^2_4 &= F^{(3)}\otimes \IdQ + \IdQ \otimes F^{(3)} \\
  A^3_4 &= F^{(2)}\otimes \IdQ \otimes \IdQ + \IdQ \otimes F^{(2)}\otimes \IdQ +
          \IdQ \otimes \IdQ \otimes F^{(2)}.
\end{align}
Using $\Phi_\rho$ and $\Phi_{\rho u}$ to construct corrections to degree 2 and 3
polynomials of $\rho$ and $\rho \vec u$ yields, again by the mixed-product rule
(\Eq{eq:kromipro}), substituting the expressions from
\Eq{eq:lbm-grouped} for $F^{(2)}f^{[2]}$ and $F^{(3)}f^{[3]}$,
and making use of \Eq{eq:rhoF2f2}--(\ref{eq:rhouF3f3}):
\begin{align}
  \left[\ddt{\rho^2}\right]'_{(4)}
  &= \Phi_\rho^{[2]} \left[\ddt{f^{[2]}}\right]'_{(4)} = \Phi_\rho^{[2]} A^2_4  f^{[4]} \nonumber \\
  &= (\Phi_\rho \otimes \Phi_\rho) \left(F^{(3)}\otimes \IdQ + \IdQ \otimes
    F^{(3)}\right) (f \otimes f \otimes f \otimes f) \nonumber \\
  &= (\Phi_\rho F^{(3)} f^{[3]})\otimes(\Phi_\rho\IdQ f) + (\Phi_\rho\IdQ
    f)\otimes(\Phi_\rho F^{(3)} f^{[3]}) \nonumber \\
  \label{eq:rho2c4}
  &= -\frac{2}{\kn\tau}du^2\rho \\
  \left[\ddt{\rho^2u^2}\right]'_{(4)}
  &= \Phi_{\rho u}^{[2]} \left[\ddt{f^{[2]}}\right]'_{(4)} = \Phi_{\rho u}^{[2]} A^2_4  f^{[4]} \nonumber \\
  &= (\Phi_{\rho u} \otimes \Phi_{\rho u}) \left(F^{(3)}\otimes \IdQ + \IdQ \otimes
    F^{(3)}\right) (f \otimes f \otimes f \otimes f) \nonumber \\
  \label{eq:rho2u2c4}
  &= 0.
\end{align}

We now derive a recursion formula for the higher-order corrections to powers of
$\rho$ and $\rho \vec u$ at arbitrary order, following steps very similar to the
$k=3$ example.  At any Carleman order $k$, corrections to $f^{[k-1]}$ and
$f^{[k]}$ arise due to the neglected terms containing $f^{[k+1]}$, which, by
inspection of \Eq{eq:Cdef}, can be written as
\begin{align}
  \left[\ddt{f^{[k-1]}}\right]'_{(k+1)} &= A^{k-1}_{k+1} f^{[k+1]} \\
  \left[\ddt{f^{[k]}}\right]'_{(k+1)} &= A^{k}_{k+1} f^{[k+1]}.
\end{align}
These equations are analogous to \Eq{eq:f2c4}--(\ref{eq:f3c4}) for the $k=3$
example.  Only these two subvectors of $V$ receive corrections due to the
diagonal band structure of \Eq{eq:Cinfdef}, whose striping is determined by
the presence of terms up to cubic degree in \Eq{eq:lbm-carleman}.  The
corresponding corrections to powers of the macroscopic fluid variables can be
derived by left multiplication with Kronecker powers of $\Phi_\rho$ and
$\Phi_{\rho u}$ in analogy to \Eq{eq:rho2c4}--(\ref{eq:rho2u2c4}):
\begin{align}
  \left[\ddt{\rho^{k-1}}\right]'_{(k+1)}
  &= \Phi_\rho^{[k-1]}A^{k-1}_{k+1}
    f^{[k+1]} \nonumber \\
  &= \left(\Phi_\rho^{[k-2]}\otimes\Phi_\rho\right)
    \left[\sum_{r=1}^{k-1} \IdQ^{[r-1]} \otimes F^{(3)} \otimes \IdQ^{[k-r-1]} \right]
    \left(f^{[k-2]}\otimes f^{[3]}\right)\nonumber \\
  &= -\frac{k-1}{\kn\tau}\left(\sum_m w_m \left[c\rho^3(\vec{e}_m\cdot\vec{u})^2 +
    d\rho^3u^2 \right]\right)\rho^{k-2} \nonumber \\
  &= O(k(\kn\tau)^{-1}\rho^{k+1}u^2) \\
  \left[\ddt{\rho^{k}}\right]'_{(k+1)}   &= \Phi_\rho^{[k]}A^{k}_{k+1} f^{[k+1]}
                                           \nonumber \\
  &= \left(\Phi_\rho^{[k-1]}\otimes\Phi_\rho\right)
    \left[\sum_{r=1}^{k} \IdQ^{[r-1]} \otimes F^{(2)} \otimes \IdQ^{[k-r]} \right]
    \left(f^{[k-1]}\otimes f^{[2]}\right)\nonumber \\
  &= \frac{2k}{\kn\tau}\left(\sum_m w_m \left[c\rho^2(\vec{e}_m\cdot\vec{u})^2 +
    d\rho^2u^2 \right]\right)\rho^{k-1} \nonumber \\
  &= O(k(\kn\tau)^{-1}\rho^{k+1}u^2) \\
  \label{eq:rhouF3fn}
  \left[\ddt{\rho^{p+q}u^{q}}\right]'_{(k+1)} &= 0, \quad q \geq p \geq 0, q > 0, p +
                                              q = k - 1 \\
  \label{eq:rhouF2fn}
  \left[\ddt{\rho^{p+q}u^{q}}\right]'_{(k+1)}   &= 0, \quad  q \geq p \geq 0, q > 0, p +
                                              q = k,
\end{align}
for integer $p$ and $q$ ($q \geq p$ is required for getting $[k-1]$th or $k$th $\poly(u, \rho)$), where \Eq{eq:rhouF3fn} and \Eq{eq:rhouF2fn} follow from
\Eq{eq:rhouF2f2} and \Eq{eq:rhouF3f3}.  Thus, the macroscopic fluid
properties follow a hierarchy where subsequent higher-order corrections are
suppressed by factors of $O(u^2) = O(\text{Ma}^2)$ relative to the
highest-order retained terms. 
Since the LBE is restricted to $\text{Ma}\ll 1$, the
Carleman truncation error and the Taylor expansion error on $1/\rho$
are small even when a low Carleman truncation order is used.  We also note that
the expression for the truncation error is independent of the system
size and Reynolds number, since the overall $(\kn\tau)^{-1}$ normalization is
common to the retained and neglected terms. This ensures that the Carleman order can be kept
constant as the system size is extended or $\text{Re}$ is increased. A numerical consistency check of the Carleman-linearized nonlinear collision term of \Eq{eq:LBMs} shows that the truncation error is on the order of $10^{-14}$ for the truncation order $k=3$ (\Fig{fig:CLBM_collision_error}). We refer to details of the LBM and CLBM simulations and the discussions in section~\ref{sec:simu}. 

\subsection{Quantum algorithm complexity}
\label{sec:qac}

In this section, we derive bounds on the complexity of applying the
\citet{Berry2017QuantumPrecision} algorithm to the Carleman-linearized
LBE of degree 3 at $n$ spatial grid points with $Q$ discrete
velocities (\Eq{eq:clbm}). For notational compactness, we
introduce the abbreviated notation
\begin{align}
  \mathcal V &= \mathcal V^{(3)}(\vec x) \\
  \mathcal C &= \mathcal C^{(3)}(\vec x).
\end{align}
Carleman order $k=3$ is explicitly chosen as we showed above that the
truncation error at $k=3$ is $O(\ma^2)$, which is sufficient to
prevent truncation error from dominating the solution error.

Before proceeding further, we summarize the properties of the physical
systems which can be simulated using our procedure, including
conditions that the systems must satisfy for the procedure to be
applicable.  We also give typical values of system parameters for a
particular application (atmospheric turbulence).  The key unrestricted
parameter of the physical application is the number of grid points
$n$.  For simulations of turbulent flows, the $n\gg 1$ behavior of the
algorithm is of utmost interest; for example, for direct simulations
of atmospheric turbulence, $n > 10^{20}$ is desirable (and
unattainable using state-of-the-art algorithms on classical computers).

\paragraph{Parameters of restricted range}
\label{sec:prop-of-phys-syst:param-restr-range}

The fluid flow to be modeled by our procedure needs to satisfy certain
conditions.  These conditions restrict two parameters of the flow, the
Mach number $\ma$ and Knudsen number $\kn$.  First, the flow needs to
be ``weakly compressible'', i.e., $\ma \ll 1$.  This is a
requirement both for the LBE to recover the NSE and for the linearized
LBE to approximate the nonlinear LBE (both of which errors scale as
$\ma^2$).  In LBM practice, $\ma < 10^{-1}$ is a typical upper bound,
but atmospheric flow satisfies $\ma \leq 10^{-2}$ in most
circumstances.  Second, the flow to be modeled must also be
``continuum flow'', i.e., macroscopic features like turbulent eddies
must be much larger than the molecular scale.  This is reflected in
Boltzmann relaxation parameter $\kn\tau\ll 1$.  Typical values in the
atmosphere are $\kn < 10^{-8}$.  The scaled relaxation $\tau$ is
restricted to the range $0.5<\tau\leq 1$ in LBM practice.
Furthermore, the relaxation
parameter restricts the number of grid points according to
\Eq{eq:kn-logn} to $\log n \ll (\kn\tau)^{-3}$.
This restriction is unlikely to have practical consequences; in the
atmospheric example, it is equivalent to $n \ll \exp(10^{24})$.
For
atmospheric flow, $\ma\ll 1$ is the most restrictive condition.

\paragraph{Free parameters}

The number of spatial grid points $n$ used to resolve the flow is
unrestricted, i.e., application science would like the largest value
possible subject to getting a result within a reasonable wallclock
time limit.  Reynolds number $\re$ is similarly unrestricted, i.e.,
application science would like the largest value possible.  Typical
values of these parameters currently attainable using state-of-the-art
classical algorithms and computers are $n\approx 10^9$ and
$\re \approx 10^5$.  The parameters $\re$, $n$, and physical integral
domain size $L$ are not independent; higher $\re$ requires finer
resolution, implying higher $n/L^d$ (in $d$ spatial dimensions).  A transformational capability in
atmospheric science would be direct numerical simulation of entire cloud systems
($L\approx 100$ km) at $\re\approx10^9$, which requires a spatial grid spacing of 1 mm (Kolmogorov length scale) or $n\approx 10^{21}$.

The physical integration time $\tilde T$ depends on specifics of the
physical problem.  It is typically determined by the time required for
the turbulence to spin up from homogeneous initial conditions and
for the turbulent eddies then to turn over several times so that
domain-mean statistics are representative of the fully developed
turbulent steady state.

\paragraph{Constants}

The number of discrete molecular velocities $Q$ is fixed, i.e., the same
value can be used for simulation of arbitrary systems (as long as
$\ma \ll 1$); values in widespread use are $Q\in\{3,9,27\}$ for 1D, 2D,
and 3D fluid flows, respectively.  The constants $a=1$, $b=3$, $c=9/2$, $d=-3/2$
are fixed for all lattice geometries, and the $w_m$ are fixed for each
specific lattice geometry.  For this reason, we consider these parameters to be constants; however, for some simulations it may be appropriate to take these to be free parameters that change as the desired simulation accuracy increases to minimize algorithmic complexity.

\paragraph{Solution error sources}

Two distinct sources of solution error appear.  The first is due to
the Carleman-linearized LBE being an approximation to the actual NSE
dynamics of the system; this error is $O(\ma^2)$, hence the $\ma\ll 1$
requirement for our procedure to be applicable.  The second is
discretization error in the time integration of $\partial\mathcal V/\partial t
= \mathcal C \mathcal V$ in the quantum algorithm; this is
parameterized by $\epsilon$, which should be chosen so it is not the
dominant error, i.e., $\epsilon=O(\ma^2)$.

\subsubsection*{Contributions to algorithm complexity of \citet{Berry2017QuantumPrecision}}
  
The complexity of solving \Eq{eq:clbm} is given by \Eq{eq:ca} in terms
of the properties of the coefficient matrix $\mathcal C$ and the solution vector
$\mathcal V$; for ease of reference, the equation is
reproduced here:
\begin{equation}
\tag{\ref{eq:ca}}
\text{gate complexity}=O(\Vert \mathcal C\Vert \kappa_\mathcal{J} g T s\cdot
\text{poly}(\log{(\kappa_\mathcal{J} s g \beta T \Vert \mathcal C \Vert N/\epsilon)})),
\end{equation}
where $\mathcal C=\mathcal J \mathcal D \mathcal J^{-1}$ is an $N\times N$ diagonalizable matrix with
eigenvalues $\mathcal D=\text{diag}(\lambda_i)$ ($i \in 1,\dots, N$) whose real part
$\mathcal{R}(\lambda_i)\leq 0\ \forall i$. The norm $\|\mathcal C\|$ is the
2-norm, the largest singular value of $\mathcal C$.  
The initial condition norm enters through
\begin{equation}
\label{eq:beta}
\beta = (\| \vert{\mathcal V}_\text{in}\rangle \| + T \norm{\ket{b}}) / \| \vert{\mathcal V}(T)\rangle \| = O(1), 
\end{equation}
which is dependent on the physical problem under consideration but independent
of other system parameters.
Here we consider the gate complexity to be the number of one and two-qubit gates needed for the simulation.  
The condition number in the above expression,
\begin{equation}
    \kappa_\mathcal{J} = \|\mathcal{J}\|\cdot\|\mathcal{J}^{-1}\|,
\end{equation}
is the condition number of the matrix $\mathcal{J}$ of
eigenvectors of $\mathcal C$.
The dissipation parameter $g$ is defined as
\begin{equation}
\label{eq:g}
g = \underset{t\in[0,T]}{\max} \Vert
\mathcal V(t)\Vert/\Vert \mathcal V(T)\Vert,
\end{equation}
and $s$ is the sparsity (number of nonzero
entries per row) of $\mathcal C$. The number of degrees of freedom $N$, which is the dimension of $\mathcal C$ and
$\mathcal V$, is given in terms of the number of grid points $n$ and discrete
velocities $Q$ by
\begin{equation}
  \label{eq:ndf}
  N = (n^3Q^3 +  n^2Q^2 + nQ).
\end{equation}
The term-by-term analysis of the factors that determine the algorithm’s complexity (\Eq{eq:ca}) is
conducted below.
\paragraph{Evolution time}

The physical evolution time $\tilde T$ is imposed by the nature of the system
being simulated.  Physical and lattice evolution time are related by the
\citet{Sterling1996StabilityMethods} factor (\Eq{eq:scale-t}),
\begin{equation}
  \label{eq:evolutionT}
  T=\tilde T/\left(\frac{L}{e_r}\right).
\end{equation}

\paragraph{Dissipation parameter $g$}

The dissipation parameter $g$ depends on the forcing of turbulence. For decaying homogeneous turbulence with $b(t) = 0$, the decay of the kinetic energy of the flow can be described by a power law 
\begin{equation}
  \label{eq:u2}
u^2 = K \chi  \tilde t^{-6/5} 
\end{equation}
based on the Kolmogorov theory, where $K$ is the dimensionless Kolmogorov constant that depends on the dimension of the flow and $\chi$ is a constant with a dimension of $[L^2T^{-4/5}]$ \citep{Saffman67, Skrbek20}. 
The corresponding decay of the norm of the velocity field can be obtained from \Eq{eq:u2} as
\begin{equation}
  \label{eq:u-norm}
  \norm{\vec{u}}= \sqrt{K \chi} \tilde t^{-3/5},
\end{equation}
which leads to the following scaling of the dissipation parameter in terms of the flow velocity 
\begin{equation}
\label{eq:gu}
  g_u (\tilde t) = \frac{\norm{\vec{u}_{\rm in}}}{\norm{\vec{u}}} = \frac{\norm{\vec{u}_{\rm in}}}{\sqrt{K \chi}}  \tilde t^{3/5}. 
\end{equation}
At $\tilde t = \tilde T$, we insert \Eq{eq:evolutionT} into \Eq{eq:gu} and obtain
\begin{equation}
\label{eq:gu-T}
  g_u (T \left(\frac{L}{e_r}\right)) = \frac{\norm{\vec{u}_{\rm in}}}{\sqrt{K \chi}} \left(\frac{L}{e_r}\right)^{3/5} T^{3/5}. 
\end{equation}
For decaying turbulence, we note that $\underset{\tilde t\in[0,\tilde T]}{\max} \norm{\vec{u}(\tilde t)} = \norm{\vec{u}_{\rm in}}$ and a decaying Maxwell distribution function $f_m^{\rm eq}$ in \Eq{eq:feq} that eventually becomes stationary with $\rho \approx 1$ as $\underset{t\rightarrow \infty}{\lim} \vec{u}=0$,
\begin{equation}
\feq_m(T) \approx w_m a \, .
\label{eq:feqT}
\end{equation}
This leads to a stationary $f_m$ and as they relax to a stationary $f_m^{\rm eq}$ following \Eq{eq:lbm_origin} at a physically meaningful evolution time $T$. 
As a result, the dissipation parameter in terms of $g_f$ also reaches a stationary state,
\begin{equation}
\label{eq:gf}
g_f=\frac{\underset{t\in[0,T]}{\max} \norm{
f(t)}}{\norm{f(T)}} \le \frac{1}{\norm{w_m a}}.
\end{equation}

With the physical reasoning in hand, we now prove \Eq{eq:gf}.
We first upper bound $\underset{t\in[0,T]}{\max} \norm{
f(t)}$. For flow field with periodic boundary condition considered here, the variance of the density field can be expressed as \citep{Li_2020}
\begin{equation}
\label{eq:rhov}
\frac{\langle \rho^2 \rangle}{\langle \rho \rangle^2} = 1 + \sigma_\rho,
\end{equation}
where $\sigma_\rho$ is the standard deviation of the density field.
We recall that the flow field considered here is weakly impressible, i.e., $\rho\approx 1$ without losing generality, which leads to $\langle \rho \rangle^2\approx 1$ and $\sigma_\rho \approx 0$. Inserting them into \Eq{eq:rhov}, we obtain
\begin{equation}
\label{eq:rho-norm}
\langle \rho^2 \rangle = \norm{\rho}^2 \approx 1.
\end{equation}
Combine \Eq{eq:rho} and \Eq{eq:rho-norm} and observe the fact that $f_m > 0$, we get
\begin{equation}
\label{eq:fm-norm}
\norm{f_m(t)} \le 1.
\end{equation}
The mass and energy conservation during the collision ensures that
$|f_m (t) - \feq_m (t)| \propto u^2$ \citep{Chen1998LatticeFlows}. 
For weakly compressible flow, we observe $u^2 \ll 1$ and \Eq{eq:feqT}, and therefore, 
\begin{equation}
  \label{eq:nfeq}
|f_m (T)| \approx |\feq_m (T)| \approx w_m a
\end{equation}
for physically meaningful dissipation time scales $T$.
The conclusion of $g_f \approx O(1)$ (\Eq{eq:gf}) is reached by inserting \Eq{eq:fm-norm} and \Eq{eq:nfeq} 
into $\frac{\underset{t\in[0,T]}{\max} \norm{
f(t)}}{\norm{f(T)}}$ and considering the fact that $f(T) \approx \feq(T)$ at the stationary state of $f(T)$. 
\Eq{eq:gf} offers a mathematically upper bound of $g_f$. For weakly compressible decaying turbulence, \Eq{eq:nfeq} would hold for any simulation time $t$, i.e., $\underset{t\in[0,T]}{\max} \norm{
f(t)} \approx w_m a$ holds, which leads to a more realistic estimate of $g_f$ 
\begin{equation}
\label{eq:gf-phy}
g_f \approx 1. 
\end{equation}
\Eq{eq:gf-phy} is also confirmed by our numerical simulation of \Eq{eq:lbm_origin} (\Fig{fig:ts_f-norm-D2Q9}), which yields $g_f = 1 \pm 0.003$. 
Even though the macroscopic flow field $\vec{u}$ decays as $\norm{u}\propto \tilde t^{3/5}$ (\Eq{eq:gu}) following the Kolmogorov theory, the particle distribution function $f_m$ ($\phi$ for $n$-point LBE) relaxes to its stationary equilibrium state. This feature offers a more favourable $g_f \propto O(1)$ for the QLSA and is again due to the inherent nature of LBE. The only assumptions to arrive at \Eq{eq:gf} and \Eq{eq:gf-phy} are periodic boundary conditions (homogeneity), weakly compressible and decaying turbulence. \citet{Liu21} showed that the dissipation parameter $g$ decays exponentially for a homogeneous quadratic nonlinear (weak nonlinearity with $R < 1$) equation. The energy cascade of NSE and the geometric nature of its LBE form refute this exponential decay.   

To bound $\|\mathcal V (T)\|$, we first revisit the solution error.
Entries of $\mathcal V(T)$, $\mathcal V_j(T)$ as the truncated Taylor series of the solution of \Eq{eq:LBMs} at time $T$, $f(T)$, must satisfy
the bound,
\begin{equation}
  \label{eq:g-nn2}
    \|f^{\otimes j}(T)- \mathcal V_j(T)\|\leq \delta\|f^{\otimes j}(T)\|\,
\end{equation}
which leads to
\begin{align}
  \label{eq:g-nn3a}
  \norm{\mathcal V_j(T)} \ge (1-\delta)\norm{f^{\otimes j}(T)} \\
  \label{eq:g-nn3b}
  \norm{\mathcal V_j(T)} \le (1+\delta)\norm{f^{\otimes j}(T)}.  
\end{align}
Therefore, for all $t$, the following holds \citep{Krovi2023improvedquantum}
\begin{align}
  \|\mathcal V(t)\|^2
        &= \sum_{j=1}^k\|\mathcal V_j(t)\|^2
        \leq \sum_{j=1}^k(1+\delta)^2\|f^{\otimes j}(t)\|^2
        \le (1+\delta)^2 k \|f(t)\|^2                         \label{eq:parallel_inequality-a}
\\
  \|\mathcal V(t)\|^2 &= \sum_{j=1}^k\|\mathcal V_j(t)\|^2 \ge \sum_{j=1}^k (1-\delta)^2\|f^{\otimes j}(t)\|^2\ge (1-\delta)^2\|f(t)\|^{2}\,.
                        \label{eq:parallel_inequality-b}
\end{align}
Combining \Eq{eq:g} and (\ref{eq:parallel_inequality-a})--(\ref{eq:parallel_inequality-b}) yields
\begin{equation}
\label{eq:g-delta}
g \le \frac{1+\delta}{1-\delta}\sqrt{k}\frac{\underset{t\in[0,T]}{\max} \norm{
f(t)}}{\norm{f(T)}}.
\end{equation}
We insert \Eq{eq:gf-phy} to \Eq{eq:g-delta} and obtain,
\begin{equation}
\label{eq:g-delta2}
g \le \frac{1+\delta}{1-\delta}\sqrt{k} g_f \approx \frac{1+\delta}{1-\delta}\sqrt{k}
\end{equation}
with an error of $O(10^{-3})$.
Taking $\delta \le 1/3$ and an even larger-than-one $g_f=3/2$, \Eq{eq:g-delta2} can be reduced to
\begin{equation}
\label{eq:g-decay}
g \le 3\sqrt{k}.
\end{equation}

\paragraph{Sparsity}

The coefficient matrices $F^{(1)}$, $F^{(2)}$, and $F^{(3)}$ and resulting transfer matrices $A^i_j$
for a single point contain $O(Q^3\times Q^3)$ elements.  In the $n$-point
Carleman matrix, locality is enforced by \Eq{eq:F_i_alpha}, so that the
nonzero elements of the collision Carleman matrix (\Eq{eq:npt-coll}) are
diagonal in $\vec x$.

The streaming operator $S$ involves the $O(Q)$ nearest (or potentially
next-to-nearest, next-to-next-to-nearest, etc., if higher-order accuracy for the
gradient operator is desired) neighbors in space. This results in bands in $S$
that are not diagonal in $\vec x$, yielding an additional $O(Q^3\times Q^3)$ off-diagonal
elements in the streaming Carleman matrix, i.e., the first term in \Eq{eq:n_point_Carleman}.

The sparsity, i.e., the total number of nonzero elements in each row or column of
$\mathcal C$, is bounded by the sum of the sparsity of the collision Carleman
matrix and the sparsity of the streaming Carleman matrix, both of
which are $O(Q^3)$; thus,
\begin{equation}
  s=O(1),\label{eq:sparsity}
\end{equation}
i.e, sparsity is independent of the number of grid points $n$ and other free parameters of the
system for the fixed discretization schemes considered here.

\paragraph{Matrix norm}
We now determine the norm of the Carleman matrix $\mathcal C$ (\Eq{eq:n_point_Carleman}).
The matrix 2-norm of a matrix $M$ is bounded by its 1-norm and its $\infty$-norm: 
\begin{equation}
  \label{eq:specnorm}
  \|M\| \leq \sqrt{\|M\|_1 \|M\|_\infty}.
\end{equation}
The 1-norm and $\infty$-norm of $\mathcal C$ can both be directly bounded, since
they equal the maximum absolute column sum and maximum absolute row sum of the
matrix, respectively.  By \Eq{eq:sparsity}, there are $O(Q^3)$ nonzero
elements in each row.  From \Eq{eq:lbm-grouped}, these elements are each
$O((\kn\tau)^{-1})$, so that $\|\mathcal C\|_\infty = O(Q^3(\kn\tau)^{-1})$.  We
note that \Eq{eq:sparsity} also gives the number of nonzero elements in each
column due to the diagonality of $\mathcal C$ in $\vec x$, so that
$\|\mathcal C\|_1$ has the same scaling as $\|\mathcal C\|_\infty$.  Therefore,
the overall scaling of the matrix norm with the free parameters of the
system is
\begin{equation}
  \label{eq:normc}
  \|\mathcal C\| = O((\kn\tau)^{-1}).
\end{equation}
In practice, $\tau = O(1)$, because $\tau > 0.5$ is required for the stability of the
LBM \cite{Sterling1996StabilityMethods}.   Thus we have
\begin{equation}
    \|\mathcal{C} \| = O(\kn^{-1}).
\end{equation}

\paragraph{Stability}

Matrix stability refers to $\mathcal{R}(\lambda)\leq 0$ and is a necessary
condition for the \citet{Berry2017QuantumPrecision} algorithm. We now show that
$\mathcal C$ satisfies that condition if the underlying physical system is linearly
stable.

Because of its block triangular structure (in Kronecker degree, not $\vec x$),
the eigenvalue problem for $\mathcal C$ reduces to finding the eigenvalues of the block-diagonal
(in Kronecker degree) elements: the transfer matrices
\begin{align}
  \mathcal A_1^1 - \mathcal B_1^1  =& \mathcal F^{(1)} - S \\
  \mathcal A_2^2 - \mathcal B_2^2 =& \IdnQ\otimes (\mathcal F^{(1)} - S) + 
                                    (\mathcal F^{(1)} - S) \otimes\IdnQ \\
  \mathcal A_3^3 - \mathcal B_3^3 =& \IdnQ^{(2)}\otimes (\mathcal F^{(1)} - S)
                                    \nonumber \\
  &+ \IdnQ\otimes (\mathcal F^{(1)} - S)\otimes\IdnQ \nonumber \\
  &+ (\mathcal F^{(1)} - S)\otimes\IdnQ^{(2)}.
\end{align}

Denote the eigenvectors and eigenvalues of $\mathcal F^{(1)} - S$ as $v_i$ and $\mu_i$
($i\in 1,\dots, nQ$).  It is trivial to show that the eigenvectors and eigenvalues
of $(\mathcal F^{(1)} -  S) \otimes \IdnQ$ and $\IdnQ \otimes (\mathcal F^{(1)} - S)$ are simply
$v^{(2)} = v \otimes v$ and $\mu^{(2)} = \mu_i + \mu_j$
($i,j\in 1,\dots, nQ$); similarly for $(\mathcal F^{(1)} - S) \otimes \IdnQ^{(2)}$, etc.
Therefore, if all eigenvalues of $(\mathcal F^{(1)} - S)$ are nonpositive, the eigenvalues of
$\mathcal C$ will also be nonpositive.

The stability of $\mathcal C$ is thus determined by the stability of
$\mathcal F^{(1)} - S$. This is equivalent to a linear stability analysis on
the lattice Boltzmann system \cite{Sterling1996StabilityMethods}.

\paragraph{Condition number}

To bound  $\kappa_{\mathcal{J}}$, we need to find the eigenvectors of $\mathcal C$.  This is
not possible analytically for arbitrary $n$; if an analytic expression for the
eigenvectors could be obtained, a quantum algorithm, or any numerical
simulation, would no longer be required to calculate the dynamics of the system
described by $\mathcal C$.  We will derive a bound on $\kappa_\mathcal{J}$ in three steps.
First, we will numerically diagonalize the single-point collision Carleman
matrix $C$ at Carleman order $k=3$ (\Eq{eq:Cdef}) for lattice structures used
to recover the NSE from the LBE; these are specific matrices with
fixed, explicitly known $a$, $b$, $c$, $d$, $w_m$, and $Q$ parameters
in \Eq{eq:lbm-grouped}, but they can solve
general flows \cite{Chen1998LatticeFlows}.  Second, we will show that the
eigenvectors of the $n$-point collision Carleman matrix $\mathcal C_c(\vec x)$,
defined in \Eq{eq:npt-coll}, can be constructed from the eigenvectors of the
single-point $C$.  Third, we will perform a perturbative expansion to include
the effect of the streaming Carleman matrix $\mathcal C_s(\vec x)$, defined in
\Eq{eq:npt-stream}, and derive an approximate expression for $\kappa_\mathcal{J}$ that
holds as long as the Knudsen number of the flow is small; this condition needs
to be satisfied for the LBE to describe the flow.  The numbered paragraphs below
provide the details of these three steps.

\begin{figure}
  \centering
  \includegraphics[width=12cm]{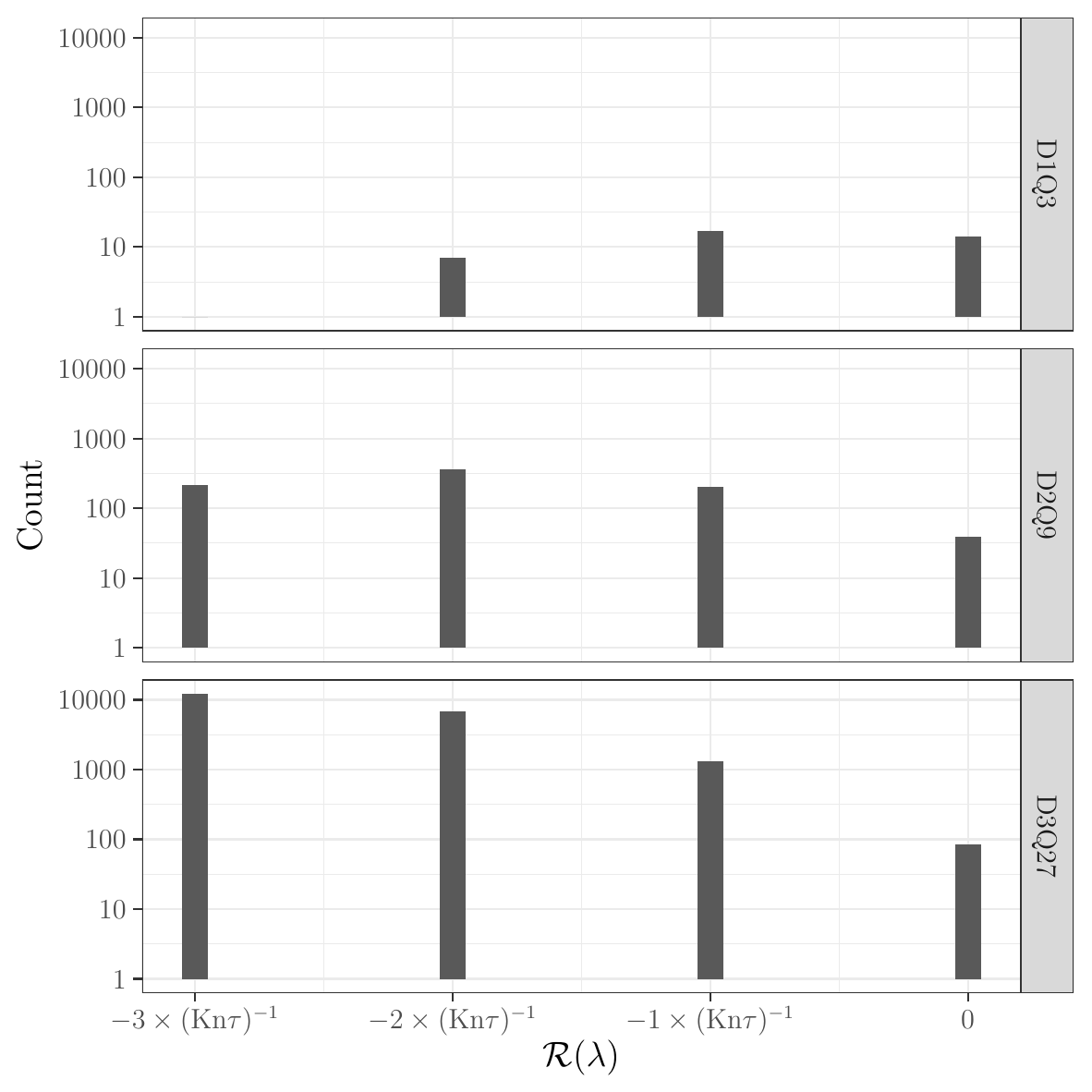} 
    \caption{Histograms of eigenvalue spectra $\mathcal{R}(\lambda)$ of 3rd-degree Carleman-linearized D1Q3, D2Q9, and
    D3Q27 LBM matrices.  Up to floating point error $<10^{-12}$ from the numerical
    diagonalization, the eigenvalues are all discrete multiples $-\{0, 1, 2, 3\}
  \times (\kn\tau)^{-1}$ of the Boltzmann relaxation scale $(\kn\tau)^{-1}$.
  Again up to numerical fuzz $<10^{-13}$, all eigenvalues are purely real.}
  \label{fig:spectral}
\end{figure}

\begin{enumerate}
\item In one, two, and three spatial dimensions, the D1Q3, D2Q9, and D3Q27
  lattice topologies are frequently used and are known to recover the NSE in the
  $\ma\ll 1$ limit \cite{Chen1998LatticeFlows} on which the present work
  focuses.  The single-point collision Carleman matrices $C$ and their
  spectral decompositions $C=JDJ^{-1}$ are provided as a data
  supplement \cite{url}. 
  By inspection of these numerical results,
  all eigenvalues of $C$ are 
  $-(\kn\tau)^{-1} \times \{0, 1, 2, 3\}$, that is, they are either degenerate or separated by gaps of
  $(\kn\tau)^{-1}$, as shown by the histogram of the eigenvalue spectra
  $\mathcal{R}(\lambda)$ in Fig.~\ref{fig:spectral}.  The eigenvector matrices
  $J$ and $J^{-1}$  are similarly constant for each choice of $Q\in\{3,9,27\}$.
  The corresponding condition numbers are $\kappa_J = 12.94$ for D1Q3, $\kappa_J
  = 85.66$ for D2Q9, and $\kappa_J = 2269$ for D3Q27.  Bounds based on
  \Eq{eq:specnorm} are $\kappa_J \le 62.41$ for D1Q3, $\kappa_J
  \le 2252$ for D2Q9, and $\kappa_J \le 1.152\times 10^5$ for D3Q27; these bounds will be used
  to generalize to $n$ grid points below.
\item Denote the eigenvectors of the single-point $C$ as $\xi_i$, $i\in
  1,\dots,Q$, and decompose
  \begin{equation}
    \xi_i = (\xi_i^{(1)}, \xi_i^{(2)}, \xi_i^{(3)})^\trans
  \end{equation}
  according to whether its elements represent 1-, 2-, or 3-form elements (powers of $f_m$ polynomials defined in \Eq{eq:lbm-carleman}) of the
  single-point Carleman vectors $V$.  Using the $\delta_\alpha$ vector from
  \Eq{eq:delta-alpha}, we can construct localized eigenvectors $\Xi_i(\vec
  x_\alpha)$ of the $n$-point collision Carleman matrix (\Eq{eq:npt-coll}):
  \begin{equation}
    \label{eq:Xi}
    \Xi_i(\vec x_\alpha) =
    (\delta_\alpha \otimes \xi_i^{(1)},
    \delta_\alpha^{[2]} \otimes \xi_i^{(2)},
    \delta_\alpha^{[3]} \otimes \xi_i^{(3)})^\trans.
  \end{equation}
  By \Eq{eq:diagonality-A}, $\Xi_i(\vec x_\alpha)$ is an eigenvector of the
  $n$-point collision matrix $\mathcal C_c^{(3)}(\vec x)$ for any
  $\alpha\in 1,\dots,n$, as the diagonality of the $n$-point transfer matrices
  in $\vec x$ ensures that the nonzero blocks of $\mathcal C_c$ are aligned with
  the nonzero blocks of $\Xi_i(\vec x_\alpha)$.  The eigenvectors of the
  $n$-point collision Carleman matrix $\mathcal C_c$ are therefore a simple
  blockwise repetition of the eigenvectors of the single-point $C$.
  Furthermore, it follows from \Eq{eq:Xi} that
  \begin{equation}
    \label{eq:Xi-ortho}
    \Xi_i(\vec x_\alpha)^\trans \Xi_j(\vec x_\beta) = \xi_i^\trans \xi_j \delta_{\alpha\beta},
  \end{equation}
  i.e., the eigenvectors at different grid points are mutually orthogonal.  The
  corresponding eigenvalues are equal for all $\alpha \in 1,\dots,n$, i.e., they
  are repetitions of the eigenvalues of $C$.

  The $\Xi_i(\vec x_\alpha)$ are stacked column-wise or row-wise to
  form the eigenvector matrices of $\mathcal C_c$.  Denote these eigenvector
  matrices as $\mathcal J_0$ and $\mathcal J_0^{-1}$.  (The subscript 0 notation
  is used in anticipation of the perturbation analysis carried out in the next
  paragraph.)  By \Eq{eq:Xi-ortho}, the eigenvectors localized at different grid points
  form blocks in $\mathcal J_0$ and $\mathcal J_0^{-1}$ that ensure that row and
  column sums on $\mathcal J_0$ and $\mathcal J_0^{-1}$ consist of the
  same $O(Q^3)$
  nonzero elements as row and column sums on the \textit{single-point constant}
  eigenvector matrices $J$ and $J^{-1}$, rather than the full
  $O(n^3Q^3)$ dimension of the $n$-point 
  matrices.  Thus, the same spectral-norm bound from
  \Eq{eq:specnorm} applies to $\|J\|$ and $\|\mathcal J_0\|$; and
  the same bound applies to $\|J^{-1}\|$ and $\|\mathcal J_0^{-1}\|$.
  Thus, $\kappa_{\mathcal{J}_0}$ is fixed (for fixed $Q$), independent
  of $n$.
  
\item Finally, we bound the effect of the streaming contribution $\mathcal C_s$
  to the eigenvectors of $\mathcal C=\mathcal C_s + \mathcal C_c$ and thence the
  effect on $\mathcal J$, $\mathcal J^{-1}$, and $\kappa_{\mathcal J}$.
  Inspection of \Eq{eq:lbm-grouped} shows that the streaming term in the LBE
  is suppressed by $\kn\ll 1$ relative to the collision term; this suggests that
  we can treat streaming using well established principles of (degenerate)
  perturbation theory \cite{Sakurai1994, bamieh2020tutorial}, with the collision Carleman matrix in
  \Eq{eq:n_point_Carleman} constituting the base matrix (with known
  eigenvectors) and the streaming term constituting the $O(\kn)\ll 1$
  perturbation.

  Item 2.\ above derived the properties of the $n$-point collision eigenvector
  matrices $\mathcal J_0$ and $\mathcal J_0^{-1}$ in terms of the single-point
  eigenvector matrices $J$ and $J^{-1}$ that had to be computed numerically.
  When the perturbation due to $\mathcal C_s$ is included, the eigenvectors are
  perturbed \cite{Sakurai1994} in proportion to the perturbation strength
  ($\kn\tau$) and the gap size of the eigenvalue spectrum of $\mathcal C_c$,
  neglecting until the next paragraph the issue of degenerate eigenvalues.  For
  notational convenience, we suppress the $\vec x_a$ notation on the
  eigenvectors of $\mathcal C_c$; the perturbed eigenvectors to first order are
  then given by $\Xi_i + \Xi_i'$, where
  \begin{equation}
    \label{eq:pertnondeg}
    \Xi_i' =  \kn\tau \sum_{k\neq i} \frac{\Xi_i^{-1} \mathcal C_s \Xi_k}{\lambda_i -
      \lambda_k} \Xi_k.
  \end{equation}
  Numerical calculation of the eigenvalue spectrum of the single-point Carleman
  matrices in item 1.\ above showed the gap between nondegenerate eigenvalues
  to be integer multiples of $(\kn\tau)^{-1} \gg 1$, so the perturbation to the eigenvectors is
  proportional to $O((\kn\tau)^{2}) \ll 1$ times the expectation values $\sum\Xi_i^{-1}
  \mathcal C_s \Xi_k$.  In the nondegenerate case, the eigenvectors
  form an orthonormal basis, and the vector norm of the $\Xi_i'$ can be bounded
  as follows:
  \begin{align}
    \vert \Xi_i'\vert^2 &= (\kn\tau)^2 \sum\limits_{k\neq i}
    \frac{\Xi_i^{-1}\mathcal C_s \Xi_k
    \Xi_k^{-1}\mathcal C_s^\dagger \Xi_i}{(\lambda_k -
                          \lambda_i)^2} \nonumber \\
    &\leq \frac{(\kn\tau)^2}{\min_k (\lambda_k - \lambda_i)^2} \sum\limits_{k\neq i}
    \Xi_i^{-1}\mathcal C_s \Xi_k \Xi_k^{-1}\mathcal C_s^\dagger
      \Xi_i \nonumber \\
    \label{eq:temp1}
    &\leq \frac{(\kn\tau)^2}{\min_k (\lambda_k - \lambda_i)^2} \left[\sum\limits_k
      \Xi_i^{-1}\mathcal C_s \Xi_k \Xi_k^{-1}\mathcal C_s^\dagger \Xi_i
    - \left\vert \Xi_i^{-1} \mathcal C_s \Xi_i\right\vert^2\right].
  \end{align}
  Using completeness and noting that the final term in \Eq{eq:temp1} is
  nonpositive-definite,
  \begin{equation}
    \vert \Xi_i'\vert^2 \leq \frac{(\kn\tau)^2}{\min_k (\lambda_k -
    \lambda_i)^2} \Xi_i^{-1} \mathcal C_s \mathcal
    C_s^\dagger \Xi_i,
  \end{equation}
  and therefore
  \begin{equation}
    \vert \Xi_i'\vert \leq \frac{\kn\tau}{\min_k (\lambda_k -
      \lambda_i)} \|\mathcal C_s \|.
  \end{equation}
  The eigenvalues of the streaming operator correspond to the spectrum of
  wavenumbers supported by the spatial discretization.  On the lattice, these
  wavenumbers are rescaled according to \Eq{eq:scale-x} so that the maximum
  wavenumber is $O(1)$, and hence
  \begin{equation}
    \label{eq:xi}
    \vert \Xi_i'\vert = O\left(\frac{\kn\tau}{\min_k (\lambda_k -
      \lambda_i)}\right).
  \end{equation}
  As we determined numerically in
  Fig.~\ref{fig:spectral}, there are up to four (only three in
  the one-dimensional LBE) discrete eigenvalues $-\{0, 1, 2, 3\}
  \times (\kn\tau)^{-1}$, so 
  \begin{equation}
    \label{eq:vecnorm-xiprime}
    \vert \Xi_i'\vert = O((\kn\tau)^2).
  \end{equation}
  
  The repetition of the single-point Carleman eigenvalues in the $n$-point
  Carleman eigenvalue spectrum means that each eigenvalue is highly degenerate.  We
  therefore need to treat the case of eigenvalue degeneracy in
  \Eq{eq:pertnondeg}.  This, too, is a textbook application of perturbation
  theory \cite{Sakurai1994}.  The procedure treats each degenerate
  subspace of $\mathcal J_0$ and chooses the eigenbasis such that it diagonalizes the perturbation within each of the subspaces.  There are up to four (only three in
  the one-dimensional LBE) such degenerate subspaces
  corresponding to the four discrete eigenvalues $-\{0, 1, 2, 3\}
  \times (\kn\tau)^{-1}$; we label these spaces $D_0$ through $D_3$.
  For each space, we define a projection operator
  \begin{equation}
    \label{eq:projection}
    P_l = \sum\limits_{\Xi_i \in D_l} \frac{\Xi_i \Xi_i^\trans}{\|\Xi_i\|^2}.
  \end{equation}
  We then diagonalize the perturbation operator $\mathcal C_s$ in each
  degenerate subspace in turn.  
  For the $l$th subspace, the
  first-order eigenvector perturbation is given by
  \begin{equation}
    \label{eq:pertdeg}
    P_l \Xi_i' = (\kn\tau)^2
    P_l \sum\limits_{j\neq i} \frac{\Xi_i}{\mu_j - \mu_i}
    \sum\limits_{k\notin D_l} \Xi_i^{-1} \mathcal C_s \Xi_k
    \frac{1}{\lambda_{D_l} - \lambda_k} \Xi_k^{-1} \mathcal C_s \Xi_j \ ,
  \end{equation}
  where the projection operator ensures that the degeneracy-breaking
  diagonalization of $\mathcal C_s$ is only performed within each
  degenerate subspace and $\mu_{i(j)}$ are the eigenvectors of
  $\mathcal C_s$; between subspaces, \Eq{eq:pertnondeg} continues
  to apply.  The scaling of the terms in \Eq{eq:pertdeg} is
  familiar except for $\sum (\mu_j - \mu_i)^{-1}$.  This term
  needs careful examination because the smallest difference in
  eigenvalues of $\mathcal C_s$ is set by the smallest difference in
  scaled wavenumbers, which scales with the number of grid points as
  $n^{-1/D}$ in $D\in\{1,2,3\}$ spatial dimensions.  In the $n\gg 1$ limit, we can
  bound this factor by
  \begin{equation}
    \label{eq:mu}
    \sum\limits_{j\neq i} \frac{1}{\mu_j - \mu_i} \propto
    \int\limits_{\vec{k}_i \neq \vec{k}_j} \frac{d^D\vec{k}_j}
    {\vert\vec{k}_j - \vec{k}_i\vert} \propto  
    \int\limits_{k =
    O\left(n^{-\frac 1 D}\right)}^{O(1)} \frac{k^{D-1}dk}{k} = O\left(\log n\right),
  \end{equation}
for the most restrictive case, $D=1$.
For $D=2$ and $D=3$, \Eq{eq:mu} is reduced to $1-n^{-1/2}$ and $\frac{1}{2} (1-n^{-1/3})$, respectively, which becomes $0$ in the $n\gg 1$ limit. 

  Combining the various factors, the degenerate perturbations to $\Xi_i$ are suppressed
  relative to the unperturbed eigenvectors (treating $D$ as a fixed
  value rather than a parameter of the problem) by $O((\kn\tau)^3\log
  n)$.  The nondegenerate perturbations are suppressed
  by $O((\kn\tau)^2)$.  The singular values of $\mathcal J$ are not
  necessarily smooth but can be evaluated from the eigenvalues of the
  square of $\mathcal J$, i.e., $\Sigma = {\mathcal{J}}^\dagger {\mathcal{J}}$.  We then have that \begin{equation}
        \|{\mathcal{J}}^\dagger {\mathcal{J}} - {\mathcal{J}_0}^\dagger {\mathcal{J}_0}\| \le (\|\mathcal{J}\| +\|\mathcal{J}_0\|)\|\mathcal{J} - \mathcal{J}_0\|.
  \end{equation}  This shows that the singular values remain smooth provided that $\mathcal{J}_0$ is non-singular and further  places
  the restrictions
  \begin{align}
    (\kn\tau)^2 \ll 1 \\
    \label{eq:kn-logn}
    (\kn\tau)^3 \log n \ll 1 
  \end{align}
  on the free parameters of the system.  The first condition is less
  restrictive than the $\kn\tau \ll 1$ condition the system must
  already satisfy, while the second is unlikely to be a practical
  impediment for reasonable values of $n$.

  Lastly, we need to bound the effect of the perturbation on
  $\|\mathcal J^{-1}\|$.  To accomplish this, we use the approximation
  \begin{equation}
    \label{eq:J}
    \mathcal J^{-1} = (\mathcal J_0 + \mathcal J')^{-1} \approx  \mathcal
    J_0^{-1} - \mathcal J_0^{-1} \mathcal J' \mathcal J_0^{-1}
  \end{equation}
  where $\mathcal J' = \mathcal J - \mathcal J_0$ is the perturbation
  to the eigenvectors; since this perturbation scales as
  $O((\kn\tau)^2)$, the neglected terms in the expansion of the
  inverse scale as $O((\kn\tau)^4)$.  By the submultiplicative property
  of the spectral norm,
  \begin{equation}
    \|\mathcal J^{-1} - \mathcal J_0^{-1}\| \le \|\mathcal J_0^{-1}\|^2 \|\mathcal J'\|.
  \end{equation}
  Using $\|\mathcal J_0^{-1}\|$
  from the block structure
  of $\mathcal C_c$ and $\max_i \vert \Xi_i'\vert = O((\kn\tau)^2)$ from
  \Eq{eq:vecnorm-xiprime} to bound $\|\mathcal J'\|$, we conclude that the
  $n$-point $\|\mathcal J^{-1}\|$ is also independent of $n$.
  Finally, we numerically bound $\kappa_{\mathcal J}$ as
  \begin{equation}
    \label{eq:kj-value}
    \kappa_{\mathcal J} \le 1.152 \times 10^5.
  \end{equation}
\end{enumerate}

\subsubsection*{Final quantum gate complexity to solve \Eq{eq:clbm}}
Finally, inserting  \Eq{eq:epsilon}, \Eq{eq:evolutionT}, \Eq{eq:g-decay},  \Eq{eq:sparsity}, \Eq{eq:normc}, and \Eq{eq:kj-value} 
into \Eq{eq:ca} yields the following result for the quantum algorithm complexity as quantified by the number of two-qubit gates needed in the simulation:
\begin{equation}
  \label{eq:complong}
  \text{gate complexity} = O(t_c^{-1}\tilde{T}\poly\log(n/\epsilon)),
\end{equation}
where $\tilde T$ is the physical evolution time.
The scaling of each term in \Eq{eq:ca} is summarized in Table~\ref{tab:scaling}.
\begin{table}
  \centering
  \begin{tabular}{lll}
    \hline\hline
    Term & Scaling & Reference \\ \hline
    $\beta$    & $O(1)$                &  \Eq{eq:beta}     \\
    $g$        & $3\sqrt{k}$                &  \Eq{eq:g-decay}        \\
    $s$        & $O(1)$              &  \Eq{eq:sparsity} \\
    $\|\mathcal C\|$    & $O(\kn^{-1})$ & \Eq{eq:normc}    \\
    $\kappa_J$ & $\le 1.152\times 10^5$ & \Eq{eq:kj-value}        \\
    \hline\hline
  \end{tabular}
  \caption{Scaling of each term in \Eq{eq:ca} with lattice Boltzmann equation parameters leading to the
    result of \Eq{eq:complong}. Note that the dissipation parameter $g$ is a constant because the Carleman linearized LBE converges at the truncation order $k=3$.  
    }
  \label{tab:scaling}
\end{table}

The solution error of using the quantum
linear system algorithm (QLSA) of
\citet{Berry2017QuantumPrecision} to solve
\Eq{eq:LBMs} come from two sources: the
Carleman truncation error and the QLSA
error. Since the solution error from QLSA
contributes $\text{poly}(\log(\epsilon)$ to
the gate complexity expressed in \Eq{eq:ca},
we focus on bounding the Carleman truncation
error, which is $O(\text{Ma}^2)$. For weakly
compressible flow, $\text{Ma}\ll1$,
therefore, the Carleman truncation error is
very small. In \citet{Liu21}, a QLSA with
gate complexity scales with $T^2$ was used
due to their time dependent Carleman matrix.
A time dependent Carleman matrix requires
discretizing time using the forward Euler
method, which contributes to the solution
error. In contrast, our Carleman matrix
$\mathcal{C}$ is constant, which affords us
to use the QLSA of
\citet{Berry2017QuantumPrecision} with a
gate complexity scales with $T$
(\Eq{eq:ca}). Overall, our method of using
QLSA to solve \Eq{turb} has yields a gate
complexity scaling with $\poly(\log{N})$ and
$T$ for any arbitrary Reynolds number
comparing to \citet{Liu21}'s complexity
scaling with $T^2$ for $R<1$.

\section{Discussion}
\ifarXiv
\newcommand{\rem}{Results}
\fi

\subsection{Trading nonlinearity for degrees of freedom}
\label{sec:trading}

The first step in our process of adapting nonlinear fluid dynamics to the requirements of
quantum algorithms is to use 
the lattice Boltzmann form of the NSE (see \rem).  The lattice
Boltzmann equation (LBE) is a nondimensionalized form of the Boltzmann equation
with discrete velocities and a collision term that often follows the Bhatnagar--Gross--Krook form
\cite{Bhatnagar1954ASystems}.  It is called ``lattice'' 
because it yields Lagrangian-like solutions when solved with first-order spatial
and temporal discretization with upwind advection \cite{Chen1998LatticeFlows} in
which the discrete-velocity particles arrive at neighboring grid points after
exactly one time step. This solution method is referred to as the lattice
Boltzmann method (LBM).
For our purpose, the essential characteristics of the LBE are the linear advection term
(inherent to the Boltzmann formulation) and the manifest form of the degree of nonlinearity
in the collision term (due to the nondimensionalization).

The advection term is linear in the kinetic-theory-motivated LBE because the discrete velocity distributions are advected in space by
constant lattice velocities, the analogue of which in the kinetic theory of
fluids is the speed of sound.  
The collision-term nonlinearity is instead $O(\text{Ma}^2)$, where $\text{Ma}$
is the Mach number, defined as the ratio between the characteristic
fluid velocity and the speed of sound.  This is because the fluid velocity $\bm{u}$ is proportional to the first moment
of the particle distribution function.  In nearly incompressible flows, $\bm{u}$
is always small relative to the speed of sound.  The nondimensional velocities,
which are scaled by the lattice velocity, make the $\vert \vec{u} \vert =
O(\ma)$ scaling explicit.

Thus, the character of the nonlinearity is fundamentally changed between the NSE
and LBE. The
nonlinearity of the NSE stems from the term $\bm{u}\cdot \nabla
\bm{u}$, and this nonlinearity is characterized by $\text{Re}$. In contrast, the
nonlinearity in the LBE stems from $\vert\bm{u}\vert^2$, and this nonlinearity is characterized
by $\text{Ma}^2$.  This change is advantageous for high-Re, low-Ma flows, the
situation in most geophysical and many engineering applications.  
However, it comes at the expense of using a large number of discrete lattice
velocities to form the quadrature ($Q$) that simulates the molecular
velocity distribution. For instance, to use LBM to simulate three-dimensional NSE,
the number of degrees of freedom must be increased by a factor of 27 in the D3Q27 formulation. Here, the number after \textit{D} stands for the dimension of the system, and that after \textit{Q} represents the number of discrete lattice velocities in the quadrature.
For $n$ grid points, the number of LBE degrees of freedom increases $Q$-fold to
$nQ$. 

The second step in our process of adapting nonlinear fluid dynamics to the requirements of
quantum algorithms is to eliminate the weakened 
  nonlinearity in the lattice Boltzmann formalism using Carleman linearization \citep{Carleman1932,Forets2017} (see \rem).
Carleman linearization leads to an infinite-dimensional system of linear
differential equations.  The system is truncated at a finite degree for
practical implementation, resulting in a  truncation
error.  In work that has pursued this avenue \citep{Liu21,
  Lloyd2020QuantumEquations}, the truncation error depends on the degree of
nonlinearity in the original equation \citep{Forets2017, Liu21}.  For turbulent
flows, \citet{Liu21} find that the Carleman-linearized Burgers' equation is
intractable for quantum algorithms at $R \ge 1$.  We show, in
contrast, that the truncation error for the Carleman-linearized
LBE is a power series in Ma, as 
expected for a system of equations whose nonlinearity is characterized by Ma; therefore, the truncation error is strongly suppressed for weakly compressible flow and becomes independent of Re (see \rem).

As discussed in the \rem, the lowest Carleman linearization degree that can
be used for the LBE is 3 (i.e., terms up to cubic degree in the distribution
functions appear in the truncated equations).  This degree is also sufficient,
as the associated truncation error is of the same order (i.e., $\ma^2$) as the LBE error in
reproducing the NSE \citep{Chen1998LatticeFlows}.

We note one further difference between our analysis and the truncation error analyses of \citet{Liu21} and
\citet{Forets2017}.  These prior analyses provide error bounds for arbitrary nonlinear differential equations
based on the coefficient-matrix norms for the linear and nonlinear terms.  Our
results provide much tighter error bounds due to two specific properties of the
LBE.  First, the symmetries of the lattice lead to exact
cancellation of certain terms, which makes the truncation error for powers of $u$
identically zero.  Second, we benefit from the velocity rescaling on the lattice, which, as discussed earlier, suppresses the nonlinear terms by $O(\text{Ma})$ relative to the linear terms \citep{LALLEMAND2021109713}.

After Carleman linearization, the final number of degrees of freedom in the
LBE is $N=n^3Q^3 + n^2Q^2 + nQ = O(n^3Q^3)$.  This is a dramatic
increase over the $n$ degrees of freedom required for NSE at the
same resolution.  However, this is a polynomial increase; if it permits us to take
advantage of $O(\log N)$ scaling, then this is a trade-off well worth making.

\subsection{Quantum speedup for turbulence}

Accurate linear approximation of the LBE in hand, we now need to determine whether the $\O(\poly\log N)$ scaling of quantum linear systems solvers can be achieved for our particular system of equations \cite{Aaronson2015,Berry2017QuantumPrecision}.

First, it must be possible to load the initial state and extract the final state
in at most $\O(\poly\log N)$ time. This constraint precludes applying this method to
initial-value sensitive problems (e.g., numerical weather prediction).  However, a vast class of
interesting physical applications (e.g., climate projections) are 
boundary-value sensitive problems; for these problems, the
initial distribution can be highly idealized (e.g., as a Gaussian distribution),
and the turbulent features can be allowed to spin up over time.  Such
an initial state can be prepared
in the required time using methods such as
\citet{kitaev2008wavefunction,grover2002creating}.
The information extracted from the final state needs to fit into a single scalar
that results from measuring the expectation value of an operator on the final
state.  This is because of the infamous output problem in quantum
computing~\cite{Aaronson2015} wherein the cost of reading the solution from a
quantum algorithm is exponentially greater than the cost of finding the quantum
solution.  This usually occurs when one wishes to learn the entire solution as a
quantum state and, thus, is forced to perform quantum state
tomography~\cite{cramer2010efficient}.  Expectation values can typically be
learned at modest cost by sampling quantities such as the domain-mean flux
across a boundary. This is a meaningful quantity that can be extracted from
sampling the state and that does not incur the exponential overhead if one is willing
to accept errors that are inverse polynomial in the system
size. We remark that our algorithm produces a state vector $\ket{\mathcal V}$ that encodes the solution without providing a detailed procedure of extracting the solution from $\ket{\mathcal V}$. Developing an efficient measurement method to ensure the end-to-end quanutm advantage will be explored in future studies.
  
Second, the quantum solver needs to be able to evolve the initial state to the final state (i.e., the state after an evolution time $\tilde T$) in at most $\O(\poly\log N)$ time.  Here, we make use of the known complexity of state-of-the-art quantum algorithms for ODEs from \citet{Berry2017QuantumPrecision}; we eschew existing approaches to nonlinear differential equations that only use forward-Euler discretizations because of our specialized analysis of the error in Carleman linearization.  
Expressing each factor in the \citet{Berry2017QuantumPrecision} complexity
formula (\Eq{eq:ca}) in terms of the properties of the Carleman-linearized LBE
results in \Eq{eq:complong}. 
The system parameters that determine the complexity are the 
number of discrete lattice velocities $Q$, the LBE collision time scale $t_c$, 
the evolution time $\tilde T$, and the number of grid points $n$. 

The number of discrete velocities is fixed and independent
of other system parameters; as discussed, $Q\leq 27 \ll n$ is widely used in three-dimensional LBM simulations.  
Evolution time is an external requirement that is determined by the
physics problem being solved. Collision time is associated with the
molecular nature of the fluid that can be decoupled from
the smallest turbulent time scale (here, the Kolmogorov time scale). The resulting complexity scales as $\poly\log(n)$.  This opens the possibility of simulating enormous domains while resolving a complex interplay between physics ranging from Kolmogorov to integral scales.

The overall complexity, therefore, can be set to be independent of the NSE nonlinearity (i.e., the Reynolds number).
This conclusion is aligned with a fundamental assumption of the NSE: the continuum assumption that the infinitesimal volume of fluid is much larger than that associated with molecules, which allows the molecular details of the fluid to be dropped. Because of this assumption, the degree of nonlinearity in the LBE formulation can be adjusted independently of that in the NSE. 
This remarkable result comes from the fundamental difference, discussed above, between the Reynolds-type nonlinearity in the original NSE and the Mach-type nonlinearity in their LBE form.

\ifarXiv
\section{Conclusion and outlook}
\fi

We have shown how to use quantum solvers to simulate turbulence.
The resulting logarithmic scaling in number of degrees of freedom compares to the polynomial scaling of the gold-standard
classical algorithms \cite{Orszag1970,Exascale2017}.  (We make no claim that the
best known classical algorithms are, in fact, the best possible \cite{Tang21}.)
This marks an important
milestone in understanding turbulence through numerical
simulations.  Much ground remains to be covered before reaching the ultimate destination
of applying this tool
to actual complex physical
systems; at the same time, previously unthinkable possibilities are now
visible on the horizon.

The specific manifestation of turbulence depends sensitively on boundary
conditions.  We have not considered these in our idealized derivations, but work on doing so 
is urgently needed.  Furthermore, many complex physical systems are
heterogeneous, consisting of multiple coupled subsystems and multiple species or thermodynamic phases.
There are well-established ways to impose complex boundary
conditions and to add phases, phase interactions, phase change, and reactions in
the LBM \citep{Zou97, Ladd2001, Guo02}.
Such schemes, combined with further development based on our results, can be
applied to challenging multiphase or reactive flow problems, such as droplet
coalescence under turbulence \cite{Chun2005,Li2018EffectDroplets} and nanoparticle
assembly under an external field that involves a fluid flow \citep{Saville1977,Boles2016}.
In essence, our method will need to be extended by further increasing the number of degrees of freedom until
not just the Kolmogorov scales but even the solid-particle or liquid-droplet microscale is
explicitly resolved. 
  A large number of applications have already been simulated using the LBM \citep{Martys01, Ladd2001,  CHEN2014210, LI201662,Liu16, HE2019160,Petersen21, SAMANTA2022111288, NOURGALIEV2003117, Aidun10, Bernaschi2019}, and these could immediately benefit from such work. 
  
More generally, many systems besides the NSE do not meet the linear and
nondissipative requirements of quantum algorithms at first glance. There is
often a trade-off between linearity and the number of explicit degrees of
freedom sampled. The full dimensionality of a molecular dynamics simulation versus
the reduced dimensionality of the nonlinear NSE hydrodynamics is a canonical example.
Based on our LBE method, it is plausible that---and it should be urgently
tested whether---many of the nonlinear multiscale transport phenomena
described by the Boltzmann transport equation would also benefit from a quantum speedup.

\section{Acknowledgments}
The capability to apply the Carleman-linearized LBE to turbulent fluid dynamics, quantum algorithm complexity analysis, and domain applications to atmospheric science were developed 
under the Laboratory Directed Research and Development Program at Pacific
Northwest National Laboratory, a multiprogram national laboratory operated by
Battelle for the United States Department of Energy (DOE). 
The code development was partially funded by the Co-design Center for Quantum Advantage (C2QA) under contract number DE-SC0012704 (PNNL FWP 76274). 
GKS and MSC were funded to consider domain applications to chemical physics and biophysics under U.S. DOE, Office of Science, Office
of Basic Energy Sciences, Division of Chemical Sciences, Geosciences, and
Bioscience, Chemical Physics and Interfacial Sciences Program FWP 16249 and U.S. National Science Foundation grant NSF-MCB 2221824, respectively.
We thank Dominic Berry for a correction to an earlier version of the preprint. 
We thank Jin-Peng Liu and Hari Krovi for insightful discussions of Carleman linearization and QLSA during the revision of this work.
We thank Jim Ang, Larry Berg, Jay Bardhan, Leo Donner, Samson Hagos,
Karol Kowalski, Ian Kraucunas, Ruby Leung, Bruce Palmer, Christina
Sackmann, Wendy Shaw, and Hui Wan for comments and discussions.

\clearpage

\appendix
\section{Single-point CLBM simulations}
\label{sec:simu}
\begin{figure*}[t!]\begin{center}
\includegraphics[width= \textwidth]{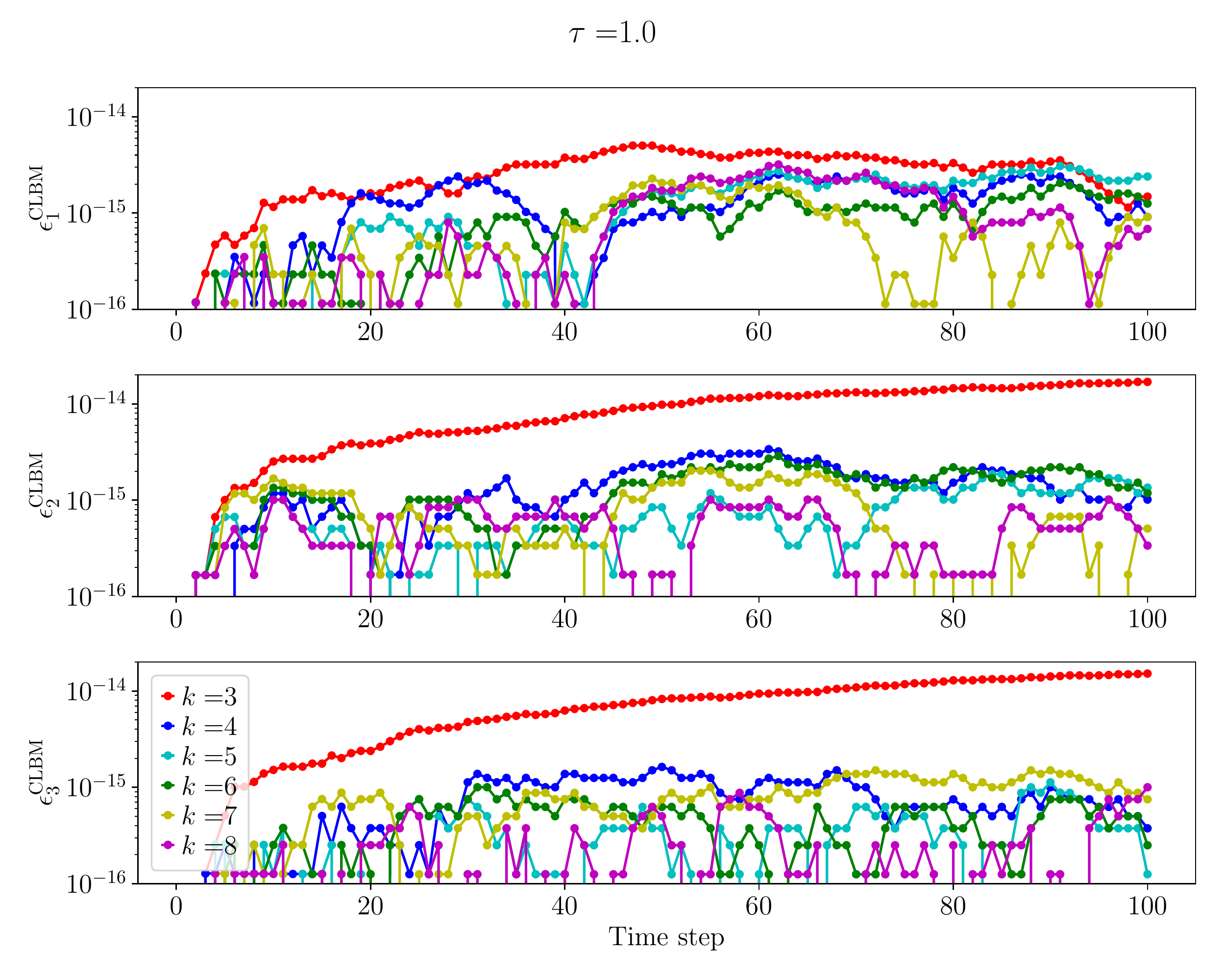}
\end{center}
\caption{Relative error $\epsilon^{\rm CLBM}_m = |f^{\rm CLBM}_m - f_m|/ f_m$ for LBM-D1Q3 with relaxation time $\tau=1.0$.}
\label{fig:CLBM_collision_error}
\end{figure*}

To test the convergence of Carleman truncation error as a function of the
truncation order, we perform LBM-D1Q3 with a single node and periodic boundary
conditions. We use a single node because the Carleman linearization is only
applied to the collision term. All simulations are initiated with an arbitrary initial
velocity.
The time step $dt=\tau/10$ is used for all
the simulations. The relative error ($\epsilon^{\rm CLBM}_m = |f^{\rm CLBM}_m -
f_m|/ f_m$) is on the order of $10^{-14}$ for the truncation order $k=3$ and on the order of
$10^{-15}$ for $k \ge 4$ (\Fig{fig:CLBM_collision_error}), which are close to the machine error. 
This numerically confirms our conclusion that the Carleman truncation error converges at order $k=3$ when Carleman linearing the LBE (\Eq{eq:LBMs}) for weakly compressible flow ($\text{Ma} \ll 1$). The close-to-machine-error $\epsilon_m^{\text{CLBM}}$ also suggests that the truncated CLBM can hardly be distinguished from the infinite CLBM (real solution) until $1/\epsilon_m^{\text{CLBM}}$. 
This remarkably low Carleman truncation error is particularly favourable to QLSA and paves our way to confidently apply QLSA to the Carleman-linearized LBE (\Eq{eq:clbm}).    

\begin{figure*}[t!]\begin{center}
\includegraphics[width=\textwidth]{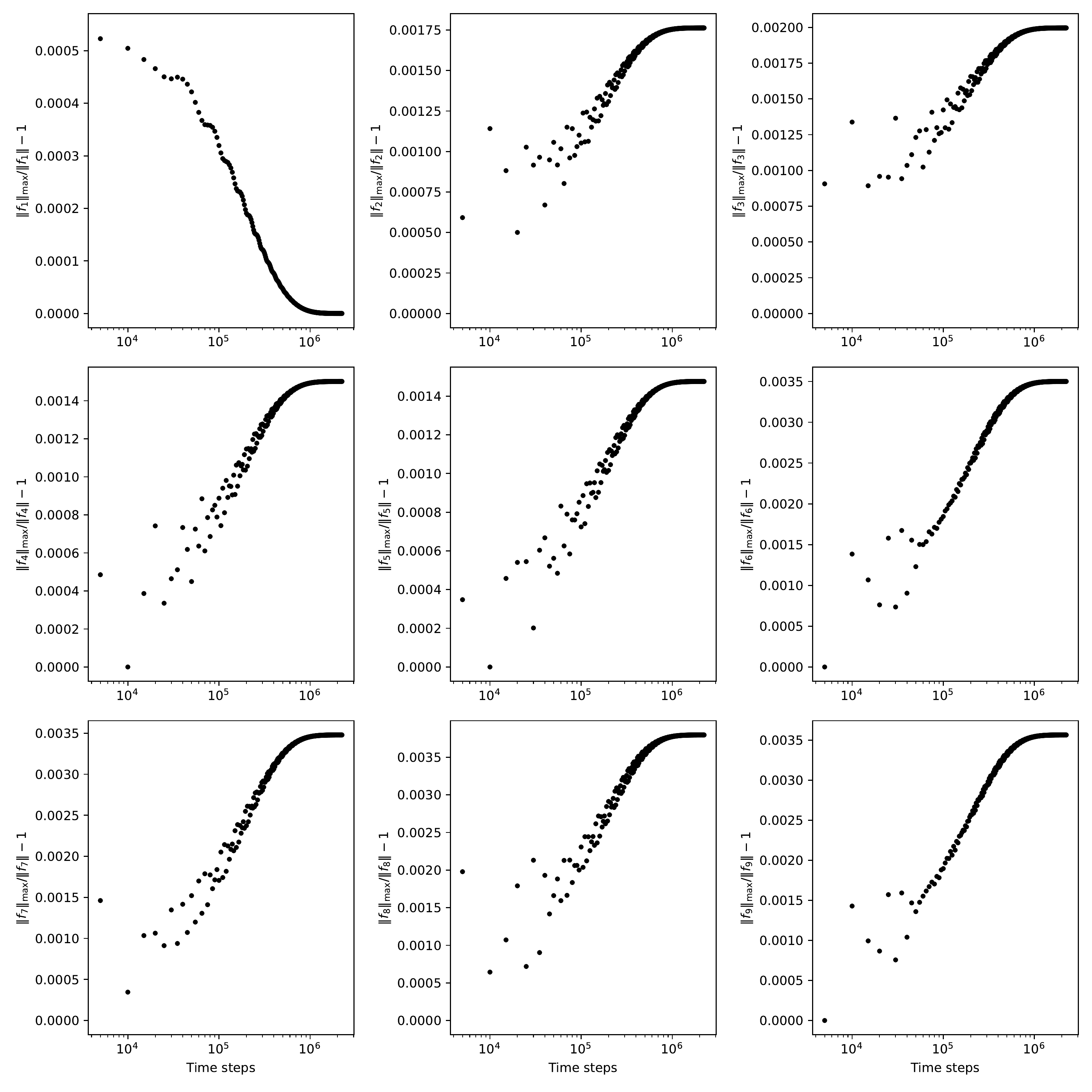}
\end{center}
\caption{$\underset{t\in[0,T]}{\max} \norm{
f(t)} / \norm{f(T)} -1$ from the D2Q9 LBM decaying simulaiton with $128^2$ grid points. The simulation is initialized with developed turbulence fields from a forced LBM simulaiton. The initial Reynolds number is $\text{Re}=400$ and Mach is $\text{Ma}=0.03$.}
\label{fig:ts_f-norm-D2Q9}
\end{figure*}

\section{An alternative algorithm}
\subsubsection*{Contributions to algorithm complexity of \citet{Krovi2023improvedquantum}}
  
We now analyse the complexity of solving \Eq{eq:clbm} using the algorithm of \citet{Krovi2023improvedquantum}, which does not require the diagonalizability of matrices compared to the algorithm of \citet{Berry2017QuantumPrecision}. The algorithm complexity of \citet{Krovi2023improvedquantum} is restated here: 
\begin{equation}
\label{eq:ca_Krovi}
\text{query complexity}=O ( g T \Vert \mathcal C\Vert E(\mathcal C)
\text{poly}(s, \log k, \log (1 + \frac{T e^2 \Vert b \Vert}{\Vert \mathcal V_T \Vert}), \log (\frac{1}{\epsilon}), \log (T \Vert \mathcal C\Vert E(\mathcal C)) )),
\end{equation}
with the gate complexity no greater than a factor of
\begin{equation}
\label{eq:ca_Krovi_gate}
O (\poly \log(1 + \frac{T e^2 \Vert b \Vert}{\Vert \mathcal V_T \Vert}, 1/\epsilon, T \Vert \mathcal C \Vert) ),
\end{equation}
where   
\begin{equation}
  \label{eq:ec-def}
  E(\mathcal C) = \underset{t\in[0,T]}{\text{sup}} \Vert \exp(\mathcal C t) \Vert,
\end{equation}
is adopted to bound $E(\mathcal C)$ without additional assumptions on $\mathcal C$, e.g., normality or diagonalizability \citep{Krovi2023improvedquantum}. $e$ is the Euler constant.

\paragraph{Norm of the matrix exponential $E(\mathcal C)$}

We now bound $E(\mathcal C)$ following the method of
\citet{Krovi2023improvedquantum} after the stability of $\mathcal C$ is ensured
as discussed in the paragraph above. 
We start with the single-point collision-only Carleman matrix $C$ of \Eq{eq:Cdef} as the collision dominates LBM.  
Similar nomenclature of
\citet{Krovi2023improvedquantum} is adopted in the following discussion. The
following quantities
\begin{align}
  \label{eq:ln-nn1a}
    &\sigma(C)=\{\lambda\,|\, \lambda \text{ is an eigenvalue of } C\}&\text{Spectrum}\\
  \label{eq:ln-nn1b}
    &\alpha(C)=\max \{\Re(\lambda) \,|\, \lambda\in \sigma(C)\}&\text{Spectral abscissa} \\
  \label{eq:ln-nn1c}
    &\mu(C)=\max\{\lambda\,|\, \lambda\in \sigma((C+ C^\dag)/2)\}&\text{Log-norm}
\end{align}
and bounds \citep{Dahlquist63} 
\begin{equation}
  \label{eq:ln-nn2}
    \exp(\alpha t)\leq \|\exp(C t)\|\leq \exp(\mu t)\leq \exp(\|C \|t)\,.
\end{equation}
are defined to facilitate our discussions.

We first reduce the 3-order LBE, \Eq{eq:lbm-carleman}, to a quadratic equation,
\begin{equation}
  \label{eq:lbm-carleman-reduce}
  \ddt{\hat f} = -S \hat f + F^{(0)} +  \hat F^{(1)} \hat f +  \hat F^{(2)}  \hat f^{[2]},
\end{equation}
where $\hat f_i = f^{[i]}$ (not to be confused with the subscript of $f_m$), $\hat f =(\hat f_1, \hat f_2, \hat f_3)^T$, 
\begin{equation}
  \label{eq:hatF1}
\hat F^{(1)} 
 = \begin{pmatrix}
A_1^1 & A_2^1 \\ 
0 & A_2^2
\end{pmatrix},
\end{equation}
and
\begin{equation}
  \label{eq:hatF2}
\hat F^{(2)} 
 = \begin{pmatrix}
   0, & 0, A_3^1 & 0 \\ 
   0, & 0, A_3^2 & A_4^2 
\end{pmatrix} .
\end{equation}
We denote $\hat{F}^{(0)} = F^0$ for consistency and ease of the following discussion.

\begin{figure}[t!]\begin{center}
\includegraphics[width=\textwidth]{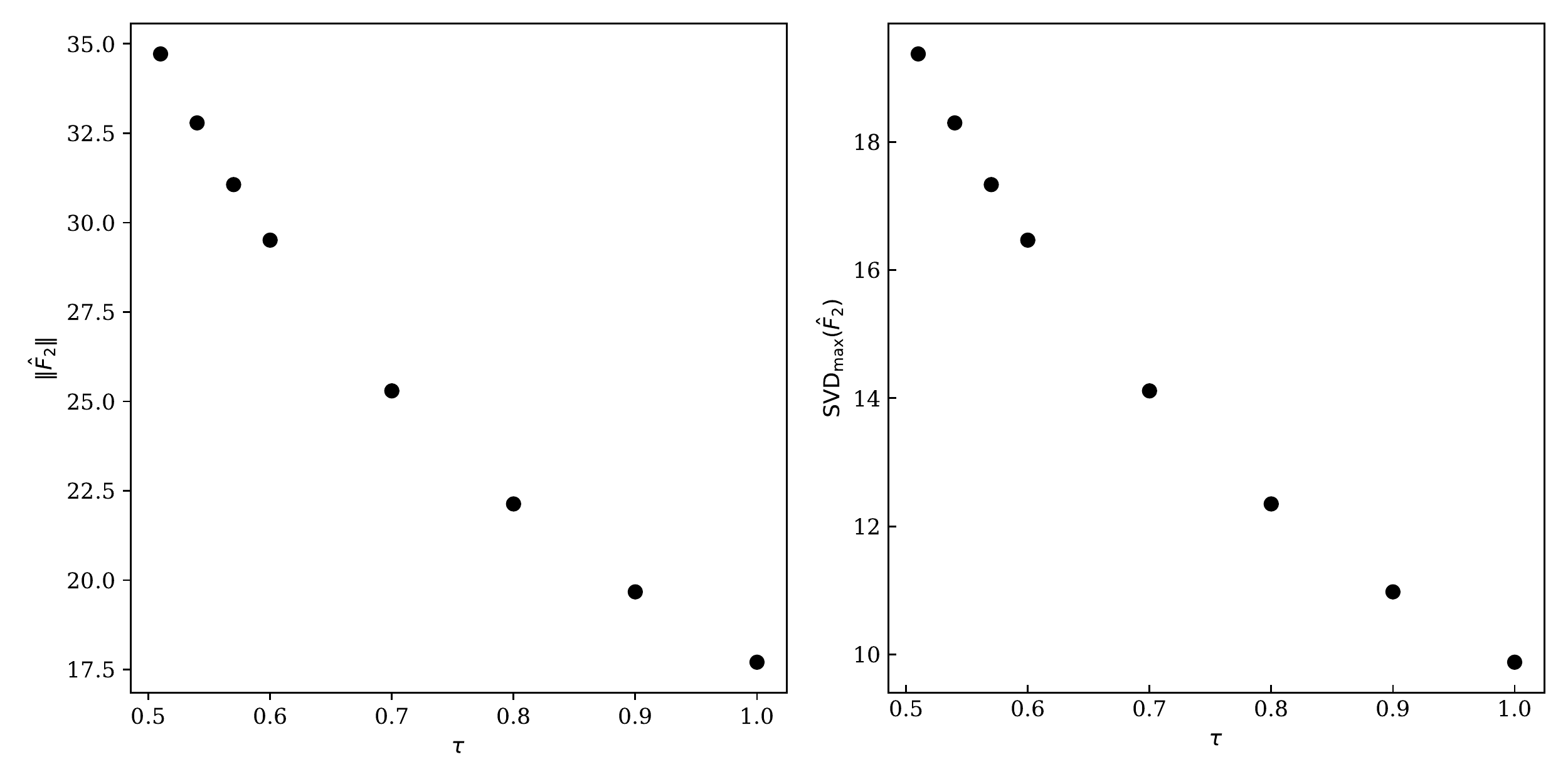}
\end{center}
\caption{$\| \hat{F}_2 \|$ and maximum SVD of $\hat{F}_2$ for the D1Q3 LBM.}
\label{fig:svd_norm_hat_F2_tau}
\end{figure}

  Following the proof of \citet{Krovi2023improvedquantum}, we first split $C = H_0+H_1+H_2$, where
\begin{align}
  \label{eq:ln-nn3a}
    H_0 &= \sum_{j=2}^k \ket{j}\bra{j-1}\otimes A^j_{j-1}\\
  \label{eq:ln-nn3b}
    H_1 &= \sum_{j=1}^k \ket{j}\bra{j}\otimes A^j_j\\
  \label{eq:ln-nn3c}
    H_2 &= \sum_{j=2}^k \ket{j}\bra{j+1}\otimes A^j_{j+1}\,.
\end{align}
The exponential can be bounded as follows.
\begin{equation}
  \label{eq:ln-nn4}
    \|e^{Ct}\|\leq e^{\mu(C)t}\,.
\end{equation}
We have
\begin{equation}
  \label{eq:ln-nn5}
    \mu(C)=\sup_{ V:\| V\|=1}\bra{ V} C\ket{ V} = \sup_{V:\| V\|=1}\bra{ V}H_1\ket{ V} + \sup_{ V:\|V\|=1}\bra{ V}(H_0+H_2)\ket{ V}\,.
\end{equation}
This gives us
\begin{equation}
  \label{eq:ln-nn6}
    \mu(C)\leq \mu(H_1) + \mu(H_0)+\mu(H_2),.
\end{equation}
The Proposition 3.3 of \citet{Forets2017} states that 
for all $i\geq 1$, $0\leq j \leq k-1$, the estimate $\|A^i_{i+j\|} \leq i \|F_{j+1}\|$ holds. 
Therefore, 
we can write $\norm{H_{0, 1, 2}}=k\norm{\hat{F}_{0, 1, 2}}$ and 
\begin{equation}
  \label{eq:ln-nn7}
    \|e^{H_{0, 1, 2}}\|=\|e^{\hat{F}_{0, 1, 2}}\|^k\,.
\end{equation}
This gives us
\begin{equation}
  \label{eq:emu}
    \|e^{C}\|\leq e^{(\mu(\hat{F}_1)+\mu(\hat{F}_0)+\mu(\hat{F}_2))k}\,,
\end{equation}
In LBM, both $\hat{F}_1$ and $\hat{F}_2$ are constant matrices determined by the LBM constants for a given relaxation time $\tau$. Therefore, $\mu(\hat{F}_{1})$ and $\mu(\hat{F}_{2})$ are also constants and can be obtained explicitly. We show that $\mu(\hat{F}_1) \le 0$.
Since $\hat{F}_2$ in \Eq{eq:hatF2} is not square, we use the largest singuler value of $\hat{F}_2$ to bound $\mu(\hat{F}_2)$, which leads to $\mu(\hat{F}_2) = O(1)$ as shown in \Fig{fig:svd_norm_hat_F2_tau}.
For decaying turbulence ($b=0$), $\hat{F}_0 = 0$, therefore, $\mu(\hat{F}_0) = 0$. \Eq{eq:emu} can be bounded as 

\begin{equation}
  \label{eq:ec}
    E(C)\leq \exp{(\varrho t)}\,,
\end{equation}
where $\varrho =O(1)$.

\subsubsection*{Final quantum gate complexity to solve \Eq{eq:clbm}}
Finally, inserting \Eq{eq:epsilon}, \Eq{eq:evolutionT}, \Eq{eq:g-decay}, \Eq{eq:sparsity}, \Eq{eq:normc}, and \Eq{eq:ec},    into
\Eq{eq:ca_Krovi} yields the following result for the quantum algorithm complexity for the decaying turbulence 
\begin{equation}
  \label{eq:final-result-Krovi}
  \text{complexity} = O(t_c^{-1}\tilde{T}3\sqrt{k} \exp{(\varrho \tilde{T} / \left(\frac{L}{e_r}\right))}  \poly\log(n/\epsilon))
\end{equation}
where $\tilde T$ is the physical evolution time, and $L$ and
$e_r$ are characteristic scales chosen for the LBE formalism.
The scaling of each term in \Eq{eq:ca_Krovi} is summarized in
Table~\ref{tab:scaling_Krovi}. For any arbitrary simulation
time. The term $\tilde T/L$ in \Eq{eq:final-result-Krovi} can be reduced to
\begin{equation}
  \label{eq:T_L}
  \tilde T/L = 1/\norm{\vec{u}} = \frac{\tilde T^{3/5}}{\sqrt{K \chi}} .    
\end{equation}
The final complexity of applying \Eq{eq:ca_Krovi} to \Eq{eq:clbm} is
\begin{equation}
  \label{eq:final-result-Krovi2}
  \text{complexity} = O(t_c^{-1}\tilde{T}\sqrt{k}  \exp{(\varrho \frac{\tilde T^{3/5}}{e_r \sqrt{K \chi}})} \poly\log(n/\epsilon)),
\end{equation}
which yields an exponential growth in evolution time $\tilde T$. There could be a more strict bound such that $\norm{\mu (C)} \le 0$, which warrants further inquiry. Nevertheless, our analysis suggest urgent needs of domain driven QLSA development to achieve quantum speedup in solving nonlinear transport problems.
More importantly, our analysis sheds light on exploring the fundamental limit of applying quantum computing to nonlinear transport problems.       

\begin{table}
  \centering
  \begin{tabular}{lll}
    \hline\hline
    Term & Scaling & Reference \\ \hline
    $g$        & $3\sqrt{k}$                &  \Eq{eq:g-decay}        \\
    $s$        & $O(1)$              &  \Eq{eq:sparsity} \\
    $\|\mathcal C\|$    & $O(\kn^{-1})$ & \Eq{eq:normc}    \\
    $E(C)$        & $\exp{(\varrho t)}$                &  \Eq{eq:ec}        \\
    \hline\hline
  \end{tabular}
  \caption{Scaling of each term in \Eq{eq:ca} with lattice Boltzmann equation parameters leading to the
    result of \Eq{eq:final-result-Krovi}. 
    }
  \label{tab:scaling_Krovi}
\end{table}

\begin{thebibliography}{82}%
\makeatletter
\providecommand \@ifxundefined [1]{%
 \@ifx{#1\undefined}
}%
\providecommand \@ifnum [1]{%
 \ifnum #1\expandafter \@firstoftwo
 \else \expandafter \@secondoftwo
 \fi
}%
\providecommand \@ifx [1]{%
 \ifx #1\expandafter \@firstoftwo
 \else \expandafter \@secondoftwo
 \fi
}%
\providecommand \natexlab [1]{#1}%
\providecommand \enquote  [1]{``#1''}%
\providecommand \bibnamefont  [1]{#1}%
\providecommand \bibfnamefont [1]{#1}%
\providecommand \citenamefont [1]{#1}%
\providecommand \href@noop [0]{\@secondoftwo}%
\providecommand \href [0]{\begingroup \@sanitize@url \@href}%
\providecommand \@href[1]{\@@startlink{#1}\@@href}%
\providecommand \@@href[1]{\endgroup#1\@@endlink}%
\providecommand \@sanitize@url [0]{\catcode `\\12\catcode `\$12\catcode
  `\&12\catcode `\#12\catcode `\^12\catcode `\_12\catcode `\%12\relax}%
\providecommand \@@startlink[1]{}%
\providecommand \@@endlink[0]{}%
\providecommand \url  [0]{\begingroup\@sanitize@url \@url }%
\providecommand \@url [1]{\endgroup\@href {#1}{\urlprefix }}%
\providecommand \urlprefix  [0]{URL }%
\providecommand \Eprint [0]{\href }%
\providecommand \doibase [0]{https://doi.org/}%
\providecommand \selectlanguage [0]{\@gobble}%
\providecommand \bibinfo  [0]{\@secondoftwo}%
\providecommand \bibfield  [0]{\@secondoftwo}%
\providecommand \translation [1]{[#1]}%
\providecommand \BibitemOpen [0]{}%
\providecommand \bibitemStop [0]{}%
\providecommand \bibitemNoStop [0]{.\EOS\space}%
\providecommand \EOS [0]{\spacefactor3000\relax}%
\providecommand \BibitemShut  [1]{\csname bibitem#1\endcsname}%
\let\auto@bib@innerbib\@empty
\bibitem [{\citenamefont {Omenn}(2006)}]{Omenn2006}%
  \BibitemOpen
  \bibfield  {author} {\bibinfo {author} {\bibfnamefont {G.~S.}\ \bibnamefont
  {Omenn}},\ }\bibfield  {title} {\bibinfo {title} {Grand challenges and great
  opportunities in science, technology, and public policy},\ }\href
  {https://doi.org/10.1126/science.1135003} {\bibfield  {journal} {\bibinfo
  {journal} {Science}\ }\textbf {\bibinfo {volume} {314}},\ \bibinfo {pages}
  {1696} (\bibinfo {year} {2006})}\BibitemShut {NoStop}%
\bibitem [{\citenamefont {{World Climate Research Programme}}(  20)}]{WCRP}%
  \BibitemOpen
  \bibfield  {author} {\bibinfo {author} {\bibnamefont {{World Climate Research
  Programme}}},\ }\href@noop {} {\bibinfo {title} {{WCRP Grand Challenges}}}
  (\bibinfo {year} {last accessed 2023--01--20}),\ \bibinfo {note}
  {\url{https://www.wcrp-climate.org/grand-challenges/grand-challenges-overview}}\BibitemShut
  {NoStop}%
\bibitem [{\citenamefont {{National Academy of
  Engineering}}(2016)}]{NASEM2016}%
  \BibitemOpen
  \bibfield  {author} {\bibinfo {author} {\bibnamefont {{National Academy of
  Engineering}}},\ }\href {https://doi.org/10.17226/23440} {\emph {\bibinfo
  {title} {Grand Challenges for Engineering: Imperatives, Prospects, and
  Priorities: Summary of a Forum}}},\ edited by\ \bibinfo {editor}
  {\bibfnamefont {S.}~\bibnamefont {Olson}}\ (\bibinfo  {publisher} {The
  National Academies Press},\ \bibinfo {address} {Washington, DC},\ \bibinfo
  {year} {2016})\BibitemShut {NoStop}%
\bibitem [{\citenamefont {Hasselmann}(1976)}]{Hasselmann1976}%
  \BibitemOpen
  \bibfield  {author} {\bibinfo {author} {\bibfnamefont {K.}~\bibnamefont
  {Hasselmann}},\ }\bibfield  {title} {\bibinfo {title} {Stochastic climate
  models .1. theory},\ }\href
  {https://doi.org/10.1111/j.2153-3490.1976.tb00696.x} {\bibfield  {journal}
  {\bibinfo  {journal} {Tellus}\ }\textbf {\bibinfo {volume} {28}},\ \bibinfo
  {pages} {473} (\bibinfo {year} {1976})}\BibitemShut {NoStop}%
\bibitem [{\citenamefont {Majda}\ and\ \citenamefont
  {Klein}(2003)}]{Majda2003}%
  \BibitemOpen
  \bibfield  {author} {\bibinfo {author} {\bibfnamefont {A.~J.}\ \bibnamefont
  {Majda}}\ and\ \bibinfo {author} {\bibfnamefont {R.}~\bibnamefont {Klein}},\
  }\bibfield  {title} {\bibinfo {title} {Systematic multiscale models for the
  tropics},\ }\href
  {https://doi.org/10.1175/1520-0469(2003)060<0393:SMMFTT>2.0.CO;2} {\bibfield
  {journal} {\bibinfo  {journal} {Journal of the Atmospheric Sciences}\
  }\textbf {\bibinfo {volume} {60}},\ \bibinfo {pages} {393} (\bibinfo {year}
  {2003})}\BibitemShut {NoStop}%
\bibitem [{\citenamefont {Bony}\ \emph {et~al.}(2015)\citenamefont {Bony},
  \citenamefont {Stevens}, \citenamefont {Frierson}, \citenamefont {Jakob},
  \citenamefont {Kageyama}, \citenamefont {Pincus}, \citenamefont {Shepherd},
  \citenamefont {Sherwood}, \citenamefont {Siebesma}, \citenamefont {Sobel},
  \citenamefont {Watanabe},\ and\ \citenamefont {Webb}}]{Bony2015}%
  \BibitemOpen
  \bibfield  {author} {\bibinfo {author} {\bibfnamefont {S.}~\bibnamefont
  {Bony}}, \bibinfo {author} {\bibfnamefont {B.}~\bibnamefont {Stevens}},
  \bibinfo {author} {\bibfnamefont {D.~M.~W.}\ \bibnamefont {Frierson}},
  \bibinfo {author} {\bibfnamefont {C.}~\bibnamefont {Jakob}}, \bibinfo
  {author} {\bibfnamefont {M.}~\bibnamefont {Kageyama}}, \bibinfo {author}
  {\bibfnamefont {R.}~\bibnamefont {Pincus}}, \bibinfo {author} {\bibfnamefont
  {T.~G.}\ \bibnamefont {Shepherd}}, \bibinfo {author} {\bibfnamefont {S.~C.}\
  \bibnamefont {Sherwood}}, \bibinfo {author} {\bibfnamefont {A.~P.}\
  \bibnamefont {Siebesma}}, \bibinfo {author} {\bibfnamefont {A.~H.}\
  \bibnamefont {Sobel}}, \bibinfo {author} {\bibfnamefont {M.}~\bibnamefont
  {Watanabe}},\ and\ \bibinfo {author} {\bibfnamefont {M.~J.}\ \bibnamefont
  {Webb}},\ }\bibfield  {title} {\bibinfo {title} {Clouds, circulation and
  climate sensitivity},\ }\href {https://doi.org/10.1038/NGEO2398} {\bibfield
  {journal} {\bibinfo  {journal} {Nature Geoscience}\ }\textbf {\bibinfo
  {volume} {8}},\ \bibinfo {pages} {261} (\bibinfo {year} {2015})}\BibitemShut
  {NoStop}%
\bibitem [{\citenamefont {Ghil}\ and\ \citenamefont
  {Lucarini}(2020)}]{Ghil2020}%
  \BibitemOpen
  \bibfield  {author} {\bibinfo {author} {\bibfnamefont {M.}~\bibnamefont
  {Ghil}}\ and\ \bibinfo {author} {\bibfnamefont {V.}~\bibnamefont
  {Lucarini}},\ }\bibfield  {title} {\bibinfo {title} {The physics of climate
  variability and climate change},\ }\href
  {https://doi.org/10.1103/RevModPhys.92.035002} {\bibfield  {journal}
  {\bibinfo  {journal} {Reviews of Modern Physics}\ }\textbf {\bibinfo {volume}
  {92}},\ \bibinfo {pages} {035002} (\bibinfo {year} {2020})}\BibitemShut
  {NoStop}%
\bibitem [{\citenamefont {Gupta}\ \emph {et~al.}(2022)\citenamefont {Gupta},
  \citenamefont {Mastrantonas}, \citenamefont {Masoller},\ and\ \citenamefont
  {Kurths}}]{Gupta2022}%
  \BibitemOpen
  \bibfield  {author} {\bibinfo {author} {\bibfnamefont {S.}~\bibnamefont
  {Gupta}}, \bibinfo {author} {\bibfnamefont {N.}~\bibnamefont {Mastrantonas}},
  \bibinfo {author} {\bibfnamefont {C.}~\bibnamefont {Masoller}},\ and\
  \bibinfo {author} {\bibfnamefont {J.}~\bibnamefont {Kurths}},\ }\bibfield
  {title} {\bibinfo {title} {Perspectives on the importance of complex systems
  in understanding our climate and climate change-the nobel prize in physics
  2021},\ }\href {https://doi.org/10.1063/5.0090222} {\bibfield  {journal}
  {\bibinfo  {journal} {Chaos}\ }\textbf {\bibinfo {volume} {32}},\ \bibinfo
  {pages} {052102} (\bibinfo {year} {2022})}\BibitemShut {NoStop}%
\bibitem [{\citenamefont {Bernaschi}\ \emph {et~al.}(2019)\citenamefont
  {Bernaschi}, \citenamefont {Melchionna},\ and\ \citenamefont
  {Succi}}]{Bernaschi2019}%
  \BibitemOpen
  \bibfield  {author} {\bibinfo {author} {\bibfnamefont {M.}~\bibnamefont
  {Bernaschi}}, \bibinfo {author} {\bibfnamefont {S.}~\bibnamefont
  {Melchionna}},\ and\ \bibinfo {author} {\bibfnamefont {S.}~\bibnamefont
  {Succi}},\ }\bibfield  {title} {\bibinfo {title} {Mesoscopic simulations at
  the physics-chemistry-biology interface},\ }\href
  {https://doi.org/10.1103/RevModPhys.91.025004} {\bibfield  {journal}
  {\bibinfo  {journal} {Reviews of Modern Physics}\ }\textbf {\bibinfo {volume}
  {91}},\ \bibinfo {pages} {025004} (\bibinfo {year} {2019})}\BibitemShut
  {NoStop}%
\bibitem [{\citenamefont {Boles}\ \emph {et~al.}(2016)\citenamefont {Boles},
  \citenamefont {Engel},\ and\ \citenamefont {Talapin}}]{Boles2016}%
  \BibitemOpen
  \bibfield  {author} {\bibinfo {author} {\bibfnamefont {M.~A.}\ \bibnamefont
  {Boles}}, \bibinfo {author} {\bibfnamefont {M.}~\bibnamefont {Engel}},\ and\
  \bibinfo {author} {\bibfnamefont {D.~V.}\ \bibnamefont {Talapin}},\
  }\bibfield  {title} {\bibinfo {title} {Self-assembly of colloidal
  nanocrystals: From intricate structures to functional materials},\ }\href
  {https://doi.org/10.1021/acs.chemrev.6b00196} {\bibfield  {journal} {\bibinfo
   {journal} {Chemical Reviews}\ }\textbf {\bibinfo {volume} {116}},\ \bibinfo
  {pages} {11220} (\bibinfo {year} {2016})}\BibitemShut {NoStop}%
\bibitem [{\citenamefont {Salzmann}\ \emph {et~al.}(2021)\citenamefont
  {Salzmann}, \citenamefont {van~der Sluijs}, \citenamefont {Soligno},\ and\
  \citenamefont {Vanmaekelbergh}}]{Salzmann2021}%
  \BibitemOpen
  \bibfield  {author} {\bibinfo {author} {\bibfnamefont {V.}~\bibnamefont
  {Salzmann}, \bibfnamefont {Bastiaan~B.}}, \bibinfo {author} {\bibfnamefont
  {M.~M.}\ \bibnamefont {van~der Sluijs}}, \bibinfo {author} {\bibfnamefont
  {G.}~\bibnamefont {Soligno}},\ and\ \bibinfo {author} {\bibfnamefont
  {D.}~\bibnamefont {Vanmaekelbergh}},\ }\bibfield  {title} {\bibinfo {title}
  {Oriented attachment: From natural crystal growth to a materials engineering
  tool},\ }\href {https://doi.org/10.1021/acs.accounts.0c00739} {\bibfield
  {journal} {\bibinfo  {journal} {Accounts of Chemical Research}\ }\textbf
  {\bibinfo {volume} {54}},\ \bibinfo {pages} {787} (\bibinfo {year}
  {2021})}\BibitemShut {NoStop}%
\bibitem [{\citenamefont {De~Yoreo}\ \emph {et~al.}(2015)\citenamefont
  {De~Yoreo}, \citenamefont {Gilbert}, \citenamefont {Sommerdijk},
  \citenamefont {Penn}, \citenamefont {Whitelam}, \citenamefont {Joester},
  \citenamefont {Zhang}, \citenamefont {Rimer}, \citenamefont {Navrotsky},
  \citenamefont {Banfield}, \citenamefont {Wallace}, \citenamefont {Michel},
  \citenamefont {Meldrum}, \citenamefont {Coelfen},\ and\ \citenamefont
  {Dove}}]{DeYoreo2015}%
  \BibitemOpen
  \bibfield  {author} {\bibinfo {author} {\bibfnamefont {J.~J.}\ \bibnamefont
  {De~Yoreo}}, \bibinfo {author} {\bibfnamefont {P.~U. P.~A.}\ \bibnamefont
  {Gilbert}}, \bibinfo {author} {\bibfnamefont {N.~A. J.~M.}\ \bibnamefont
  {Sommerdijk}}, \bibinfo {author} {\bibfnamefont {R.~L.}\ \bibnamefont
  {Penn}}, \bibinfo {author} {\bibfnamefont {S.}~\bibnamefont {Whitelam}},
  \bibinfo {author} {\bibfnamefont {D.}~\bibnamefont {Joester}}, \bibinfo
  {author} {\bibfnamefont {H.}~\bibnamefont {Zhang}}, \bibinfo {author}
  {\bibfnamefont {J.~D.}\ \bibnamefont {Rimer}}, \bibinfo {author}
  {\bibfnamefont {A.}~\bibnamefont {Navrotsky}}, \bibinfo {author}
  {\bibfnamefont {J.~F.}\ \bibnamefont {Banfield}}, \bibinfo {author}
  {\bibfnamefont {A.~F.}\ \bibnamefont {Wallace}}, \bibinfo {author}
  {\bibfnamefont {F.~M.}\ \bibnamefont {Michel}}, \bibinfo {author}
  {\bibfnamefont {F.~C.}\ \bibnamefont {Meldrum}}, \bibinfo {author}
  {\bibfnamefont {H.}~\bibnamefont {Coelfen}},\ and\ \bibinfo {author}
  {\bibfnamefont {P.~M.}\ \bibnamefont {Dove}},\ }\bibfield  {title} {\bibinfo
  {title} {Crystallization by particle attachment in synthetic, biogenic, and
  geologic environments},\ }\href {https://doi.org/10.1126/science.aaa6760}
  {\bibfield  {journal} {\bibinfo  {journal} {Science}\ }\textbf {\bibinfo
  {volume} {349}},\ \bibinfo {pages} {aaa6760} (\bibinfo {year}
  {2015})}\BibitemShut {NoStop}%
\bibitem [{\citenamefont {De~Yoreo}\ \emph {et~al.}(2022)\citenamefont
  {De~Yoreo}, \citenamefont {Nakouzi}, \citenamefont {Jin}, \citenamefont
  {Chun},\ and\ \citenamefont {Mundy}}]{DeYoreo2022}%
  \BibitemOpen
  \bibfield  {author} {\bibinfo {author} {\bibfnamefont {J.~J.}\ \bibnamefont
  {De~Yoreo}}, \bibinfo {author} {\bibfnamefont {E.}~\bibnamefont {Nakouzi}},
  \bibinfo {author} {\bibfnamefont {B.}~\bibnamefont {Jin}}, \bibinfo {author}
  {\bibfnamefont {J.}~\bibnamefont {Chun}},\ and\ \bibinfo {author}
  {\bibfnamefont {C.~J.}\ \bibnamefont {Mundy}},\ }\bibfield  {title} {\bibinfo
  {title} {Spiers memorial lecture: Assembly-based pathways of
  crystallization},\ }\href {https://doi.org/10.1039/d2fd00061j} {\bibfield
  {journal} {\bibinfo  {journal} {Faraday Discussions}\ }\textbf {\bibinfo
  {volume} {235}},\ \bibinfo {pages} {9} (\bibinfo {year} {2022})}\BibitemShut
  {NoStop}%
\bibitem [{\citenamefont {Russel}\ \emph {et~al.}(1991)\citenamefont {Russel},
  \citenamefont {Russel}, \citenamefont {Saville},\ and\ \citenamefont
  {Schowalter}}]{Russel}%
  \BibitemOpen
  \bibfield  {author} {\bibinfo {author} {\bibfnamefont {W.~B.}\ \bibnamefont
  {Russel}}, \bibinfo {author} {\bibfnamefont {W.}~\bibnamefont {Russel}},
  \bibinfo {author} {\bibfnamefont {D.~A.}\ \bibnamefont {Saville}},\ and\
  \bibinfo {author} {\bibfnamefont {W.~R.}\ \bibnamefont {Schowalter}},\
  }\href@noop {} {\emph {\bibinfo {title} {Colloidal dispersions}}}\ (\bibinfo
  {publisher} {{Cambridge University Press}},\ \bibinfo {year}
  {1991})\BibitemShut {NoStop}%
\bibitem [{\citenamefont {Pope}(2000)}]{pope2000turbulent}%
  \BibitemOpen
  \bibfield  {author} {\bibinfo {author} {\bibfnamefont {S.~B.}\ \bibnamefont
  {Pope}},\ }\href@noop {} {\emph {\bibinfo {title} {Turbulent flows}}}\
  (\bibinfo  {publisher} {Cambridge university press},\ \bibinfo {year}
  {2000})\BibitemShut {NoStop}%
\bibitem [{\citenamefont {Bodenschatz}\ \emph {et~al.}(2010)\citenamefont
  {Bodenschatz}, \citenamefont {Malinowski}, \citenamefont {Shaw},\ and\
  \citenamefont {Stratmann}}]{Bodenschatz2010}%
  \BibitemOpen
  \bibfield  {author} {\bibinfo {author} {\bibfnamefont {E.}~\bibnamefont
  {Bodenschatz}}, \bibinfo {author} {\bibfnamefont {S.~P.}\ \bibnamefont
  {Malinowski}}, \bibinfo {author} {\bibfnamefont {R.~A.}\ \bibnamefont
  {Shaw}},\ and\ \bibinfo {author} {\bibfnamefont {F.}~\bibnamefont
  {Stratmann}},\ }\bibfield  {title} {\bibinfo {title} {Can we understand
  clouds without turbulence?},\ }\href
  {https://doi.org/10.1126/science.1185138} {\bibfield  {journal} {\bibinfo
  {journal} {Science}\ }\textbf {\bibinfo {volume} {327}},\ \bibinfo {pages}
  {970} (\bibinfo {year} {2010})}\BibitemShut {NoStop}%
\bibitem [{\citenamefont {Sherwood}\ \emph {et~al.}(2014)\citenamefont
  {Sherwood}, \citenamefont {Bony},\ and\ \citenamefont
  {Dufresne}}]{Sherwood2014}%
  \BibitemOpen
  \bibfield  {author} {\bibinfo {author} {\bibfnamefont {S.~C.}\ \bibnamefont
  {Sherwood}}, \bibinfo {author} {\bibfnamefont {S.}~\bibnamefont {Bony}},\
  and\ \bibinfo {author} {\bibfnamefont {J.-L.}\ \bibnamefont {Dufresne}},\
  }\bibfield  {title} {\bibinfo {title} {Spread in model climate sensitivity
  traced to atmospheric convective mixing},\ }\href
  {https://doi.org/10.1038/nature12829} {\bibfield  {journal} {\bibinfo
  {journal} {Nature}\ }\textbf {\bibinfo {volume} {505}},\ \bibinfo {pages}
  {37} (\bibinfo {year} {2014})}\BibitemShut {NoStop}%
\bibitem [{\citenamefont {Bretherton}(2015)}]{Bretherton2015}%
  \BibitemOpen
  \bibfield  {author} {\bibinfo {author} {\bibfnamefont {C.~S.}\ \bibnamefont
  {Bretherton}},\ }\bibfield  {title} {\bibinfo {title} {Insights into
  low-latitude cloud feedbacks from high-resolution models},\ }\href
  {https://doi.org/10.1098/rsta.2014.0415} {\bibfield  {journal} {\bibinfo
  {journal} {Philosophical Transactions of the Royal Society A-Mathematical
  Physical and Engineering Sciences}\ }\textbf {\bibinfo {volume} {373}},\
  \bibinfo {pages} {20140415} (\bibinfo {year} {2015})}\BibitemShut {NoStop}%
\bibitem [{\citenamefont {Donner}\ \emph {et~al.}(2016)\citenamefont {Donner},
  \citenamefont {O'Brien}, \citenamefont {Rieger}, \citenamefont {Vogel},\ and\
  \citenamefont {Cooke}}]{Donner2016}%
  \BibitemOpen
  \bibfield  {author} {\bibinfo {author} {\bibfnamefont {L.~J.}\ \bibnamefont
  {Donner}}, \bibinfo {author} {\bibfnamefont {T.~A.}\ \bibnamefont {O'Brien}},
  \bibinfo {author} {\bibfnamefont {D.}~\bibnamefont {Rieger}}, \bibinfo
  {author} {\bibfnamefont {B.}~\bibnamefont {Vogel}},\ and\ \bibinfo {author}
  {\bibfnamefont {W.~F.}\ \bibnamefont {Cooke}},\ }\bibfield  {title} {\bibinfo
  {title} {Are atmospheric updrafts a key to unlocking climate forcing and
  sensitivity?},\ }\href {https://doi.org/10.5194/acp-16-12983-2016} {\bibfield
   {journal} {\bibinfo  {journal} {Atmospheric Chemistry and Physics}\ }\textbf
  {\bibinfo {volume} {16}},\ \bibinfo {pages} {12983} (\bibinfo {year}
  {2016})}\BibitemShut {NoStop}%
\bibitem [{\citenamefont {Sprague}\ \emph {et~al.}(2017)\citenamefont
  {Sprague}, \citenamefont {Boldyrev}, \citenamefont {Fischer}, \citenamefont
  {Grout}, \citenamefont {Gustafson},\ and\ \citenamefont
  {Moser}}]{Exascale2017}%
  \BibitemOpen
  \bibfield  {author} {\bibinfo {author} {\bibfnamefont {M.~A.}\ \bibnamefont
  {Sprague}}, \bibinfo {author} {\bibfnamefont {S.}~\bibnamefont {Boldyrev}},
  \bibinfo {author} {\bibfnamefont {P.}~\bibnamefont {Fischer}}, \bibinfo
  {author} {\bibfnamefont {R.}~\bibnamefont {Grout}}, \bibinfo {author}
  {\bibfnamefont {W.~I.}\ \bibnamefont {Gustafson}, \bibfnamefont {Jr.}},\ and\
  \bibinfo {author} {\bibfnamefont {R.}~\bibnamefont {Moser}},\ }\href
  {https://doi.org/10.2172/1338668} {\bibinfo {title} {Turbulent flow
  simulation at the exascale: Opportunities and challenges workshop: August
  4-5, 2015, {Washington, D.C.}}} (\bibinfo {year} {2017})\BibitemShut
  {NoStop}%
\bibitem [{\citenamefont {Schneider}\ \emph {et~al.}(2017)\citenamefont
  {Schneider}, \citenamefont {Teixeira}, \citenamefont {Bretherton},
  \citenamefont {Brient}, \citenamefont {Pressel}, \citenamefont
  {Sch{\"{a}}r},\ and\ \citenamefont {Siebesma}}]{Schneider2017ClimateClouds}%
  \BibitemOpen
  \bibfield  {author} {\bibinfo {author} {\bibfnamefont {T.}~\bibnamefont
  {Schneider}}, \bibinfo {author} {\bibfnamefont {J.}~\bibnamefont {Teixeira}},
  \bibinfo {author} {\bibfnamefont {C.~S.}\ \bibnamefont {Bretherton}},
  \bibinfo {author} {\bibfnamefont {F.}~\bibnamefont {Brient}}, \bibinfo
  {author} {\bibfnamefont {K.~G.}\ \bibnamefont {Pressel}}, \bibinfo {author}
  {\bibfnamefont {C.}~\bibnamefont {Sch{\"{a}}r}},\ and\ \bibinfo {author}
  {\bibfnamefont {A.~P.}\ \bibnamefont {Siebesma}},\ }\href
  {https://doi.org/10.1038/nclimate3190} {\bibinfo {title} {{Climate goals and
  computing the future of clouds}}} (\bibinfo {year} {2017})\BibitemShut
  {NoStop}%
\bibitem [{\citenamefont {Berkooz}\ \emph {et~al.}(1993)\citenamefont
  {Berkooz}, \citenamefont {Holmes},\ and\ \citenamefont
  {Lumley}}]{berkooz1993proper}%
  \BibitemOpen
  \bibfield  {author} {\bibinfo {author} {\bibfnamefont {G.}~\bibnamefont
  {Berkooz}}, \bibinfo {author} {\bibfnamefont {P.}~\bibnamefont {Holmes}},\
  and\ \bibinfo {author} {\bibfnamefont {J.~L.}\ \bibnamefont {Lumley}},\
  }\bibfield  {title} {\bibinfo {title} {The proper orthogonal decomposition in
  the analysis of turbulent flows},\ }\href@noop {} {\bibfield  {journal}
  {\bibinfo  {journal} {Annual review of fluid mechanics}\ }\textbf {\bibinfo
  {volume} {25}},\ \bibinfo {pages} {539} (\bibinfo {year} {1993})}\BibitemShut
  {NoStop}%
\bibitem [{\citenamefont {Podvin}\ and\ \citenamefont
  {Fraigneau}(2017)}]{podvin2017few}%
  \BibitemOpen
  \bibfield  {author} {\bibinfo {author} {\bibfnamefont {B.}~\bibnamefont
  {Podvin}}\ and\ \bibinfo {author} {\bibfnamefont {Y.}~\bibnamefont
  {Fraigneau}},\ }\bibfield  {title} {\bibinfo {title} {A few thoughts on
  proper orthogonal decomposition in turbulence},\ }\href@noop {} {\bibfield
  {journal} {\bibinfo  {journal} {Physics of Fluids}\ }\textbf {\bibinfo
  {volume} {29}},\ \bibinfo {pages} {020709} (\bibinfo {year}
  {2017})}\BibitemShut {NoStop}%
\bibitem [{\citenamefont {Tennekes}\ and\ \citenamefont
  {Lumley}(1972)}]{tennekes1972first}%
  \BibitemOpen
  \bibfield  {author} {\bibinfo {author} {\bibfnamefont {H.}~\bibnamefont
  {Tennekes}}\ and\ \bibinfo {author} {\bibfnamefont {J.~L.}\ \bibnamefont
  {Lumley}},\ }\href@noop {} {\emph {\bibinfo {title} {A first course in
  turbulence}}}\ (\bibinfo  {publisher} {MIT press},\ \bibinfo {year}
  {1972})\BibitemShut {NoStop}%
\bibitem [{\citenamefont {Durbin}(2018)}]{Durbin18}%
  \BibitemOpen
  \bibfield  {author} {\bibinfo {author} {\bibfnamefont {P.~A.}\ \bibnamefont
  {Durbin}},\ }\bibfield  {title} {\bibinfo {title} {Some recent developments
  in turbulence closure modeling},\ }\href
  {https://doi.org/10.1146/annurev-fluid-122316-045020} {\bibfield  {journal}
  {\bibinfo  {journal} {Annual Review of Fluid Mechanics}\ }\textbf {\bibinfo
  {volume} {50}},\ \bibinfo {pages} {77} (\bibinfo {year} {2018})},\ \Eprint
  {https://arxiv.org/abs/https://doi.org/10.1146/annurev-fluid-122316-045020}
  {https://doi.org/10.1146/annurev-fluid-122316-045020} \BibitemShut {NoStop}%
\bibitem [{\citenamefont {Artime}\ and\ \citenamefont
  {De~Domenico}(2022)}]{Artime2022}%
  \BibitemOpen
  \bibfield  {author} {\bibinfo {author} {\bibfnamefont {O.}~\bibnamefont
  {Artime}}\ and\ \bibinfo {author} {\bibfnamefont {M.}~\bibnamefont
  {De~Domenico}},\ }\bibfield  {title} {\bibinfo {title} {From the origin of
  life to pandemics: emergent phenomena in complex systems},\ }\href
  {https://doi.org/10.1098/rsta.2020.0410} {\bibfield  {journal} {\bibinfo
  {journal} {Philosophical Transactions of the Royal Society A-Mathematical
  Physical and Engineering Sciences}\ }\textbf {\bibinfo {volume} {380}},\
  \bibinfo {pages} {20200410} (\bibinfo {year} {2022})}\BibitemShut {NoStop}%
\bibitem [{\citenamefont {Feingold}\ \emph {et~al.}(2016)\citenamefont
  {Feingold}, \citenamefont {McComiskey}, \citenamefont {Yamaguchi},
  \citenamefont {Johnson}, \citenamefont {Carslaw},\ and\ \citenamefont
  {Schmidt}}]{Feingold2016}%
  \BibitemOpen
  \bibfield  {author} {\bibinfo {author} {\bibfnamefont {G.}~\bibnamefont
  {Feingold}}, \bibinfo {author} {\bibfnamefont {A.}~\bibnamefont
  {McComiskey}}, \bibinfo {author} {\bibfnamefont {T.}~\bibnamefont
  {Yamaguchi}}, \bibinfo {author} {\bibfnamefont {J.~S.}\ \bibnamefont
  {Johnson}}, \bibinfo {author} {\bibfnamefont {K.~S.}\ \bibnamefont
  {Carslaw}},\ and\ \bibinfo {author} {\bibfnamefont {K.~S.}\ \bibnamefont
  {Schmidt}},\ }\bibfield  {title} {\bibinfo {title} {New approaches to
  quantifying aerosol influence on the cloud radiative effect},\ }\href
  {https://doi.org/10.1073/pnas.1514035112} {\bibfield  {journal} {\bibinfo
  {journal} {Proceedings of the National Academy of Sciences of the United
  States of America}\ }\textbf {\bibinfo {volume} {113}},\ \bibinfo {pages}
  {5812} (\bibinfo {year} {2016})}\BibitemShut {NoStop}%
\bibitem [{\citenamefont {Wheatcraft}\ and\ \citenamefont
  {Cushman}(1991)}]{Wheatcraft1991}%
  \BibitemOpen
  \bibfield  {author} {\bibinfo {author} {\bibfnamefont {S.}~\bibnamefont
  {Wheatcraft}}\ and\ \bibinfo {author} {\bibfnamefont {J.}~\bibnamefont
  {Cushman}},\ }\bibfield  {title} {\bibinfo {title} {Hierarchical approaches
  to transport in heterogeneous porous media},\ }\href
  {https://doi.org/10.1002/rog.1991.29.s1.263} {\bibfield  {journal} {\bibinfo
  {journal} {Reviews of Geophysics}\ }\textbf {\bibinfo {volume} {29}},\
  \bibinfo {pages} {263} (\bibinfo {year} {1991})}\BibitemShut {NoStop}%
\bibitem [{\citenamefont {Carslaw}\ \emph {et~al.}(2018)\citenamefont
  {Carslaw}, \citenamefont {Lee}, \citenamefont {Regayre},\ and\ \citenamefont
  {Johnson}}]{Carslaw2018}%
  \BibitemOpen
  \bibfield  {author} {\bibinfo {author} {\bibfnamefont {K.~S.}\ \bibnamefont
  {Carslaw}}, \bibinfo {author} {\bibfnamefont {L.~A.}\ \bibnamefont {Lee}},
  \bibinfo {author} {\bibfnamefont {L.~A.}\ \bibnamefont {Regayre}},\ and\
  \bibinfo {author} {\bibfnamefont {J.~S.}\ \bibnamefont {Johnson}},\
  }\bibfield  {title} {\bibinfo {title} {Climate models are uncertain, but we
  can do something about it},\ }\bibfield  {journal} {\bibinfo  {journal}
  {Eos}\ }\textbf {\bibinfo {volume} {99}},\ \href
  {https://doi.org/10.1029/2018EO093757} {10.1029/2018EO093757} (\bibinfo
  {year} {2018})\BibitemShut {NoStop}%
\bibitem [{\citenamefont {Muelmenstaedt}\ and\ \citenamefont
  {Feingold}(2018)}]{Muelmenstaedt2018}%
  \BibitemOpen
  \bibfield  {author} {\bibinfo {author} {\bibfnamefont {J.}~\bibnamefont
  {Muelmenstaedt}}\ and\ \bibinfo {author} {\bibfnamefont {G.}~\bibnamefont
  {Feingold}},\ }\bibfield  {title} {\bibinfo {title} {The radiative forcing of
  aerosol-cloud interactions in liquid clouds: Wrestling and embracing
  uncertainty},\ }\href {https://doi.org/10.1007/s40641-018-0089-y} {\bibfield
  {journal} {\bibinfo  {journal} {Current Climate Change Reports}\ }\textbf
  {\bibinfo {volume} {4}},\ \bibinfo {pages} {23} (\bibinfo {year}
  {2018})}\BibitemShut {NoStop}%
\bibitem [{\citenamefont {Ge}\ \emph {et~al.}(2019)\citenamefont {Ge},
  \citenamefont {Chang}, \citenamefont {Li},\ and\ \citenamefont
  {Wang}}]{Ge2019}%
  \BibitemOpen
  \bibfield  {author} {\bibinfo {author} {\bibfnamefont {W.}~\bibnamefont
  {Ge}}, \bibinfo {author} {\bibfnamefont {Q.}~\bibnamefont {Chang}}, \bibinfo
  {author} {\bibfnamefont {C.}~\bibnamefont {Li}},\ and\ \bibinfo {author}
  {\bibfnamefont {J.}~\bibnamefont {Wang}},\ }\bibfield  {title} {\bibinfo
  {title} {Multiscale structures in particle-fluid systems: Characterization,
  modeling, and simulation},\ }\href
  {https://doi.org/10.1016/j.ces.2018.12.037} {\bibfield  {journal} {\bibinfo
  {journal} {Chemical Engineering Science}\ }\textbf {\bibinfo {volume}
  {198}},\ \bibinfo {pages} {198} (\bibinfo {year} {2019})}\BibitemShut
  {NoStop}%
\bibitem [{\citenamefont {Sherwood}\ \emph {et~al.}(2020)\citenamefont
  {Sherwood}, \citenamefont {Webb}, \citenamefont {Annan}, \citenamefont
  {Armour}, \citenamefont {Forster}, \citenamefont {Hargreaves}, \citenamefont
  {Hegerl}, \citenamefont {Klein}, \citenamefont {Marvel}, \citenamefont
  {Rohling}, \citenamefont {Watanabe}, \citenamefont {Andrews}, \citenamefont
  {Braconnot}, \citenamefont {Bretherton}, \citenamefont {Foster},
  \citenamefont {Hausfather}, \citenamefont {Heydt}, \citenamefont {Knutti},
  \citenamefont {Mauritsen}, \citenamefont {Norris}, \citenamefont
  {Proistosescu}, \citenamefont {Rugenstein}, \citenamefont {Schmidt},
  \citenamefont {Tokarska},\ and\ \citenamefont {Zelinka}}]{Sherwood2020}%
  \BibitemOpen
  \bibfield  {author} {\bibinfo {author} {\bibfnamefont {S.~C.}\ \bibnamefont
  {Sherwood}}, \bibinfo {author} {\bibfnamefont {M.~J.}\ \bibnamefont {Webb}},
  \bibinfo {author} {\bibfnamefont {J.~D.}\ \bibnamefont {Annan}}, \bibinfo
  {author} {\bibfnamefont {K.~C.}\ \bibnamefont {Armour}}, \bibinfo {author}
  {\bibfnamefont {P.~M.}\ \bibnamefont {Forster}}, \bibinfo {author}
  {\bibfnamefont {J.~C.}\ \bibnamefont {Hargreaves}}, \bibinfo {author}
  {\bibfnamefont {G.}~\bibnamefont {Hegerl}}, \bibinfo {author} {\bibfnamefont
  {S.~A.}\ \bibnamefont {Klein}}, \bibinfo {author} {\bibfnamefont {K.~D.}\
  \bibnamefont {Marvel}}, \bibinfo {author} {\bibfnamefont {E.~J.}\
  \bibnamefont {Rohling}}, \bibinfo {author} {\bibfnamefont {M.}~\bibnamefont
  {Watanabe}}, \bibinfo {author} {\bibfnamefont {T.}~\bibnamefont {Andrews}},
  \bibinfo {author} {\bibfnamefont {P.}~\bibnamefont {Braconnot}}, \bibinfo
  {author} {\bibfnamefont {C.~S.}\ \bibnamefont {Bretherton}}, \bibinfo
  {author} {\bibfnamefont {G.~L.}\ \bibnamefont {Foster}}, \bibinfo {author}
  {\bibfnamefont {Z.}~\bibnamefont {Hausfather}}, \bibinfo {author}
  {\bibfnamefont {A.~S.}\ \bibnamefont {Heydt}}, \bibinfo {author}
  {\bibfnamefont {R.}~\bibnamefont {Knutti}}, \bibinfo {author} {\bibfnamefont
  {T.}~\bibnamefont {Mauritsen}}, \bibinfo {author} {\bibfnamefont {J.~R.}\
  \bibnamefont {Norris}}, \bibinfo {author} {\bibfnamefont {C.}~\bibnamefont
  {Proistosescu}}, \bibinfo {author} {\bibfnamefont {M.}~\bibnamefont
  {Rugenstein}}, \bibinfo {author} {\bibfnamefont {G.~A.}\ \bibnamefont
  {Schmidt}}, \bibinfo {author} {\bibfnamefont {K.~B.}\ \bibnamefont
  {Tokarska}},\ and\ \bibinfo {author} {\bibfnamefont {M.~D.}\ \bibnamefont
  {Zelinka}},\ }\bibfield  {title} {\bibinfo {title} {An assessment of earth's
  climate sensitivity using multiple lines of evidence},\ }\href
  {https://doi.org/10.1029/2019RG000678} {\bibfield  {journal} {\bibinfo
  {journal} {Reviews of Geophysics}\ }\textbf {\bibinfo {volume} {58}},\
  \bibinfo {pages} {e2019RG000678} (\bibinfo {year} {2020})}\BibitemShut
  {NoStop}%
\bibitem [{\citenamefont {Bellouin}\ \emph {et~al.}(2020)\citenamefont
  {Bellouin}, \citenamefont {Quaas}, \citenamefont {Gryspeerdt}, \citenamefont
  {Kinne}, \citenamefont {Stier}, \citenamefont {Watson-Parris}, \citenamefont
  {Boucher}, \citenamefont {Carslaw}, \citenamefont {Christensen},
  \citenamefont {Daniau}, \citenamefont {Dufresne}, \citenamefont {Feingold},
  \citenamefont {Fiedler}, \citenamefont {Forster}, \citenamefont {Gettelman},
  \citenamefont {Haywood}, \citenamefont {Lohmann}, \citenamefont {Malavelle},
  \citenamefont {Mauritsen}, \citenamefont {McCoy}, \citenamefont {Myhre},
  \citenamefont {Muelmenstaedt}, \citenamefont {Neubauer}, \citenamefont
  {Possner}, \citenamefont {Rugenstein}, \citenamefont {Sato}, \citenamefont
  {Schulz}, \citenamefont {Schwartz}, \citenamefont {Sourdeval}, \citenamefont
  {Storelvmo}, \citenamefont {Toll}, \citenamefont {Winker},\ and\
  \citenamefont {Stevens}}]{Bellouin2020}%
  \BibitemOpen
  \bibfield  {author} {\bibinfo {author} {\bibfnamefont {N.}~\bibnamefont
  {Bellouin}}, \bibinfo {author} {\bibfnamefont {J.}~\bibnamefont {Quaas}},
  \bibinfo {author} {\bibfnamefont {E.}~\bibnamefont {Gryspeerdt}}, \bibinfo
  {author} {\bibfnamefont {S.}~\bibnamefont {Kinne}}, \bibinfo {author}
  {\bibfnamefont {P.}~\bibnamefont {Stier}}, \bibinfo {author} {\bibfnamefont
  {D.}~\bibnamefont {Watson-Parris}}, \bibinfo {author} {\bibfnamefont
  {O.}~\bibnamefont {Boucher}}, \bibinfo {author} {\bibfnamefont {K.~S.}\
  \bibnamefont {Carslaw}}, \bibinfo {author} {\bibfnamefont {M.}~\bibnamefont
  {Christensen}}, \bibinfo {author} {\bibfnamefont {A.-L.}\ \bibnamefont
  {Daniau}}, \bibinfo {author} {\bibfnamefont {J.-L.}\ \bibnamefont
  {Dufresne}}, \bibinfo {author} {\bibfnamefont {G.}~\bibnamefont {Feingold}},
  \bibinfo {author} {\bibfnamefont {S.}~\bibnamefont {Fiedler}}, \bibinfo
  {author} {\bibfnamefont {P.}~\bibnamefont {Forster}}, \bibinfo {author}
  {\bibfnamefont {A.}~\bibnamefont {Gettelman}}, \bibinfo {author}
  {\bibfnamefont {J.~M.}\ \bibnamefont {Haywood}}, \bibinfo {author}
  {\bibfnamefont {U.}~\bibnamefont {Lohmann}}, \bibinfo {author} {\bibfnamefont
  {F.}~\bibnamefont {Malavelle}}, \bibinfo {author} {\bibfnamefont
  {T.}~\bibnamefont {Mauritsen}}, \bibinfo {author} {\bibfnamefont {D.~T.}\
  \bibnamefont {McCoy}}, \bibinfo {author} {\bibfnamefont {G.}~\bibnamefont
  {Myhre}}, \bibinfo {author} {\bibfnamefont {J.}~\bibnamefont
  {Muelmenstaedt}}, \bibinfo {author} {\bibfnamefont {D.}~\bibnamefont
  {Neubauer}}, \bibinfo {author} {\bibfnamefont {A.}~\bibnamefont {Possner}},
  \bibinfo {author} {\bibfnamefont {M.}~\bibnamefont {Rugenstein}}, \bibinfo
  {author} {\bibfnamefont {Y.}~\bibnamefont {Sato}}, \bibinfo {author}
  {\bibfnamefont {M.}~\bibnamefont {Schulz}}, \bibinfo {author} {\bibfnamefont
  {S.~E.}\ \bibnamefont {Schwartz}}, \bibinfo {author} {\bibfnamefont
  {O.}~\bibnamefont {Sourdeval}}, \bibinfo {author} {\bibfnamefont
  {T.}~\bibnamefont {Storelvmo}}, \bibinfo {author} {\bibfnamefont
  {V.}~\bibnamefont {Toll}}, \bibinfo {author} {\bibfnamefont {D.}~\bibnamefont
  {Winker}},\ and\ \bibinfo {author} {\bibfnamefont {B.}~\bibnamefont
  {Stevens}},\ }\bibfield  {title} {\bibinfo {title} {Bounding global aerosol
  radiative forcing of climate change},\ }\href
  {https://doi.org/10.1029/2019RG000660} {\bibfield  {journal} {\bibinfo
  {journal} {Reviews of Geophysics}\ }\textbf {\bibinfo {volume} {58}},\
  \bibinfo {pages} {e2019RG000660} (\bibinfo {year} {2020})}\BibitemShut
  {NoStop}%
\bibitem [{\citenamefont {Moum}(2021)}]{Moum2021}%
  \BibitemOpen
  \bibfield  {author} {\bibinfo {author} {\bibfnamefont {J.~N.}\ \bibnamefont
  {Moum}},\ }\bibfield  {title} {\bibinfo {title} {Variations in ocean mixing
  from seconds to years},\ }\href
  {https://doi.org/10.1146/annurev-marine-031920-122846} {\bibfield  {journal}
  {\bibinfo  {journal} {Annual Review of Marine Science}\ }\textbf {\bibinfo
  {volume} {13}},\ \bibinfo {pages} {201} (\bibinfo {year} {2021})}\BibitemShut
  {NoStop}%
\bibitem [{\citenamefont {Chen}\ and\ \citenamefont
  {Doolen}(1998)}]{Chen1998LatticeFlows}%
  \BibitemOpen
  \bibfield  {author} {\bibinfo {author} {\bibfnamefont {S.}~\bibnamefont
  {Chen}}\ and\ \bibinfo {author} {\bibfnamefont {G.~D.}\ \bibnamefont
  {Doolen}},\ }\bibfield  {title} {\bibinfo {title} {{Lattice boltzmann method
  for fluid flows}},\ }\bibfield  {journal} {\bibinfo  {journal} {Annual Review
  of Fluid Mechanics}\ }\textbf {\bibinfo {volume} {30}},\ \href
  {https://doi.org/10.1146/annurev.fluid.30.1.329}
  {10.1146/annurev.fluid.30.1.329} (\bibinfo {year} {1998})\BibitemShut
  {NoStop}%
\bibitem [{\citenamefont {Orszag}\ and\ \citenamefont
  {Yakhot}(1986)}]{Orszag1986}%
  \BibitemOpen
  \bibfield  {author} {\bibinfo {author} {\bibfnamefont {S.}~\bibnamefont
  {Orszag}}\ and\ \bibinfo {author} {\bibfnamefont {V.}~\bibnamefont
  {Yakhot}},\ }\bibfield  {title} {\bibinfo {title} {Reynolds-number scaling of
  cellular-automation hydrodynamics},\ }\href
  {https://doi.org/10.1103/PhysRevLett.56.1691} {\bibfield  {journal} {\bibinfo
   {journal} {Physical Review Letters}\ }\textbf {\bibinfo {volume} {56}},\
  \bibinfo {pages} {1691} (\bibinfo {year} {1986})}\BibitemShut {NoStop}%
\bibitem [{\citenamefont {Harrow}\ \emph {et~al.}(2009)\citenamefont {Harrow},
  \citenamefont {Hassidim},\ and\ \citenamefont {Lloyd}}]{Harrow09}%
  \BibitemOpen
  \bibfield  {author} {\bibinfo {author} {\bibfnamefont {A.~W.}\ \bibnamefont
  {Harrow}}, \bibinfo {author} {\bibfnamefont {A.}~\bibnamefont {Hassidim}},\
  and\ \bibinfo {author} {\bibfnamefont {S.}~\bibnamefont {Lloyd}},\ }\bibfield
   {title} {\bibinfo {title} {Quantum algorithm for linear systems of
  equations},\ }\href {https://doi.org/10.1103/PhysRevLett.103.150502}
  {\bibfield  {journal} {\bibinfo  {journal} {Phys. Rev. Lett.}\ }\textbf
  {\bibinfo {volume} {103}},\ \bibinfo {pages} {150502} (\bibinfo {year}
  {2009})}\BibitemShut {NoStop}%
\bibitem [{\citenamefont {Berry}(2014)}]{berry2014high}%
  \BibitemOpen
  \bibfield  {author} {\bibinfo {author} {\bibfnamefont {D.~W.}\ \bibnamefont
  {Berry}},\ }\bibfield  {title} {\bibinfo {title} {High-order quantum
  algorithm for solving linear differential equations},\ }\href@noop {}
  {\bibfield  {journal} {\bibinfo  {journal} {Journal of Physics A:
  Mathematical and Theoretical}\ }\textbf {\bibinfo {volume} {47}},\ \bibinfo
  {pages} {105301} (\bibinfo {year} {2014})}\BibitemShut {NoStop}%
\bibitem [{\citenamefont {Berry}\ \emph {et~al.}(2017)\citenamefont {Berry},
  \citenamefont {Childs}, \citenamefont {Ostrander},\ and\ \citenamefont
  {Wang}}]{Berry2017QuantumPrecision}%
  \BibitemOpen
  \bibfield  {author} {\bibinfo {author} {\bibfnamefont {D.~W.}\ \bibnamefont
  {Berry}}, \bibinfo {author} {\bibfnamefont {A.~M.}\ \bibnamefont {Childs}},
  \bibinfo {author} {\bibfnamefont {A.}~\bibnamefont {Ostrander}},\ and\
  \bibinfo {author} {\bibfnamefont {G.}~\bibnamefont {Wang}},\ }\bibfield
  {title} {\bibinfo {title} {{Quantum Algorithm for Linear Differential
  Equations with Exponentially Improved Dependence on Precision}},\ }\href
  {https://doi.org/10.1007/S00220-017-3002-Y} {\bibfield  {journal} {\bibinfo
  {journal} {Communications in Mathematical Physics}\ }\textbf {\bibinfo
  {volume} {356}},\ \bibinfo {pages} {1057} (\bibinfo {year}
  {2017})}\BibitemShut {NoStop}%
\bibitem [{\citenamefont {Childs}\ \emph {et~al.}(2017)\citenamefont {Childs},
  \citenamefont {Kothari},\ and\ \citenamefont {Somma}}]{childs2017quantum}%
  \BibitemOpen
  \bibfield  {author} {\bibinfo {author} {\bibfnamefont {A.~M.}\ \bibnamefont
  {Childs}}, \bibinfo {author} {\bibfnamefont {R.}~\bibnamefont {Kothari}},\
  and\ \bibinfo {author} {\bibfnamefont {R.~D.}\ \bibnamefont {Somma}},\
  }\bibfield  {title} {\bibinfo {title} {Quantum algorithm for systems of
  linear equations with exponentially improved dependence on precision},\
  }\href@noop {} {\bibfield  {journal} {\bibinfo  {journal} {SIAM Journal on
  Computing}\ }\textbf {\bibinfo {volume} {46}},\ \bibinfo {pages} {1920}
  (\bibinfo {year} {2017})}\BibitemShut {NoStop}%
\bibitem [{\citenamefont {Leyton}\ and\ \citenamefont
  {Osborne}(2008)}]{leyton2008quantum}%
  \BibitemOpen
  \bibfield  {author} {\bibinfo {author} {\bibfnamefont {S.~K.}\ \bibnamefont
  {Leyton}}\ and\ \bibinfo {author} {\bibfnamefont {T.~J.}\ \bibnamefont
  {Osborne}},\ }\href {https://doi.org/10.48550/arXiv.0812.4423} {\bibinfo
  {title} {A quantum algorithm to solve nonlinear differential equations}}
  (\bibinfo {year} {2008}),\ \Eprint {https://arxiv.org/abs/0812.4423}
  {arXiv:0812.4423 [quant-ph]} \BibitemShut {NoStop}%
\bibitem [{\citenamefont {Lloyd}\ \emph {et~al.}(2020)\citenamefont {Lloyd},
  \citenamefont {De~Palma}, \citenamefont {Gokler}, \citenamefont {Kiani},
  \citenamefont {Liu}, \citenamefont {Marvian}, \citenamefont {Tennie},\ and\
  \citenamefont {Palmer}}]{Lloyd2020QuantumEquations}%
  \BibitemOpen
  \bibfield  {author} {\bibinfo {author} {\bibfnamefont {S.}~\bibnamefont
  {Lloyd}}, \bibinfo {author} {\bibfnamefont {G.}~\bibnamefont {De~Palma}},
  \bibinfo {author} {\bibfnamefont {C.}~\bibnamefont {Gokler}}, \bibinfo
  {author} {\bibfnamefont {B.}~\bibnamefont {Kiani}}, \bibinfo {author}
  {\bibfnamefont {Z.-W.}\ \bibnamefont {Liu}}, \bibinfo {author} {\bibfnamefont
  {M.}~\bibnamefont {Marvian}}, \bibinfo {author} {\bibfnamefont
  {F.}~\bibnamefont {Tennie}},\ and\ \bibinfo {author} {\bibfnamefont
  {T.}~\bibnamefont {Palmer}},\ }\bibfield  {title} {\bibinfo {title} {{Quantum
  algorithm for nonlinear differential equations}}\ }\href
  {https://doi.org/10.48550/arXiv.2011.06571} {10.48550/arXiv.2011.06571}
  (\bibinfo {year} {2020})\BibitemShut {NoStop}%
\bibitem [{\citenamefont {{Liu}}\ \emph {et~al.}(2021)\citenamefont {{Liu}},
  \citenamefont {{Kolden}}, \citenamefont {{Krovi}}, \citenamefont
  {{Loureiro}}, \citenamefont {{Trivisa}},\ and\ \citenamefont
  {{Childs}}}]{Liu21}%
  \BibitemOpen
  \bibfield  {author} {\bibinfo {author} {\bibfnamefont {J.-P.}\ \bibnamefont
  {{Liu}}}, \bibinfo {author} {\bibfnamefont {H.~{\O}.}\ \bibnamefont
  {{Kolden}}}, \bibinfo {author} {\bibfnamefont {H.~K.}\ \bibnamefont
  {{Krovi}}}, \bibinfo {author} {\bibfnamefont {N.~F.}\ \bibnamefont
  {{Loureiro}}}, \bibinfo {author} {\bibfnamefont {K.}~\bibnamefont
  {{Trivisa}}},\ and\ \bibinfo {author} {\bibfnamefont {A.~M.}\ \bibnamefont
  {{Childs}}},\ }\bibfield  {title} {\bibinfo {title} {{Efficient quantum
  algorithm for dissipative nonlinear differential equations}},\ }\href
  {https://doi.org/10.1073/pnas.2026805118} {\bibfield  {journal} {\bibinfo
  {journal} {Proceedings of the National Academy of Science}\ }\textbf
  {\bibinfo {volume} {118}},\ \bibinfo {eid} {e2026805118} (\bibinfo {year}
  {2021})}\BibitemShut {NoStop}%
\bibitem [{Note1()}]{Note1}%
  \BibitemOpen
  \bibinfo {note} {$R$ is a parameter characterizing the ratio of the
  nonlinearity and forcing to the linear dissipation and is analogous to the
  Reynolds number}\BibitemShut {NoStop}%
\bibitem [{\citenamefont {Siebert}\ \emph {et~al.}(2006)\citenamefont
  {Siebert}, \citenamefont {Lehmann},\ and\ \citenamefont
  {Wendisch}}]{Siebert06}%
  \BibitemOpen
  \bibfield  {author} {\bibinfo {author} {\bibfnamefont {H.}~\bibnamefont
  {Siebert}}, \bibinfo {author} {\bibfnamefont {K.}~\bibnamefont {Lehmann}},\
  and\ \bibinfo {author} {\bibfnamefont {M.}~\bibnamefont {Wendisch}},\
  }\bibfield  {title} {\bibinfo {title} {Observations of small-scale turbulence
  and energy dissipation rates in the cloudy boundary layer},\ }\href
  {https://doi.org/10.1175/JAS3687.1} {\bibfield  {journal} {\bibinfo
  {journal} {Journal of the Atmospheric Sciences}\ }\textbf {\bibinfo {volume}
  {63}},\ \bibinfo {pages} {1451 } (\bibinfo {year} {2006})}\BibitemShut
  {NoStop}%
\bibitem [{\citenamefont {Grabowski}\ and\ \citenamefont
  {Wang}(2013)}]{Grabowski2013GrowthEnvironment}%
  \BibitemOpen
  \bibfield  {author} {\bibinfo {author} {\bibfnamefont {W.~W.}\ \bibnamefont
  {Grabowski}}\ and\ \bibinfo {author} {\bibfnamefont {L.~P.}\ \bibnamefont
  {Wang}},\ }\bibfield  {title} {\bibinfo {title} {{Growth of cloud droplets in
  a turbulent environment}},\ }\href
  {https://doi.org/10.1146/annurev-fluid-011212-140750} {\bibfield  {journal}
  {\bibinfo  {journal} {Annual Review of Fluid Mechanics}\ }\textbf {\bibinfo
  {volume} {45}},\ \bibinfo {pages} {293} (\bibinfo {year} {2013})}\BibitemShut
  {NoStop}%
\bibitem [{\citenamefont {Bhatnagar}\ \emph {et~al.}(1954)\citenamefont
  {Bhatnagar}, \citenamefont {Gross},\ and\ \citenamefont
  {Krook}}]{Bhatnagar1954ASystems}%
  \BibitemOpen
  \bibfield  {author} {\bibinfo {author} {\bibfnamefont {P.~L.}\ \bibnamefont
  {Bhatnagar}}, \bibinfo {author} {\bibfnamefont {E.~P.}\ \bibnamefont
  {Gross}},\ and\ \bibinfo {author} {\bibfnamefont {M.}~\bibnamefont {Krook}},\
  }\bibfield  {title} {\bibinfo {title} {{A Model for Collision Processes in
  Gases. I. Small Amplitude Processes in Charged and Neutral One-Component
  Systems}},\ }\href {https://doi.org/10.1103/PhysRev.94.511} {\bibfield
  {journal} {\bibinfo  {journal} {Physical Review}\ }\textbf {\bibinfo {volume}
  {94}},\ \bibinfo {pages} {511} (\bibinfo {year} {1954})}\BibitemShut
  {NoStop}%
\bibitem [{\citenamefont {Reider}\ and\ \citenamefont
  {Sterling}(1995)}]{Sterling95}%
  \BibitemOpen
  \bibfield  {author} {\bibinfo {author} {\bibfnamefont {M.~B.}\ \bibnamefont
  {Reider}}\ and\ \bibinfo {author} {\bibfnamefont {J.~D.}\ \bibnamefont
  {Sterling}},\ }\bibfield  {title} {\bibinfo {title} {Accuracy of
  discrete-velocity bgk models for the simulation of the incompressible
  navier-stokes equations},\ }\href
  {https://doi.org/https://doi.org/10.1016/0045-7930(94)00037-Y} {\bibfield
  {journal} {\bibinfo  {journal} {Computers \& Fluids}\ }\textbf {\bibinfo
  {volume} {24}},\ \bibinfo {pages} {459} (\bibinfo {year} {1995})}\BibitemShut
  {NoStop}%
\bibitem [{\citenamefont {Toschi}\ and\ \citenamefont
  {Succi}(2005)}]{Toschi05}%
  \BibitemOpen
  \bibfield  {author} {\bibinfo {author} {\bibfnamefont {F.}~\bibnamefont
  {Toschi}}\ and\ \bibinfo {author} {\bibfnamefont {S.}~\bibnamefont {Succi}},\
  }\bibfield  {title} {\bibinfo {title} {Lattice boltzmann method at finite
  knudsen numbers},\ }\href {https://doi.org/10.1209/epl/i2004-10393-0}
  {\bibfield  {journal} {\bibinfo  {journal} {Europhysics Letters}\ }\textbf
  {\bibinfo {volume} {69}},\ \bibinfo {pages} {549} (\bibinfo {year}
  {2005})}\BibitemShut {NoStop}%
\bibitem [{\citenamefont {Forets}\ and\ \citenamefont
  {Pouly}(2017)}]{Forets2017}%
  \BibitemOpen
  \bibfield  {author} {\bibinfo {author} {\bibfnamefont {M.}~\bibnamefont
  {Forets}}\ and\ \bibinfo {author} {\bibfnamefont {A.}~\bibnamefont {Pouly}},\
  }\bibfield  {title} {\bibinfo {title} {Explicit error bounds for carleman
  linearization}\ }\href {https://doi.org/10.48550/ARXIV.1711.02552}
  {10.48550/ARXIV.1711.02552} (\bibinfo {year} {2017})\BibitemShut {NoStop}%
\bibitem [{\citenamefont {Carleman}(1932)}]{Carleman1932}%
  \BibitemOpen
  \bibfield  {author} {\bibinfo {author} {\bibfnamefont {T.}~\bibnamefont
  {Carleman}},\ }\bibfield  {title} {\bibinfo {title} {Application of the
  theory of linear integration equations to nonlinear systems of differential
  equations},\ }\href {https://doi.org/10.1007/BF02546499} {\bibfield
  {journal} {\bibinfo  {journal} {Acta Mathematica}\ }\textbf {\bibinfo
  {volume} {59}},\ \bibinfo {pages} {63} (\bibinfo {year} {1932})}\BibitemShut
  {NoStop}%
\bibitem [{\citenamefont {Sterling}\ and\ \citenamefont
  {Chen}(1996)}]{Sterling1996StabilityMethods}%
  \BibitemOpen
  \bibfield  {author} {\bibinfo {author} {\bibfnamefont {J.~D.}\ \bibnamefont
  {Sterling}}\ and\ \bibinfo {author} {\bibfnamefont {S.}~\bibnamefont
  {Chen}},\ }\bibfield  {title} {\bibinfo {title} {{Stability Analysis of
  Lattice Boltzmann Methods}},\ }\href {https://doi.org/10.1006/JCPH.1996.0016}
  {\bibfield  {journal} {\bibinfo  {journal} {Journal of Computational
  Physics}\ }\textbf {\bibinfo {volume} {123}},\ \bibinfo {pages} {196}
  (\bibinfo {year} {1996})}\BibitemShut {NoStop}%
\bibitem [{\citenamefont {Saffman}(1967)}]{Saffman67}%
  \BibitemOpen
  \bibfield  {author} {\bibinfo {author} {\bibfnamefont {P.~G.}\ \bibnamefont
  {Saffman}},\ }\bibfield  {title} {\bibinfo {title} {{Note on Decay of
  Homogeneous Turbulence}},\ }\href {https://doi.org/10.1063/1.1762284}
  {\bibfield  {journal} {\bibinfo  {journal} {The Physics of Fluids}\ }\textbf
  {\bibinfo {volume} {10}},\ \bibinfo {pages} {1349} (\bibinfo {year}
  {1967})},\ \Eprint
  {https://arxiv.org/abs/https://pubs.aip.org/aip/pfl/article-pdf/10/6/1349/12566054/1349\_1\_online.pdf}
  {https://pubs.aip.org/aip/pfl/article-pdf/10/6/1349/12566054/1349\_1\_online.pdf}
  \BibitemShut {NoStop}%
\bibitem [{\citenamefont {Skrbek}\ and\ \citenamefont
  {Stalp}(2000)}]{Skrbek20}%
  \BibitemOpen
  \bibfield  {author} {\bibinfo {author} {\bibfnamefont {L.}~\bibnamefont
  {Skrbek}}\ and\ \bibinfo {author} {\bibfnamefont {S.~R.}\ \bibnamefont
  {Stalp}},\ }\bibfield  {title} {\bibinfo {title} {{On the decay of
  homogeneous isotropic turbulence}},\ }\href
  {https://doi.org/10.1063/1.870447} {\bibfield  {journal} {\bibinfo  {journal}
  {Physics of Fluids}\ }\textbf {\bibinfo {volume} {12}},\ \bibinfo {pages}
  {1997} (\bibinfo {year} {2000})},\ \Eprint
  {https://arxiv.org/abs/https://pubs.aip.org/aip/pof/article-pdf/12/8/1997/12696548/1997\_1\_online.pdf}
  {https://pubs.aip.org/aip/pof/article-pdf/12/8/1997/12696548/1997\_1\_online.pdf}
  \BibitemShut {NoStop}%
\bibitem [{\citenamefont {Li}\ and\ \citenamefont {Mattsson}(2020)}]{Li_2020}%
  \BibitemOpen
  \bibfield  {author} {\bibinfo {author} {\bibfnamefont {X.-Y.}\ \bibnamefont
  {Li}}\ and\ \bibinfo {author} {\bibfnamefont {L.}~\bibnamefont {Mattsson}},\
  }\bibfield  {title} {\bibinfo {title} {Dust growth by accretion of molecules
  in supersonic interstellar turbulence},\ }\href
  {https://doi.org/10.3847/1538-4357/abb9ad} {\bibfield  {journal} {\bibinfo
  {journal} {The Astrophysical Journal}\ }\textbf {\bibinfo {volume} {903}},\
  \bibinfo {pages} {148} (\bibinfo {year} {2020})}\BibitemShut {NoStop}%
\bibitem [{\citenamefont {Krovi}(2023)}]{Krovi2023improvedquantum}%
  \BibitemOpen
  \bibfield  {author} {\bibinfo {author} {\bibfnamefont {H.}~\bibnamefont
  {Krovi}},\ }\bibfield  {title} {\bibinfo {title} {Improved quantum algorithms
  for linear and nonlinear differential equations},\ }\href
  {https://doi.org/10.22331/q-2023-02-02-913} {\bibfield  {journal} {\bibinfo
  {journal} {{Quantum}}\ }\textbf {\bibinfo {volume} {7}},\ \bibinfo {pages}
  {913} (\bibinfo {year} {2023})}\BibitemShut {NoStop}%
\bibitem [{url(2023)}]{url}%
  \BibitemOpen
  \href {https://doi.org/10.5281/zenodo.7776286} {\bibinfo {title} {Code
  location for {LBM}}} (\bibinfo {year} {2023})\BibitemShut {NoStop}%
\bibitem [{\citenamefont {Sakurai}(1994)}]{Sakurai1994}%
  \BibitemOpen
  \bibfield  {author} {\bibinfo {author} {\bibfnamefont {J.~J.}\ \bibnamefont
  {Sakurai}},\ }\href@noop {} {\emph {\bibinfo {title} {{Modern quantum
  mechanics; rev. ed.}}}}\ (\bibinfo  {publisher} {Addison-Wesley},\ \bibinfo
  {address} {Reading, MA},\ \bibinfo {year} {1994})\BibitemShut {NoStop}%
\bibitem [{\citenamefont {Bamieh}(2020)}]{bamieh2020tutorial}%
  \BibitemOpen
  \bibfield  {author} {\bibinfo {author} {\bibfnamefont {B.}~\bibnamefont
  {Bamieh}},\ }\bibfield  {title} {\bibinfo {title} {A tutorial on matrix
  perturbation theory (using compact matrix notation)},\ }\href@noop {}
  {\bibfield  {journal} {\bibinfo  {journal} {arXiv preprint arXiv:2002.05001}\
  } (\bibinfo {year} {2020})}\BibitemShut {NoStop}%
\bibitem [{\citenamefont {Lallemand}\ \emph {et~al.}(2021)\citenamefont
  {Lallemand}, \citenamefont {Luo}, \citenamefont {Krafczyk},\ and\
  \citenamefont {Yong}}]{LALLEMAND2021109713}%
  \BibitemOpen
  \bibfield  {author} {\bibinfo {author} {\bibfnamefont {P.}~\bibnamefont
  {Lallemand}}, \bibinfo {author} {\bibfnamefont {L.-S.}\ \bibnamefont {Luo}},
  \bibinfo {author} {\bibfnamefont {M.}~\bibnamefont {Krafczyk}},\ and\
  \bibinfo {author} {\bibfnamefont {W.-A.}\ \bibnamefont {Yong}},\ }\bibfield
  {title} {\bibinfo {title} {The lattice boltzmann method for nearly
  incompressible flows},\ }\href {https://doi.org/10.1016/j.jcp.2020.109713}
  {\bibfield  {journal} {\bibinfo  {journal} {Journal of Computational
  Physics}\ }\textbf {\bibinfo {volume} {431}},\ \bibinfo {pages} {109713}
  (\bibinfo {year} {2021})}\BibitemShut {NoStop}%
\bibitem [{\citenamefont {Aaronson}(2015)}]{Aaronson2015}%
  \BibitemOpen
  \bibfield  {author} {\bibinfo {author} {\bibfnamefont {S.}~\bibnamefont
  {Aaronson}},\ }\bibfield  {title} {\bibinfo {title} {Read the fine print},\
  }\href {https://doi.org/10.1038/nphys3272} {\bibfield  {journal} {\bibinfo
  {journal} {Nature Physics}\ }\textbf {\bibinfo {volume} {11}},\ \bibinfo
  {pages} {291} (\bibinfo {year} {2015})}\BibitemShut {NoStop}%
\bibitem [{\citenamefont {Kitaev}\ and\ \citenamefont
  {Webb}(2008)}]{kitaev2008wavefunction}%
  \BibitemOpen
  \bibfield  {author} {\bibinfo {author} {\bibfnamefont {A.}~\bibnamefont
  {Kitaev}}\ and\ \bibinfo {author} {\bibfnamefont {W.~A.}\ \bibnamefont
  {Webb}},\ }\bibfield  {title} {\bibinfo {title} {Wavefunction preparation and
  resampling using a quantum computer},\ }\href@noop {} {\bibfield  {journal}
  {\bibinfo  {journal} {arXiv preprint arXiv:0801.0342}\ } (\bibinfo {year}
  {2008})}\BibitemShut {NoStop}%
\bibitem [{\citenamefont {Grover}\ and\ \citenamefont
  {Rudolph}(2002)}]{grover2002creating}%
  \BibitemOpen
  \bibfield  {author} {\bibinfo {author} {\bibfnamefont {L.}~\bibnamefont
  {Grover}}\ and\ \bibinfo {author} {\bibfnamefont {T.}~\bibnamefont
  {Rudolph}},\ }\bibfield  {title} {\bibinfo {title} {Creating superpositions
  that correspond to efficiently integrable probability distributions},\
  }\href@noop {} {\bibfield  {journal} {\bibinfo  {journal} {arXiv preprint
  quant-ph/0208112}\ } (\bibinfo {year} {2002})}\BibitemShut {NoStop}%
\bibitem [{\citenamefont {Cramer}\ \emph {et~al.}(2010)\citenamefont {Cramer},
  \citenamefont {Plenio}, \citenamefont {Flammia}, \citenamefont {Somma},
  \citenamefont {Gross}, \citenamefont {Bartlett}, \citenamefont
  {Landon-Cardinal}, \citenamefont {Poulin},\ and\ \citenamefont
  {Liu}}]{cramer2010efficient}%
  \BibitemOpen
  \bibfield  {author} {\bibinfo {author} {\bibfnamefont {M.}~\bibnamefont
  {Cramer}}, \bibinfo {author} {\bibfnamefont {M.~B.}\ \bibnamefont {Plenio}},
  \bibinfo {author} {\bibfnamefont {S.~T.}\ \bibnamefont {Flammia}}, \bibinfo
  {author} {\bibfnamefont {R.}~\bibnamefont {Somma}}, \bibinfo {author}
  {\bibfnamefont {D.}~\bibnamefont {Gross}}, \bibinfo {author} {\bibfnamefont
  {S.~D.}\ \bibnamefont {Bartlett}}, \bibinfo {author} {\bibfnamefont
  {O.}~\bibnamefont {Landon-Cardinal}}, \bibinfo {author} {\bibfnamefont
  {D.}~\bibnamefont {Poulin}},\ and\ \bibinfo {author} {\bibfnamefont {Y.-K.}\
  \bibnamefont {Liu}},\ }\bibfield  {title} {\bibinfo {title} {Efficient
  quantum state tomography},\ }\href@noop {} {\bibfield  {journal} {\bibinfo
  {journal} {Nature communications}\ }\textbf {\bibinfo {volume} {1}},\
  \bibinfo {pages} {1} (\bibinfo {year} {2010})}\BibitemShut {NoStop}%
\bibitem [{\citenamefont {Orszag}(1970)}]{Orszag1970}%
  \BibitemOpen
  \bibfield  {author} {\bibinfo {author} {\bibfnamefont {S.}~\bibnamefont
  {Orszag}},\ }\bibfield  {title} {\bibinfo {title} {Analytical theories of
  turbulence},\ }\href {https://doi.org/10.1017/S0022112070000642} {\bibfield
  {journal} {\bibinfo  {journal} {Journal of Fluid Mechanics}\ }\textbf
  {\bibinfo {volume} {41}},\ \bibinfo {pages} {363} (\bibinfo {year}
  {1970})}\BibitemShut {NoStop}%
\bibitem [{\citenamefont {Tang}(2021)}]{Tang21}%
  \BibitemOpen
  \bibfield  {author} {\bibinfo {author} {\bibfnamefont {E.}~\bibnamefont
  {Tang}},\ }\bibfield  {title} {\bibinfo {title} {Quantum principal component
  analysis only achieves an exponential speedup because of its state
  preparation assumptions},\ }\href
  {https://doi.org/10.1103/PhysRevLett.127.060503} {\bibfield  {journal}
  {\bibinfo  {journal} {Phys. Rev. Lett.}\ }\textbf {\bibinfo {volume} {127}},\
  \bibinfo {pages} {060503} (\bibinfo {year} {2021})}\BibitemShut {NoStop}%
\bibitem [{\citenamefont {Zou}\ and\ \citenamefont {He}(1997)}]{Zou97}%
  \BibitemOpen
  \bibfield  {author} {\bibinfo {author} {\bibfnamefont {Q.}~\bibnamefont
  {Zou}}\ and\ \bibinfo {author} {\bibfnamefont {X.}~\bibnamefont {He}},\
  }\bibfield  {title} {\bibinfo {title} {On pressure and velocity boundary
  conditions for the lattice boltzmann bgk model},\ }\href
  {https://doi.org/10.1063/1.869307} {\bibfield  {journal} {\bibinfo  {journal}
  {Physics of Fluids}\ }\textbf {\bibinfo {volume} {9}},\ \bibinfo {pages}
  {1591} (\bibinfo {year} {1997})}\BibitemShut {NoStop}%
\bibitem [{\citenamefont {Ladd}\ and\ \citenamefont
  {Verberg}(2001)}]{Ladd2001}%
  \BibitemOpen
  \bibfield  {author} {\bibinfo {author} {\bibfnamefont {A.~J.~C.}\
  \bibnamefont {Ladd}}\ and\ \bibinfo {author} {\bibfnamefont {R.}~\bibnamefont
  {Verberg}},\ }\bibfield  {title} {\bibinfo {title} {Lattice-boltzmann
  simulations of particle-fluid suspensions},\ }\href
  {https://doi.org/10.1023/A:1010414013942} {\bibfield  {journal} {\bibinfo
  {journal} {Journal of Statistical Physics}\ }\textbf {\bibinfo {volume}
  {104}},\ \bibinfo {pages} {1191} (\bibinfo {year} {2001})}\BibitemShut
  {NoStop}%
\bibitem [{\citenamefont {Guo}\ \emph {et~al.}(2002)\citenamefont {Guo},
  \citenamefont {Zheng},\ and\ \citenamefont {Shi}}]{Guo02}%
  \BibitemOpen
  \bibfield  {author} {\bibinfo {author} {\bibfnamefont {Z.}~\bibnamefont
  {Guo}}, \bibinfo {author} {\bibfnamefont {C.}~\bibnamefont {Zheng}},\ and\
  \bibinfo {author} {\bibfnamefont {B.}~\bibnamefont {Shi}},\ }\bibfield
  {title} {\bibinfo {title} {Discrete lattice effects on the forcing term in
  the lattice boltzmann method},\ }\href
  {https://doi.org/10.1103/PhysRevE.65.046308} {\bibfield  {journal} {\bibinfo
  {journal} {Phys. Rev. E}\ }\textbf {\bibinfo {volume} {65}},\ \bibinfo
  {pages} {046308} (\bibinfo {year} {2002})}\BibitemShut {NoStop}%
\bibitem [{\citenamefont {Chun}\ \emph {et~al.}(2005)\citenamefont {Chun},
  \citenamefont {Koch}, \citenamefont {Rani}, \citenamefont {Ahluwalia},\ and\
  \citenamefont {Collins}}]{Chun2005}%
  \BibitemOpen
  \bibfield  {author} {\bibinfo {author} {\bibfnamefont {J.}~\bibnamefont
  {Chun}}, \bibinfo {author} {\bibfnamefont {D.~L.}\ \bibnamefont {Koch}},
  \bibinfo {author} {\bibfnamefont {S.~L.}\ \bibnamefont {Rani}}, \bibinfo
  {author} {\bibfnamefont {A.}~\bibnamefont {Ahluwalia}},\ and\ \bibinfo
  {author} {\bibfnamefont {L.~R.}\ \bibnamefont {Collins}},\ }\bibfield
  {title} {\bibinfo {title} {Clustering of aerosol particles in isotropic
  turbulence},\ }\href
  {https://doi.org/https://doi.org/10.1017/S0022112005004568} {\bibfield
  {journal} {\bibinfo  {journal} {Journal of Fluid Mechanics}\ }\textbf
  {\bibinfo {volume} {536}},\ \bibinfo {pages} {219} (\bibinfo {year}
  {2005})}\BibitemShut {NoStop}%
\bibitem [{\citenamefont {Li}\ \emph {et~al.}(2018)\citenamefont {Li},
  \citenamefont {Brandenburg}, \citenamefont {Svensson}, \citenamefont
  {Haugen}, \citenamefont {Mehlig},\ and\ \citenamefont
  {Rogachevskii}}]{Li2018EffectDroplets}%
  \BibitemOpen
  \bibfield  {author} {\bibinfo {author} {\bibfnamefont {X.~Y.}\ \bibnamefont
  {Li}}, \bibinfo {author} {\bibfnamefont {A.}~\bibnamefont {Brandenburg}},
  \bibinfo {author} {\bibfnamefont {G.}~\bibnamefont {Svensson}}, \bibinfo
  {author} {\bibfnamefont {N.~E.}\ \bibnamefont {Haugen}}, \bibinfo {author}
  {\bibfnamefont {B.}~\bibnamefont {Mehlig}},\ and\ \bibinfo {author}
  {\bibfnamefont {I.}~\bibnamefont {Rogachevskii}},\ }\bibfield  {title}
  {\bibinfo {title} {{Effect of turbulence on collisional growth of cloud
  droplets}},\ }\bibfield  {journal} {\bibinfo  {journal} {Journal of the
  Atmospheric Sciences}\ }\textbf {\bibinfo {volume} {75}},\ \href
  {https://doi.org/10.1175/JAS-D-18-0081.1} {10.1175/JAS-D-18-0081.1} (\bibinfo
  {year} {2018})\BibitemShut {NoStop}%
\bibitem [{\citenamefont {Saviile}(1977)}]{Saville1977}%
  \BibitemOpen
  \bibfield  {author} {\bibinfo {author} {\bibfnamefont {D.~A.}\ \bibnamefont
  {Saviile}},\ }\bibfield  {title} {\bibinfo {title} {Electrokinetic effects
  with small particles},\ }\href
  {https://doi.org/https://doi.org/10.1146/annurev.fl.09.010177.001541}
  {\bibfield  {journal} {\bibinfo  {journal} {Annual review of fluid
  mechanics}\ }\textbf {\bibinfo {volume} {9}},\ \bibinfo {pages} {321}
  (\bibinfo {year} {1977})}\BibitemShut {NoStop}%
\bibitem [{\citenamefont {Martys}\ and\ \citenamefont
  {Douglas}(2001)}]{Martys01}%
  \BibitemOpen
  \bibfield  {author} {\bibinfo {author} {\bibfnamefont {N.~S.}\ \bibnamefont
  {Martys}}\ and\ \bibinfo {author} {\bibfnamefont {J.~F.}\ \bibnamefont
  {Douglas}},\ }\bibfield  {title} {\bibinfo {title} {Critical properties and
  phase separation in lattice boltzmann fluid mixtures},\ }\href
  {https://doi.org/10.1103/PhysRevE.63.031205} {\bibfield  {journal} {\bibinfo
  {journal} {Phys. Rev. E}\ }\textbf {\bibinfo {volume} {63}},\ \bibinfo
  {pages} {031205} (\bibinfo {year} {2001})}\BibitemShut {NoStop}%
\bibitem [{\citenamefont {Chen}\ \emph {et~al.}(2014)\citenamefont {Chen},
  \citenamefont {Kang}, \citenamefont {Mu}, \citenamefont {He},\ and\
  \citenamefont {Tao}}]{CHEN2014210}%
  \BibitemOpen
  \bibfield  {author} {\bibinfo {author} {\bibfnamefont {L.}~\bibnamefont
  {Chen}}, \bibinfo {author} {\bibfnamefont {Q.}~\bibnamefont {Kang}}, \bibinfo
  {author} {\bibfnamefont {Y.}~\bibnamefont {Mu}}, \bibinfo {author}
  {\bibfnamefont {Y.-L.}\ \bibnamefont {He}},\ and\ \bibinfo {author}
  {\bibfnamefont {W.-Q.}\ \bibnamefont {Tao}},\ }\bibfield  {title} {\bibinfo
  {title} {A critical review of the pseudopotential multiphase lattice
  boltzmann model: Methods and applications},\ }\href
  {https://doi.org/https://doi.org/10.1016/j.ijheatmasstransfer.2014.04.032}
  {\bibfield  {journal} {\bibinfo  {journal} {International Journal of Heat and
  Mass Transfer}\ }\textbf {\bibinfo {volume} {76}},\ \bibinfo {pages} {210}
  (\bibinfo {year} {2014})}\BibitemShut {NoStop}%
\bibitem [{\citenamefont {Li}\ \emph {et~al.}(2016)\citenamefont {Li},
  \citenamefont {Luo}, \citenamefont {Kang}, \citenamefont {He}, \citenamefont
  {Chen},\ and\ \citenamefont {Liu}}]{LI201662}%
  \BibitemOpen
  \bibfield  {author} {\bibinfo {author} {\bibfnamefont {Q.}~\bibnamefont
  {Li}}, \bibinfo {author} {\bibfnamefont {K.}~\bibnamefont {Luo}}, \bibinfo
  {author} {\bibfnamefont {Q.}~\bibnamefont {Kang}}, \bibinfo {author}
  {\bibfnamefont {Y.}~\bibnamefont {He}}, \bibinfo {author} {\bibfnamefont
  {Q.}~\bibnamefont {Chen}},\ and\ \bibinfo {author} {\bibfnamefont
  {Q.}~\bibnamefont {Liu}},\ }\bibfield  {title} {\bibinfo {title} {Lattice
  boltzmann methods for multiphase flow and phase-change heat transfer},\
  }\href {https://doi.org/https://doi.org/10.1016/j.pecs.2015.10.001}
  {\bibfield  {journal} {\bibinfo  {journal} {Progress in Energy and Combustion
  Science}\ }\textbf {\bibinfo {volume} {52}},\ \bibinfo {pages} {62} (\bibinfo
  {year} {2016})}\BibitemShut {NoStop}%
\bibitem [{\citenamefont {Liu}\ \emph {et~al.}(2016)\citenamefont {Liu},
  \citenamefont {Kang}, \citenamefont {Leonardi}, \citenamefont {Schmieschek},
  \citenamefont {Narv{\'a}ez}, \citenamefont {Jones}, \citenamefont {Williams},
  \citenamefont {Valocchi},\ and\ \citenamefont {Harting}}]{Liu16}%
  \BibitemOpen
  \bibfield  {author} {\bibinfo {author} {\bibfnamefont {H.}~\bibnamefont
  {Liu}}, \bibinfo {author} {\bibfnamefont {Q.}~\bibnamefont {Kang}}, \bibinfo
  {author} {\bibfnamefont {C.~R.}\ \bibnamefont {Leonardi}}, \bibinfo {author}
  {\bibfnamefont {S.}~\bibnamefont {Schmieschek}}, \bibinfo {author}
  {\bibfnamefont {A.}~\bibnamefont {Narv{\'a}ez}}, \bibinfo {author}
  {\bibfnamefont {B.~D.}\ \bibnamefont {Jones}}, \bibinfo {author}
  {\bibfnamefont {J.~R.}\ \bibnamefont {Williams}}, \bibinfo {author}
  {\bibfnamefont {A.~J.}\ \bibnamefont {Valocchi}},\ and\ \bibinfo {author}
  {\bibfnamefont {J.}~\bibnamefont {Harting}},\ }\bibfield  {title} {\bibinfo
  {title} {Multiphase lattice boltzmann simulations for porous media
  applications},\ }\href {https://doi.org/10.1007/s10596-015-9542-3} {\bibfield
   {journal} {\bibinfo  {journal} {Computational Geosciences}\ }\textbf
  {\bibinfo {volume} {20}},\ \bibinfo {pages} {777} (\bibinfo {year}
  {2016})}\BibitemShut {NoStop}%
\bibitem [{\citenamefont {He}\ \emph {et~al.}(2019)\citenamefont {He},
  \citenamefont {Liu}, \citenamefont {Li},\ and\ \citenamefont
  {Tao}}]{HE2019160}%
  \BibitemOpen
  \bibfield  {author} {\bibinfo {author} {\bibfnamefont {Y.-L.}\ \bibnamefont
  {He}}, \bibinfo {author} {\bibfnamefont {Q.}~\bibnamefont {Liu}}, \bibinfo
  {author} {\bibfnamefont {Q.}~\bibnamefont {Li}},\ and\ \bibinfo {author}
  {\bibfnamefont {W.-Q.}\ \bibnamefont {Tao}},\ }\bibfield  {title} {\bibinfo
  {title} {Lattice boltzmann methods for single-phase and solid-liquid
  phase-change heat transfer in porous media: A review},\ }\href
  {https://doi.org/https://doi.org/10.1016/j.ijheatmasstransfer.2018.08.135}
  {\bibfield  {journal} {\bibinfo  {journal} {International Journal of Heat and
  Mass Transfer}\ }\textbf {\bibinfo {volume} {129}},\ \bibinfo {pages} {160}
  (\bibinfo {year} {2019})}\BibitemShut {NoStop}%
\bibitem [{\citenamefont {Petersen}\ and\ \citenamefont
  {Brinkerhoff}(2021)}]{Petersen21}%
  \BibitemOpen
  \bibfield  {author} {\bibinfo {author} {\bibfnamefont {K.~J.}\ \bibnamefont
  {Petersen}}\ and\ \bibinfo {author} {\bibfnamefont {J.~R.}\ \bibnamefont
  {Brinkerhoff}},\ }\bibfield  {title} {\bibinfo {title} {On the lattice
  boltzmann method and its application to turbulent, multiphase flows of
  various fluids including cryogens: A review},\ }\href
  {https://doi.org/10.1063/5.0046938} {\bibfield  {journal} {\bibinfo
  {journal} {Physics of Fluids}\ }\textbf {\bibinfo {volume} {33}},\ \bibinfo
  {pages} {041302} (\bibinfo {year} {2021})}\BibitemShut {NoStop}%
\bibitem [{\citenamefont {Samanta}\ \emph {et~al.}(2022)\citenamefont
  {Samanta}, \citenamefont {Chattopadhyay},\ and\ \citenamefont
  {Guha}}]{SAMANTA2022111288}%
  \BibitemOpen
  \bibfield  {author} {\bibinfo {author} {\bibfnamefont {R.}~\bibnamefont
  {Samanta}}, \bibinfo {author} {\bibfnamefont {H.}~\bibnamefont
  {Chattopadhyay}},\ and\ \bibinfo {author} {\bibfnamefont {C.}~\bibnamefont
  {Guha}},\ }\bibfield  {title} {\bibinfo {title} {A review on the application
  of lattice boltzmann method for melting and solidification problems},\ }\href
  {https://doi.org/https://doi.org/10.1016/j.commatsci.2022.111288} {\bibfield
  {journal} {\bibinfo  {journal} {Computational Materials Science}\ }\textbf
  {\bibinfo {volume} {206}},\ \bibinfo {pages} {111288} (\bibinfo {year}
  {2022})}\BibitemShut {NoStop}%
\bibitem [{\citenamefont {Nourgaliev}\ \emph {et~al.}(2003)\citenamefont
  {Nourgaliev}, \citenamefont {Dinh}, \citenamefont {Theofanous},\ and\
  \citenamefont {Joseph}}]{NOURGALIEV2003117}%
  \BibitemOpen
  \bibfield  {author} {\bibinfo {author} {\bibfnamefont {R.}~\bibnamefont
  {Nourgaliev}}, \bibinfo {author} {\bibfnamefont {T.}~\bibnamefont {Dinh}},
  \bibinfo {author} {\bibfnamefont {T.}~\bibnamefont {Theofanous}},\ and\
  \bibinfo {author} {\bibfnamefont {D.}~\bibnamefont {Joseph}},\ }\bibfield
  {title} {\bibinfo {title} {The lattice boltzmann equation method: theoretical
  interpretation, numerics and implications},\ }\href
  {https://doi.org/https://doi.org/10.1016/S0301-9322(02)00108-8} {\bibfield
  {journal} {\bibinfo  {journal} {International Journal of Multiphase Flow}\
  }\textbf {\bibinfo {volume} {29}},\ \bibinfo {pages} {117} (\bibinfo {year}
  {2003})}\BibitemShut {NoStop}%
\bibitem [{\citenamefont {Aidun}\ and\ \citenamefont
  {Clausen}(2010)}]{Aidun10}%
  \BibitemOpen
  \bibfield  {author} {\bibinfo {author} {\bibfnamefont {C.~K.}\ \bibnamefont
  {Aidun}}\ and\ \bibinfo {author} {\bibfnamefont {J.~R.}\ \bibnamefont
  {Clausen}},\ }\bibfield  {title} {\bibinfo {title} {Lattice-boltzmann method
  for complex flows},\ }\href
  {https://doi.org/10.1146/annurev-fluid-121108-145519} {\bibfield  {journal}
  {\bibinfo  {journal} {Annual Review of Fluid Mechanics}\ }\textbf {\bibinfo
  {volume} {42}},\ \bibinfo {pages} {439} (\bibinfo {year} {2010})}\BibitemShut
  {NoStop}%
\bibitem [{\citenamefont {Dahlquist}(1963)}]{Dahlquist63}%
  \BibitemOpen
  \bibfield  {author} {\bibinfo {author} {\bibfnamefont {G.~G.}\ \bibnamefont
  {Dahlquist}},\ }\bibfield  {title} {\bibinfo {title} {A special stability
  problem for linear multistep methods},\ }\href
  {https://doi.org/10.1007/BF01963532} {\bibfield  {journal} {\bibinfo
  {journal} {BIT Numerical Mathematics}\ }\textbf {\bibinfo {volume} {3}},\
  \bibinfo {pages} {27} (\bibinfo {year} {1963})}\BibitemShut {NoStop}%
\end{thebibliography}
\end{document}